\documentclass[a4paper,11pt]{article}
\usepackage[english]{babel}
\usepackage{jheppub}
\pdfoutput=1
\usepackage[table]{xcolor}
\usepackage[T1]{fontenc}
\usepackage{bbold}
\usepackage{float}
\usepackage{amsmath}
\usepackage{empheq}
\usepackage{booktabs}
\usepackage{color}
\usepackage[utf8]{inputenc}
\usepackage{xspace}
\usepackage{scalerel}
\usepackage[most]{tcolorbox}
\usepackage{multirow}
\usepackage{scalefnt}
\usepackage{bold-extra}
\usepackage[shortlabels]{enumitem}
\usepackage[tikz]{bclogo}
\usepackage{subcaption}
\usepackage{cancel}
\usepackage{verbatim}
\usetikzlibrary{arrows,shapes}
\usepackage{afterpage}
\usepackage{graphicx,grffile}
\usepackage{xspace}
\usepackage[normalem]{ulem}
\usepackage{makecell}
\usepackage{tabularx,colortbl}

\newcommand{\br}[1]{{\mathrm{BR}(#1)}}

\usepackage{xcolor}

\newtcolorbox{empheqboxed}{colback=white!35,
 colframe=black,
 width=\textwidth,
 sharpish corners,
 top=-2mm, 
 bottom=0pt
}

\title{Rare $b$ decays meet high-mass Drell-Yan}

\author[a,b]{Admir Greljo,}
\author[a,b]{Jakub Salko,}
\author[a]{Aleks Smolkovi\v c,}
\author[a,c]{Peter Stangl}

\emailAdd{greljo@itp.unibe.ch}
\emailAdd{salko@itp.unibe.ch}
\emailAdd{smolkovic@itp.unibe.ch}
\emailAdd{stangl@itp.unibe.ch}

\affiliation[a]{Albert Einstein Center for Fundamental Physics, Institut f\"{u}r Theoretische Physik, Universit\"{a}t Bern, Sidlerstrasse 5, CH-3012 Bern, Switzerland.}

\affiliation[b]{Department of Physics, University of Basel, Klingelbergstrasse 82,  CH-4056 Basel, Switzerland}

\affiliation[c]{CERN, Theoretical Physics Department, CH-1211 Geneva 23, Switzerland}

\abstract{
Rare $b$ hadron decays are considered excellent probes of new semileptonic four-fermion interactions of microscopic origin. However, the same interactions also correct the high-mass Drell-Yan tails. In this work, we revisit the first statement in the context of this complementarity and chart the space of short-distance new physics that could show up in rare $b$ decays.
We analyze the latest $b \to q \ell^+ \ell^-$ measurements, where $q = d$ or $s$ and $\ell = e$ or $\mu$, including the most recent LHCb $R_{K^{(*)}}$ update, together with the latest charged and neutral current high-mass Drell-Yan data, $p p \to \ell \nu$ and $p p \to \ell^+ \ell^-$. We implement a sophisticated interpretation pipeline within the {\tt flavio} framework, allowing us to investigate the multidimensional SMEFT parameter space thoroughly and efficiently. To showcase the new functionalities of {\tt flavio}, we construct several explicit models featuring either a $Z'$ or a leptoquark, which can explain the tension in $b \to s \mu^+ \mu^-$ angular distributions and branching fractions while predicting lepton flavor universality (LFU) ratios to be SM-like, $R_{K^{(*)}} \approx R^{{\rm SM}}_{K^{(*)}}$, as indicated by the recent data. Those models are then confronted against the global likelihood, including the high-mass Drell-Yan, either finding tensions or compatibility.
}


\preprint{CERN-TH-2023-037}

\begin{document}

\maketitle

\newpage

\section{Introduction}
\label{sec:intro}

Rare $b$ hadron decays to light leptons, with underlying quark level transition $b \to q \ell^+ \ell^-$ where $q=d,s$ and $\ell = e, \mu$, are generated at the one loop in the Standard Model (SM) and further suppressed by the small quark mixing parameters. They serve as excellent tests of the theory, both QCD and electroweak, and are also considered sensitive probes of new physics (NP). These decays could, for example, be generated by tree-level exchanges of leptoquarks~\cite{Dorsner:2016wpm}, which contribute to four-quark and four-lepton transitions only at the one-loop level. Therefore, it is quite possible for leptoquarks to uncover themselves first in rare $b$ decays while avoiding neutral meson mixing and charged lepton flavor violation. Other hypothetical tree-level mediators of $b \to q \ell^+ \ell^-$ transitions include colorless vectors or scalars. In addition, various NP scenarios contribute at the one-loop level, which could also leave a sizeable effect.

Thanks to the LHCb experiment, the knowledge of rare $b$ decays has significantly advanced in the last decade, and more progress is expected in the future~\cite{Altmannshofer:2022hfs,DiCanto:2022icc,Guadagnoli:2022oxk}. Interestingly, some puzzling discrepancies between the theory and the experiment emerged in $b \to s \mu \mu$ decays~\cite{LHCb:2020lmf,LHCb:2020gog,LHCb:2020zud,LHCb:2021awg,LHCb:2021vsc,LHCb:2014cxe,LHCb:2015wdu,LHCb:2016ykl,LHCb:2021zwz}. The case looks interesting and requires further scrutiny, however, it is still premature to declare NP. The latest LHCb update~\cite{LHCb:2022qnv,LHCb:2022zom} on the lepton flavor universality (LFU) ratios is in agreement with the SM prediction, while the optimized angular observables and branching ratios in $b \to s \mu \mu$ decays are in tension. The SM theory prediction for the latter has been under debate, see e.g.~\cite{Matias:2012xw,Descotes-Genon:2013vna,Jager:2014rwa,Horgan:2013hoa,Gubernari:2020eft,Lyon:2014hpa,Ciuchini:2015qxb}. It is, however, unclear how the strong dynamics could explain the full effect, see e.g.~\cite{Gubernari:2022hxn}. More theoretical and experimental work will help resolve the puzzle. While LHCb will continue to play the leading role on the experimental side, also Belle II has promising prospects~\cite{Belle-II:2018jsg} and will deliver indispensable new results in the future.

The theoretical framework for interpreting rare $b$ decays is the weak effective theory (WET)~\cite{Buras:2020xsm}: the low-energy limit of the SM effective field theory (SMEFT)~\cite{Brivio:2017vri} below the electroweak scale. The SMEFT, on the other hand, is the low-energy limit of a general microscopic new physics with the linear realization of the electroweak symmetry breaking. Data support this framework, in particular, by the absence of new physics in the direct searches suggesting the mass gap and the measurements of the Higgs boson properties, which agree with the SM to $\mathcal{O}(10\%)$ level. The SMEFT Lagrangian is organized as a power series in the inverted NP scale. The leading baryon number conserving NP corrections arise at the mass dimension 6, and there are 2499 (59) independent operators for three families (single family) of SM fermions~\cite{Grzadkowski:2010es}. The vast number of independent theory parameters, another facet of the flavor problem, introduces complexity in the data interpretation. The organizing principle is found in flavor symmetries and breaking patterns which helps to reduce the number of relevant parameters by charting the space of theories beyond the SM into the universality classes~\cite{Faroughy:2020ina,Greljo:2022cah}. Nevertheless, the multidimensional space of the SMEFT Wilson coefficients (WC) requires a global approach in which flavor data plays a key role.

Rare $b$ decays are crucial to probe many directions in the SMEFT parameter space. Given the vast number of observables and theory parameters, model-independent data interpretation is a complex problem. The predictions of physical observables starting from a set of WC in the SMEFT evaluated at the high-energy scale (where an NP model is matched onto the SMEFT) is done in non-trivial steps: evolution of the WC down to the $b$ hadron scale through the procedure of running~\cite{Alonso:2013hga,Jenkins:2013wua,Jenkins:2013zja,Jenkins:2017dyc,Machado:2022ozb,Kumar:2021yod} and matching~\cite{Jenkins:2017jig,Dekens:2019ept}, and the evaluation of the hadronic matrix elements. The construction of the global likelihood requires a proper treatment of the available data, including systematic uncertainties and correlations. Efficient numerical tools and algorithms are needed to carry out the entire program. The EFT interpretation of rare $b$ decays has been one of the central goals of the {\tt flavio} package~\cite{Straub:2018kue}. Global fits of $b \to s \ell^+ \ell^-$ is a mature subject~\cite{Geng:2021nhg,Hurth:2021nsi,Alguero:2021anc,Ciuchini:2021smi,Gubernari:2022hxn} with {\tt flavio} playing a prominent role~\cite{Altmannshofer:2021qrr,Aebischer:2019mlg}. However, a detailed EFT interpretation of $b \to d \ell^+ \ell^-$ decays has received very little attention so far, with the notable exception of Ref.~\cite{Bause:2022rrs} focusing on muons.  In Section~\ref{sec:low_energy}, we review the inner workings of {\tt flavio} and extend the package by including $i)$ $b \to d \ell^+ \ell^-$ data and $ii)$ the latest measurements of $b \to s \ell^+ \ell^-$. As a result, we present the very first EFT study of $b \to d e^+ e^-$ and compare those to the study of $b \to d \mu^+ \mu^-$. In passing, we also update the $b \to s \ell^+ \ell^-$ fit of particular WET scenarios following the latest release of the LHCb $R_{K^{(*)}}$ measurements~\cite{LHCb:2022qnv} and the CMS $B^0_{(s)} \to \mu^+ \mu^-$ measurements~\cite{CMS:2022mgd}.

A microscopic new physics, whose infrared (IR) effects are captured by the SMEFT,  typically also gives correlations in complementary particle physics processes. For example, a four-fermion semileptonic operator in the SMEFT contributing to $b \to q \ell^+ \ell^-$ decays, will, by crossing symmetry, also contribute to $p p \to \ell^+ \ell^-$~\cite{Greljo:2017vvb} due to the presence of heavy flavor parton density functions inside a high-energy proton. In such scenarios, rare $b$ decays are most directly correlated with the high-mass Drell-Yan tails. The effect in the tails will be enhanced at high energies; the scattering amplitude ratio $\mathcal{A}_{{\rm EFT}} / \mathcal{A}_{{\rm SM}} \propto E^2 / \Lambda^2$, where $E$ is the relevant energy scale in the tails, and $\Lambda$ is the NP mass scale. Depending on the quark flavor structure, there could also be additional partonic level channels (besides $\bar b q \to \ell^+ \ell^- + $h.c.) which could further enhance the signal. The high-mass Drell-Yan production in $pp$ collisions has been exquisitely measured at ATLAS and CMS experiments~\cite{ATLAS:2020yat,CMS:2021ctt,ATLAS:2019lsy,CMS:2022yjm}. These measurements will significantly improve moving forward toward the high-luminosity phase. The Drell-Yan production in the tails is a well-known probe of the SMEFT effects. The complementarity between low-energy flavor physics and the high-mass Drell-Yan tails has been a flourishing research direction~\cite{Cirigliano:2012ab, Gonzalez-Alonso:2016etj, Faroughy:2016osc, Greljo:2017vvb, Cirigliano:2018dyk, Greljo:2018tzh,Bansal:2018eha,Angelescu:2020uug, Raj:2016aky, Schmaltz:2018nls, Brooijmans:2020yij,Fuentes-Martin:2020lea,Marzocca:2020ueu,Afik:2019htr,Alves:2018krf,Allwicher:2022mcg,Allwicher:2022gkm,Afik:2018nlr,ATLAS:2021mla,Afik:2020cvr}. The central theme of this work is to systematically explore the interplay of rare $b$ decays versus the high-mass Drell-Yan production. With this global approach, we want to chart the space of possible short-distance NP that could show up in rare $b$ decays.

Similarly to the rare $b$ decays, interpreting the inclusive high-mass Drell-Yan data in the SMEFT is challenging. In Section~\ref{sec:implementation}, we implement a new module in {\tt flavio} for predicting the neutral and charged currents Drell-Yan in the SMEFT for all dimension-6 four-fermion interactions at the tree level. We include the most relevant recent ATLAS and CMS Drell-Yan measurements and construct their likelihoods. The technical details and the validation procedure are described in Appendix~\ref{app:implementation}. The most challenging aspect of this work was optimizing the pipeline to allow for an efficient multidimensional scan of the SMEFT parameter space while keeping the theory predictions precise enough. The critical question concerns the validity of the EFT interpretation in the high-mass tails, to which we devote Section~\ref{sec:validity}. The new {\tt flavio} functionalities presented in this work are valuable additions to the toolbox of theoretical interpretation of the global data in the SMEFT. This will facilitate testing arbitrary short-distance NP models against the experiment.

With such a tool in hand, we are in a position to thoroughly explore the SMEFT parameter space with rare $b$ decays and high-mass Drell-Yan data, see Section~\ref{sec:pheno}. We consider an exhaustive set of operators and various flavor structures to identify interesting phenomenological cases that could occur in the presence of heavy new physics. We start by considering minimalistic flavor scenarios where only a single entry is present in the flavor matrix and directly compare the bounds from the two complementary data sets. We also consider more realistic flavor structures, such as minimal flavor violation (MFV), and identify the interplay and exciting correlations. These SMEFT studies allow us to draw general lessons about classes of NP models.

Finally, to be concrete and exemplify the usage of our toolbox and the application of our SMEFT results, we construct several explicit model examples in Section~\ref{sec:models}. Our model-building exercise is guided by the current trends in $b \to q \ell^+ \ell^-$ data, with tensions reported in $b \to s \mu^+ \mu^-$ decay observables while the LFU ratios are observed to be SM-like. We consider both types of tree-level mediators, $Z'$ and leptoquarks, and for each case, we distinguish the quark flavor couplings that control the production of the high-mass Drell-Yan. All models predict LFU, which, for leptoquarks, requires clever use of the global flavor symmetries. The models are matched to the SMEFT and then confronted against the global data using the {\tt flavio} framework, either finding tensions or compatibility. For those models which can reconcile with the observation, the preferred parameter space is identified for future study. We conclude in Section~\ref{sec:conc}.

\section{Rare $b$ hadron decays in {\tt flavio}}
\label{sec:low_energy}

In this Section, we discuss the implementation of rare $b$ hadron decays in the {\tt flavio} framework (Section~\ref{sec:2.1}) and extract limits on the weak effective theory coefficients for $b\to s$ transitions (Section~\ref{sec:WETbsll}) and $ b \to d $ transitions (Section~\ref{sec:WETbdee}).

\subsection{$b\to q \ell \ell$ in {\tt flavio}}
\label{sec:2.1}

{\tt Flavio} is an open source \texttt{python} package striving to significantly simplify phenomenological analyses in the Standard Model and beyond. It is built in a modular way: firstly, there is a part dedicated to implementing various flavor and other precision observables, allowing for their predictions both in the SM and in dimension 6 EFTs --- the WET below and the SMEFT above the electroweak scale. Secondly, it contains an extensive database of experimental measurements of the implemented observables, which allows for comparisons of the theoretical predictions to the data. Lastly, it contains a statistics submodule that defines many non-trivial probability distribution functions and allows the construction of complex likelihoods, which can take both theoretical and experimental uncertainties into consideration, in general with correlations and non-Gaussianities.

In this work, at low energies, we focus on leptonic and semileptonic $B$-meson decays with the underlying $b\to q \ell \ell$ transitions with $q=d,s$ and $\ell = e, \mu$. In general, these can be classified according to the final state as $B\to \ell \ell$, $B\to P \ell \ell$ and $B\to V \ell \ell$ decays, with $B$ denoting any charged/neutral $B$ meson and $P(V)$ denoting a pseudoscalar (vector) final state meson. Observables belonging to each of these classes are implemented in a general way in the \texttt{flavio.physics.bdecays} submodule, from (differential) branching ratios, to various CP-violating and angular observables. The short-distance contributions to each observable include the SM contributions, as well as the model-independent contributions in the WET at the scale of $\mu=4.8~\mathrm{GeV}$, with the weak effective Hamiltonian defined as
\begin{equation}
\mathcal{H}_\mathrm{eff} = \mathcal{H}_\mathrm{eff}^\mathrm{SM} - \frac{4 G_F}{\sqrt{2}} \frac{e^2}{16\pi^2} \sum_{q=s,d} \sum_{\ell=e, \mu} \sum_{i=9,10,S,P} V_{tb} V_{tq}^\ast(C_i^{bq\ell\ell} O_i^{bq\ell\ell} + C_i^{\prime bq\ell\ell} O_i^{\prime bq\ell\ell}) + \mathrm{h.c.}\,.
\label{eq:WET}
\end{equation}
The semileptonic operators of interest are defined as
\begin{align}
O_9^{bq\ell\ell} &=
(\bar{q} \gamma_{\mu} P_{L} b)(\bar{\ell} \gamma^\mu \ell)\,,
&
O_9^{\prime bq\ell\ell} &=
(\bar{q} \gamma_{\mu} P_{R} b)(\bar{\ell} \gamma^\mu \ell)\,,\label{eq:O9}
\\
O_{10}^{bq\ell\ell} &=
(\bar{q} \gamma_{\mu} P_{L} b)( \bar{\ell} \gamma^\mu \gamma_5 \ell)\,,
&
O_{10}^{\prime bq\ell\ell} &=
(\bar{q} \gamma_{\mu} P_{R} b)( \bar{\ell} \gamma^\mu \gamma_5 \ell)\,,\label{eq:O10}
\\
O_{S}^{bq\ell\ell} &= m_b
(\bar{q} P_{R} b)( \bar{\ell}  \ell)\,,
&
O_{S}^{\prime bq\ell\ell} &= m_b
(\bar{q}  P_{L} b)( \bar{\ell}  \ell)\,,\label{eq:OS}
\\
O_{P}^{bq\ell\ell} &= m_b
(\bar{q} P_{R} b)( \bar{\ell} \gamma_5 \ell)\,,
&
O_{P}^{\prime bq\ell\ell} &= m_b
(\bar{q}  P_{L} b)( \bar{\ell} \gamma_5 \ell)\,.\label{eq:OP}
\end{align}
The contributions of the four-quark operators $O_{1,2}$ and penguin operators $O_{3\ldots6}$ are absorbed in the usual way into the effective coefficients $C_{7,8,9}^\mathrm{eff}(q^2)$. We assume they do not receive NP contributions and are hence part of $\mathcal{H}_\mathrm{eff}^\mathrm{SM}$. Furthermore, we do not consider NP in the dipole operators $O_{7,8}$. As for the non-perturbative quantities, the meson decay constants and the form factor fit parameters are defined in the \texttt{flavio} database of theory parameters. In contrast, the functional forms of the various form factor parameterizations are defined in the same sub-module as the predictions themselves.

Next, we summarise the $b\to q \ell \ell$ observables of interest in this analysis. The $b \to s \mu \mu$ sector contains by far the most experimental and theoretical activity in recent years, fostered by the so-called $B$-anomalies in various branching ratios of $B\to K^{(\ast)} \mu \mu$, $B_s \to \phi \mu \mu$, $\Lambda_b \to \Lambda \mu \mu$ and $B_s \to \mu \mu$, as well as in angular observables such as $P_5^\prime$, and the LFU ratios $R_{K^{(\ast)}}$ (recently resolved in~\cite{LHCb:2022qnv,LHCb:2022zom}). In $b\to s e e$ there are only a few measurements available: the upper limit on branching ratio of the leptonic decay $B_s \to e e$ by LHCb~\cite{LHCb:2020pcv}, the inclusive differential branching ratio measurement of $B\to X_s e e$ by BaBar~\cite{BaBar:2013qry} and measurement of $B\to K^\ast e e$ at very low $q^2$ by LHCb~\cite{LHCb:2020dof}. The last one is particularly sensitive to effects of the dipole operator $O_7$ and we do not consider it further. In $b\to d \mu \mu$ there are upper limits on the branching ratio of $B^0\to \mu \mu$ reported by LHCb \cite{LHCb:2021awg, LHCb:2021vsc}, CMS \cite{CMS:2022mgd} and ATLAS~\cite{ATLAS:2018cur}, as well as the LHCb measurements of the differential branching ratio of $B^+\to \pi^+ \mu \mu$~\cite{LHCb:2015hsa} and a total branching ratio of $B_s \to K^{\ast 0} \mu \mu$~\cite{LHCb:2018rym}. In $b\to d e e$ there are only two measurements available: the upper limit on $B^0 \to e e$ by LHCb~\cite{LHCb:2020pcv} and the upper limit on $B \to \pi e e$ by Belle~\cite{Belle:2008tjs}.

Among the measurements reported above, only a few were missing in \texttt{flavio}, namely, we added the measurements of $B\to \pi \mu \mu$ (in the bins of $q^2 = [2,4], [4,6], [15,22]~\mathrm{GeV^2}$), $B_s \to K^{\ast 0} \mu \mu$ and $B \to \pi e e$. As for theoretical predictions of these observables, they were straightforward to implement thanks to the aforementioned general implementation of the $B\to P,V$ decays in \texttt{flavio}. Moreover, we implement the latest available $B\to \pi$ form factors from Ref.~\cite{Leljak:2021vte} where a combined fit to LCSR and lattice data was performed. We follow closely Ref.~\cite{Bause:2022rrs} for the treatment of resonant regions in $B_s \to K^{\ast 0} \mu \mu$, which adds an additional source of theoretical uncertainty at the level of $8\%$ (see Appendix of Ref.~\cite{Bause:2022rrs} for details).

As for model-independent analyses, $b\to s \mu \mu$ and $b\to s e e$ have been analyzed in great detail~\cite{Altmannshofer:2021qrr, Geng:2017svp, Alguero:2019ptt, Ciuchini:2019usw, Alok:2019ufo, Datta:2019zca}. The $b\to d \mu \mu$ sector has been recently analyzed in a model-independent way in Ref.~\cite{Bause:2022rrs} and we have been able to reproduce their bounds on various $C_i^{bd\mu\mu}$. These types of analyses can be done efficiently with \texttt{flavio} -- we will demonstrate this firstly by presenting an updated global analysis of $b\to s \mu \mu$ in light of the new $R_{K^{(\ast)}}$ measurement by LHCb~\cite{LHCb:2022qnv,LHCb:2022zom} and secondly by studying $b\to d e e$ transitions, commenting on similarities and differences with respect to $b\to d \mu \mu$ transitions. In all cases, we consider only real Wilson coefficients, see e.g.~\cite{Altmannshofer:2021qrr, Becirevic:2020ssj,Alok:2017sui,Kosnik:2021wyp,Carvunis:2021jga, Descotes-Genon:2020tnz,2212.09575} for discussions on CP violating effects.

\subsection{Model-independent bounds from $b\to s \ell \ell$}
\label{sec:WETbsll}

Rare $B$ decays based on the $b\to s \ell \ell$ transitions have received a lot of attention over the past years because in these decays a sizeable number of experimental measurements have shown deviations from the SM predictions.
In particular, LHCb has found discrepancies in several observables that contain only muons in the final state, namely in branching fractions of $B\to K \mu\mu$, $B\to K^*\mu\mu$, and $B_s\to \phi \mu\mu$~\cite{LHCb:2014cxe,LHCb:2015wdu,LHCb:2016ykl,LHCb:2021zwz} as well as in angular observables of $B\to K^*\mu\mu$~\cite{LHCb:2020lmf,LHCb:2020gog} and $B_s\to \phi \mu\mu$~\cite{LHCb:2021xxq}.
In addition to these so-called $b\to s \mu\mu$ anomalies, also ratios of branching fractions with different leptons in the final states previously showed tensions with SM predictions in the $\mu/e$ LFU observables
\begin{equation}
    R_{K} = \frac{\text{BR}(B \to K \mu^+\mu^-)}{\text{BR}(B \to K e^+e^-)}\qquad\text{and}\qquad R_{K^{*}} = \frac{\text{BR}(B \to K^{*} \mu^+\mu^-)}{\text{BR}(B \to K^{*} e^+e^-)} ~.
\end{equation}
Interestingly, both the $b\to s \mu\mu$ anomalies and the hints for $\mu/e$ LFU violation could be consistently explained by new physics contributions to a linear combination of the Wilson coefficients $C_9^{bs\mu\mu}$ and $C_{10}^{bs\mu\mu}$ (cf.~Eqs.~\eqref{eq:O9} and \eqref{eq:O10}) as shown in global fits performed by several groups~\cite{Altmannshofer:2021qrr,Geng:2021nhg, Alguero:2021anc, Hurth:2021nsi, Ciuchini:2021smi,Gubernari:2022hxn}.

Recently, LHCb has announced a combined analysis of $R_{K}$ and $R_{K^{\ast}}$~\cite{LHCb:2022qnv,LHCb:2022zom}, which takes into account the full LHC Run II data and supersedes their previous results. They report the values
\begin{equation}
\begin{split}
    0.1<q^2<1.1: \begin{cases}
        R_K &= 0.994~^{+0.090}_{-0.082} (\mathrm{stat}) ^{+0.029}_{-0.027} (\mathrm{syst}), \\
        R_{K^{\ast}} &= 0.927~^{+0.093}_{-0.087} (\mathrm{stat}) ^{+0.036}_{-0.035}(\mathrm{syst}),
   \end{cases} \\
   1.1<q^2<6.0:
   \begin{cases}
        R_K &= 0.949~^{+0.042}_{-0.041} (\mathrm{stat}) ^{+0.022}_{-0.022} (\mathrm{syst}), \\
        R_{K^{\ast}} &=1.027~^{+0.072}_{-0.068} (\mathrm{stat}) ^{+0.027}_{-0.026} (\mathrm{syst}),
   \end{cases}
\end{split}
\end{equation}
while also providing correlations between $R_K$ and $R_{K^\ast}$, which we do not list here but take into account in our analysis.
These updated results are fully compatible with the SM predictions and no longer provide evidence of a $\mu/e$ universality violation.
This raises the question of whether the $b\to s\mu\mu$ anomalies can still be consistently combined with the stringent constraints on NP provided by the new $R_{K^{(\ast)}}$ measurement.
To answer this question, we perform global fits in the WET in several scenarios. Our analysis is based on~\cite{Altmannshofer:2021qrr} and, in particular, considers its treatment of the NP dependence of the correlated theory uncertainties.

\begin{figure}[t]
     \centering
     \begin{subfigure}[b]{0.6\textwidth}
         \centering
         \includegraphics[width=\textwidth]{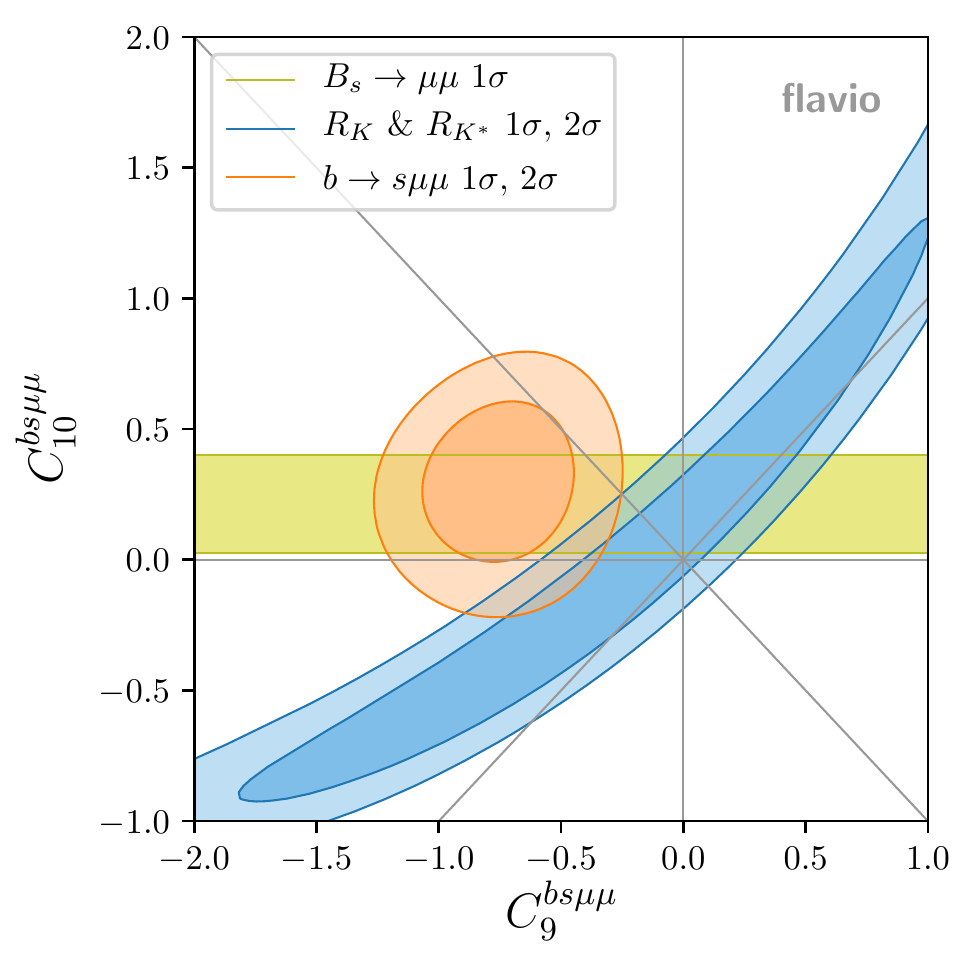}
     \end{subfigure}~
        \caption{Constraints on $(C_{9}^{bs\mu\mu}, C_{10}^{bs\mu\mu})$, including the recently updated measurement of $R_{K^{(\ast)}}$ by LHCb~\cite{LHCb:2022qnv,LHCb:2022zom}. For details see Section~\ref{sec:WETbsll}.}
        \label{fig:C9C10_bsmumu}
\end{figure}

Shown in Fig.~\ref{fig:C9C10_bsmumu} is the result of an updated fit in the two-dimensional scenario\linebreak $(C_9^{bs\mu\mu}, C_{10}^{bs\mu\mu})$ assuming no NP in the electron channel. Here we can observe a (slight) tension between the best-fit regions preferred by the LFU ratios $R_{K^{(*)}}$ (in blue) and the $b\to s \mu \mu$ observables (in orange). The fit to the branching fraction of the leptonic decay $B_s\to \mu\mu$ (in yellow) is separately compatible with both $R_{K^{(*)}}$ and the $b\to s \mu \mu$ observables and takes into account the recent measurement by CMS~\cite{CMS:2022mgd} as well as the results from LHCb~\cite{LHCb:2021awg, LHCb:2021vsc} and ATLAS~\cite{ATLAS:2018cur}, see Appendix~\ref{appendix:Bsmumu}.

A large class of WET scenarios with a single non-zero WC are considered in Appendix~\ref{appendix:WEFTfit}. Regarding 1D scenarios in Table~\ref{tab:1d}, the best performing case with NP only in muons is $C_9^{bs\mu\mu}$ where the tension between the $b\to s \mu \mu$ observables and the LFU ratios $R_{K^{(*)}}$ is $\approx 2\sigma$.

This slight tension can be resolved in the presence of LFU NP, which contributes only to the $b\to s \mu \mu$ observables but not to $R_{K^{(*)}}$. The best performing LFU 1D case is $C_9^{bs\ell\ell} \equiv C_9^{\text{univ.}}\approx-0.8$ with $3.7\sigma$ pull, cf.~Table~\ref{tab:1d}.
In principle, a shift in $C_9^{\text{univ.}}$ could be mimicked by QCD effects. Whether such a large shift can be due to underestimated non-local hadronic contributions is a matter of ongoing extensive discussions, see e.g.~\cite{Ciuchini:2022wbq,Gubernari:2022hxn}.
\begin{figure}[t]
     \centering
     \begin{subfigure}[b]{0.5\textwidth}
         \centering
         \includegraphics[width=\textwidth]{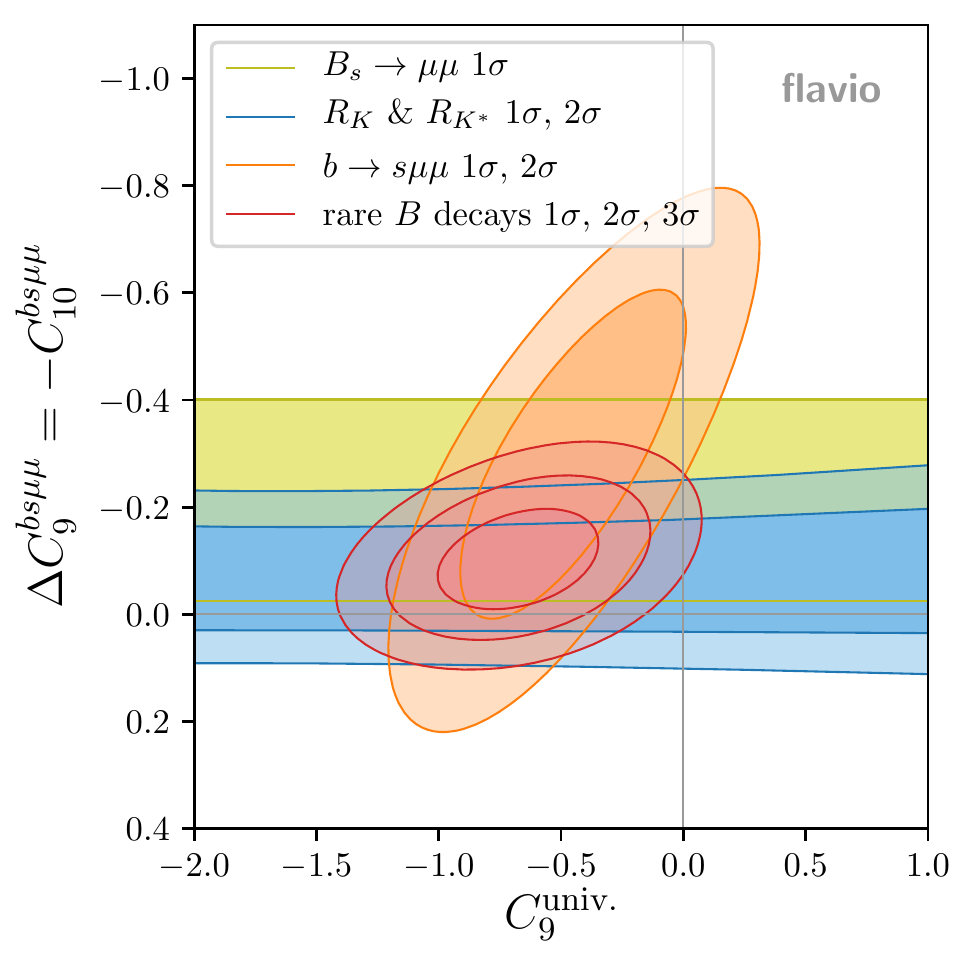}
     \end{subfigure}~
     \begin{subfigure}[b]{0.5\textwidth}
         \centering
         \includegraphics[width=\textwidth]{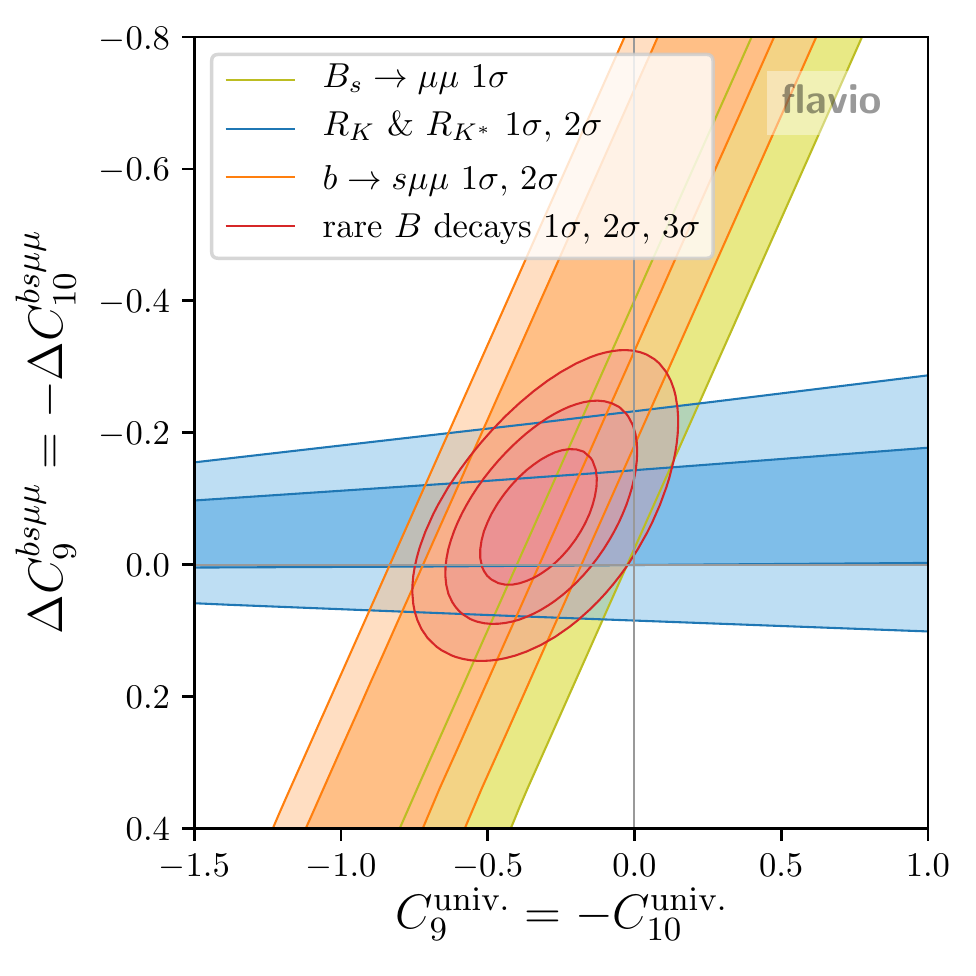}
     \end{subfigure}
        \caption{Constraints on two different WET scenarios including the recently updated measurement of $R_{K^{(\ast)}}$ by LHCb~\cite{LHCb:2022qnv,LHCb:2022zom}. For details see Section~\ref{sec:WETbsll}.}
        \label{fig:C9C10_mu_vs_univ}
\end{figure}
Interesting 2D scenarios are:
\begin{itemize}
    \item $(C_9^{\text{univ.}}, \Delta C_9^{bs\mu\mu} = -C_{10}^{bs\mu\mu})$, where $C_9^{bs\mu\mu} = C_9^{\text{univ.}} + \Delta C_9^{bs\mu\mu}$ and $C_9^{bsee} = C_9^{\text{univ.}}$.
    This scenario was previously found to be well suited to explain tensions between $R_{K^{(*)}}$ and $b\to s \mu \mu$ observables~\cite{Aebischer:2019mlg,Alguero:2019ptt}. Furthermore, it is motivated by the fact that $C_9^{\text{univ.}}$ can be generated through RGE effects in the WET~\cite{Bobeth:2011st}, the SMEFT~\cite{Aebischer:2019mlg}, and in UV models~\cite{Crivellin:2018yvo}.\footnote{\label{footnote:RGE_C9univ}%
    NP could also generate $b \to s c\bar c$ transitions which then lead to $C_9^{{\rm eff}}$~\cite{Jager:2017gal,Jager:2019bgk,Kumar:2022rcf}.  Moreover, $C_9^{\text{univ.}}$ could be generated through RGE mixing of four-quark operators in the SMEFT~\cite{Aebischer:2019mlg} (e.g. from a leptophobic $Z'$), which could potentially be probed by searches for a dijet tails/resonances~\cite{Bordone:2021cca}. Another option is to generate large $b\to s \tau \tau$ transitions which through RGE also give $C_9^{{\rm eff}}$~\cite{Bobeth:2011st,Crivellin:2018yvo,Aebischer:2019mlg,Alguero:2019ptt}. The complementary constraint at high-$p_T$ is a non-resonant deviation in the high-mass $\tau \tau$ tails~\cite{Faroughy:2016osc}.}
    The results of a fit in this scenario are shown in the left panel of Fig.~\ref{fig:C9C10_mu_vs_univ}. The fit shows a clear preference for non-zero $C_9^{\text{univ.}}$, which can fully remove the tension between $R_{K^{(*)}}$ and the $b\to s \mu \mu$ observables. For the ``rare $B$ decays'' global fit, the Gaussian approximation at the best-fit point is
    \begin{equation}
    \begin{aligned}
     C_9^{\text{univ.}} &= -0.64 \pm 0.22\,,\\
     \Delta C_9^{bs\mu\mu} = -C_{10}^{bs\mu\mu} &= -0.11 \pm 0.06\,,
    \end{aligned}
    \end{equation}
    with a correlation coefficient $\rho=-0.33$.

    \item $(C_9^{\text{univ.}} = -C_{10}^{\text{univ.}}, \Delta C_9^{bs\mu\mu} = - \Delta C_{10}^{bs\mu\mu})$, where $C_{9,10}^{bs\mu\mu} = C_{9,10}^{\text{univ.}} + \Delta C_{9,10}^{bs\mu\mu}$ and $C_{9,10}^{bsee} = C_{9,10}^{\text{univ.}}$.
    This scenario corresponds to NP coupling purely to left-handed SM fields. We find that a non-zero $C_9^{\text{univ.}} = -C_{10}^{\text{univ.}}$ can consistently explain the $b\to s\mu\mu$ anomalies, while the LFU violating purely muonic contribution to $\Delta C_9^{bs\mu\mu} = - \Delta C_{10}^{bs\mu\mu}$ is compatible with zero at the one sigma level. It is worth noting that $B_s \to \mu\mu$ preferred parameter space is compatible with $b \to s \mu \mu$ at $1\sigma$. For the ``rare $B$ decays'' global fit, the Gaussian approximation at the best-fit point is
    \begin{equation}
    \begin{aligned}
     C_9^{\text{univ.}} = -C_{10}^{\text{univ.}} &= -0.29 \pm 0.13\,,\\
     \Delta C_9^{bs\mu\mu} = - \Delta C_{10}^{bs\mu\mu} &= -0.08 \pm 0.07\,,
    \end{aligned}
    \end{equation}
    with a correlation coefficient $\rho=-0.54$.
\end{itemize}

The (slight) tension between $R_{K^{(*)}}$ and $b\to s \mu \mu$ observables and its resolution through LFU NP is the motivation for discussing manifestly LFU models in Section~\ref{sec:models}.\footnote{Our models generate LFU $b \to s \ell^+ \ell^-$ transitions at the tree level at the UV matching scale (in contrast to models discussed in footnote~\ref{footnote:RGE_C9univ}).
One could also imagine loop-level UV models with new scalars and fermions running in the box diagram. Such models should also use the flavor symmetries to enforce LFU as in Section~\ref{sec:modelIIb} for the tree-level leptoquark model. Loop-level models are more difficult to hide from direct resonance searches at the LHC since the implied mass scale is lower.}

\subsection{Model-independent bounds from $b\to d \ell \ell$}
\label{sec:WETbdee}

\begin{figure}[t]
     \centering
     \begin{subfigure}[b]{0.43\textwidth}
         \centering
         \includegraphics[width=\textwidth]{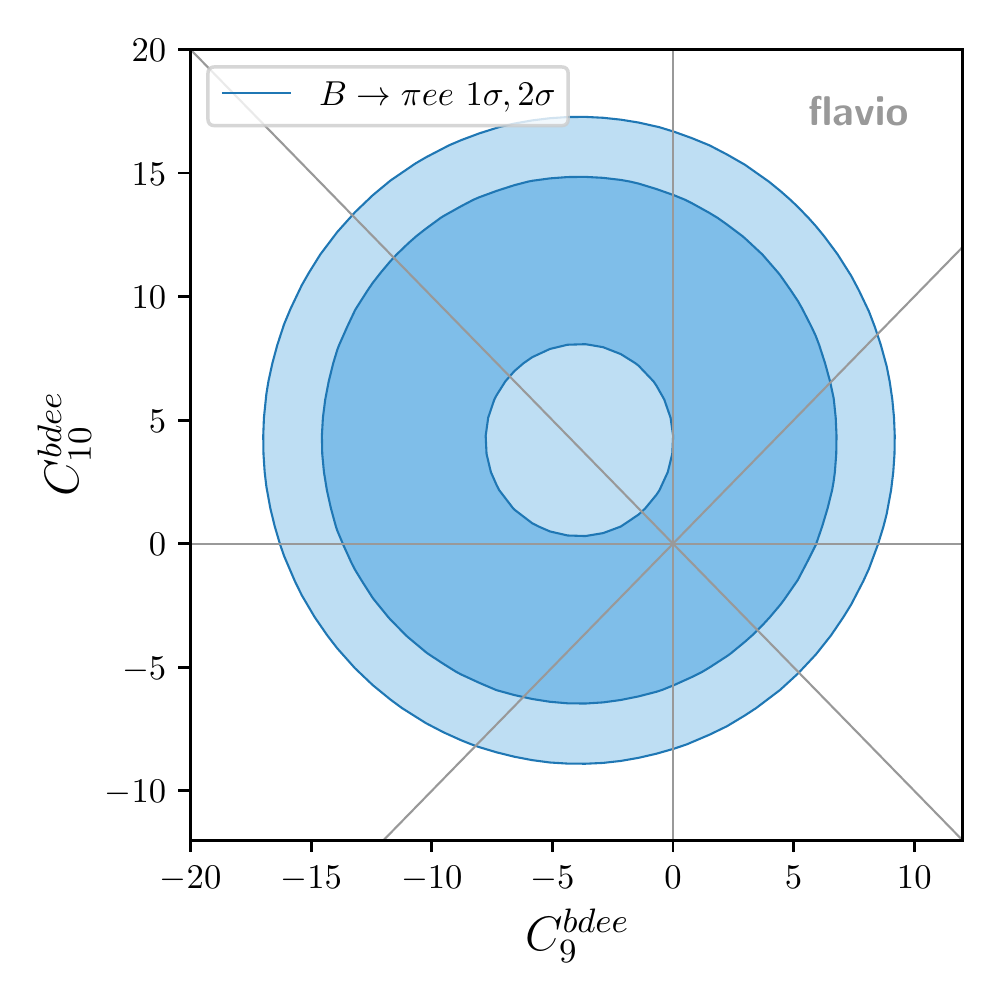}
     \end{subfigure}~
     \begin{subfigure}[b]{0.43\textwidth}
         \centering
         \includegraphics[width=\textwidth]{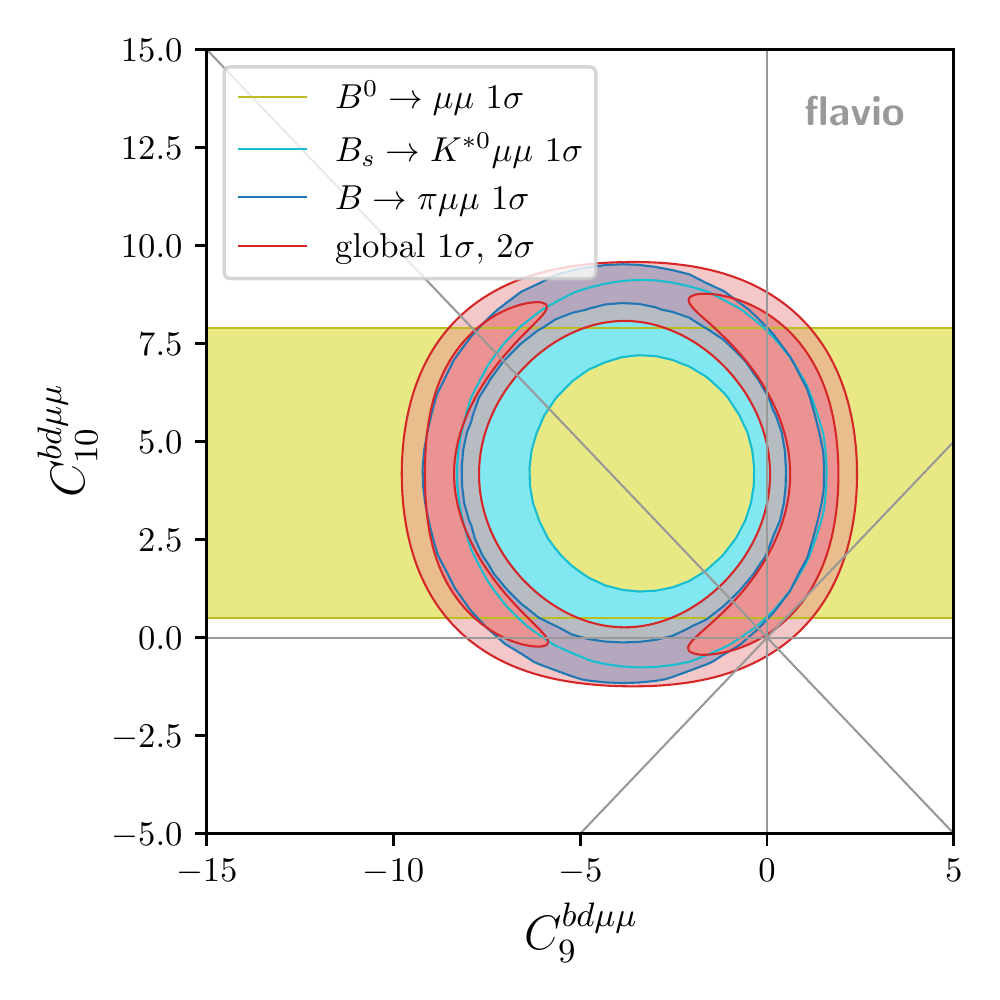}
     \end{subfigure}
     \\
      \begin{subfigure}[b]{0.43\textwidth}
         \centering
         \includegraphics[width=\textwidth]{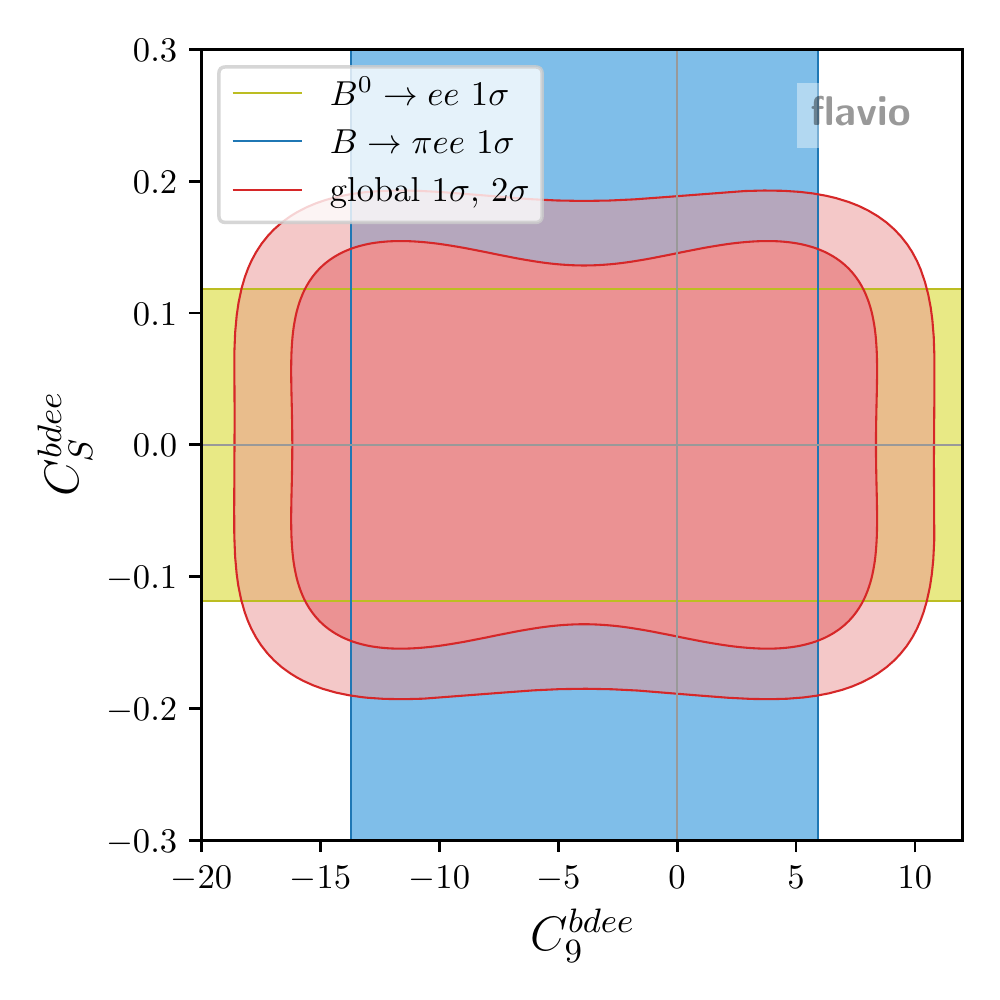}
     \end{subfigure}~
     \begin{subfigure}[b]{0.43\textwidth}
         \centering
         \includegraphics[width=\textwidth]{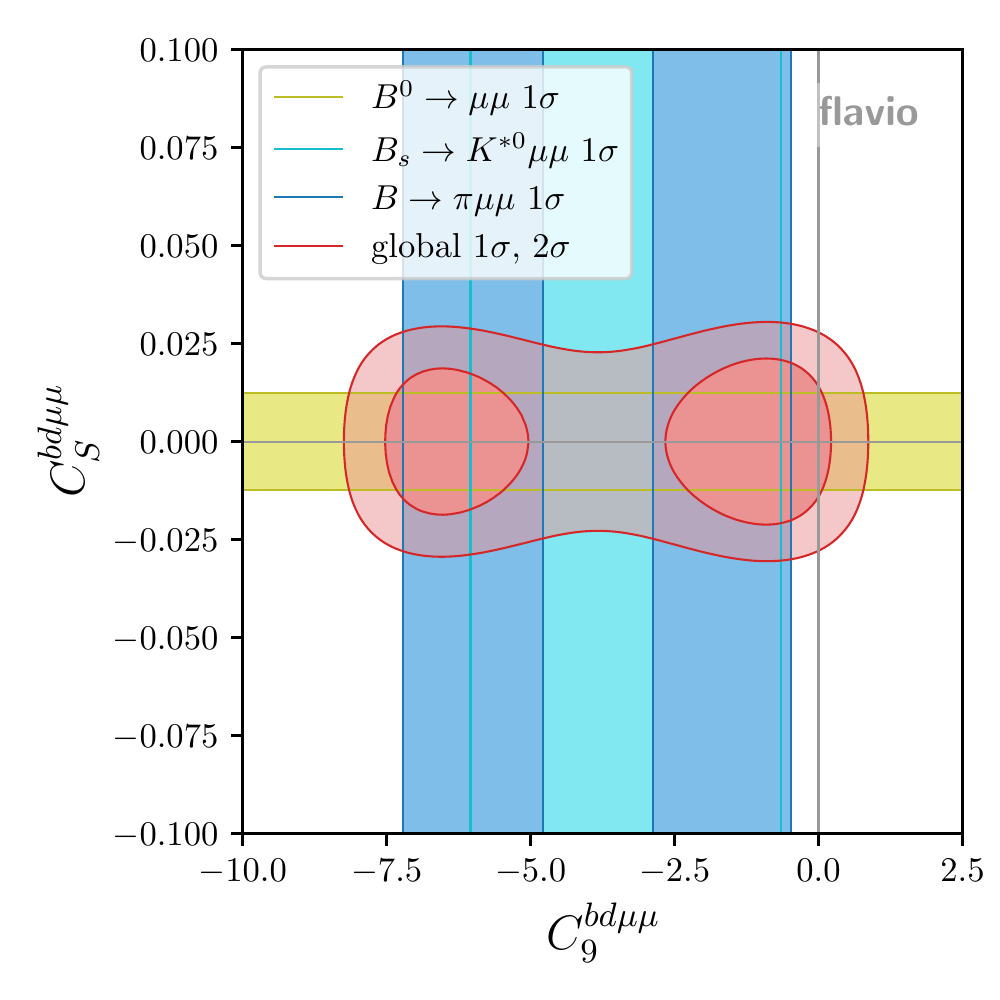}
     \end{subfigure}
     \\
     \begin{subfigure}[b]{0.43\textwidth}
         \centering
         \includegraphics[width=\textwidth]{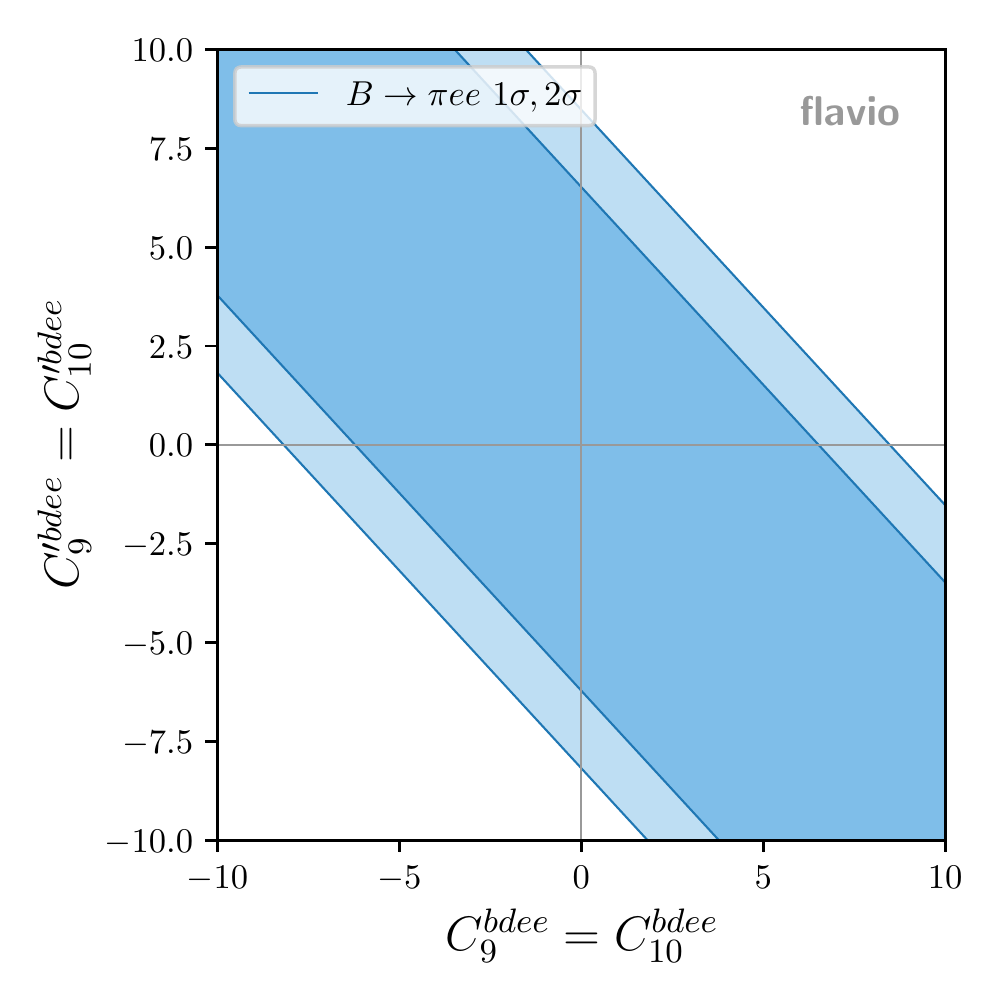}
     \end{subfigure}~
     \begin{subfigure}[b]{0.43\textwidth}
         \centering
         \includegraphics[width=\textwidth]{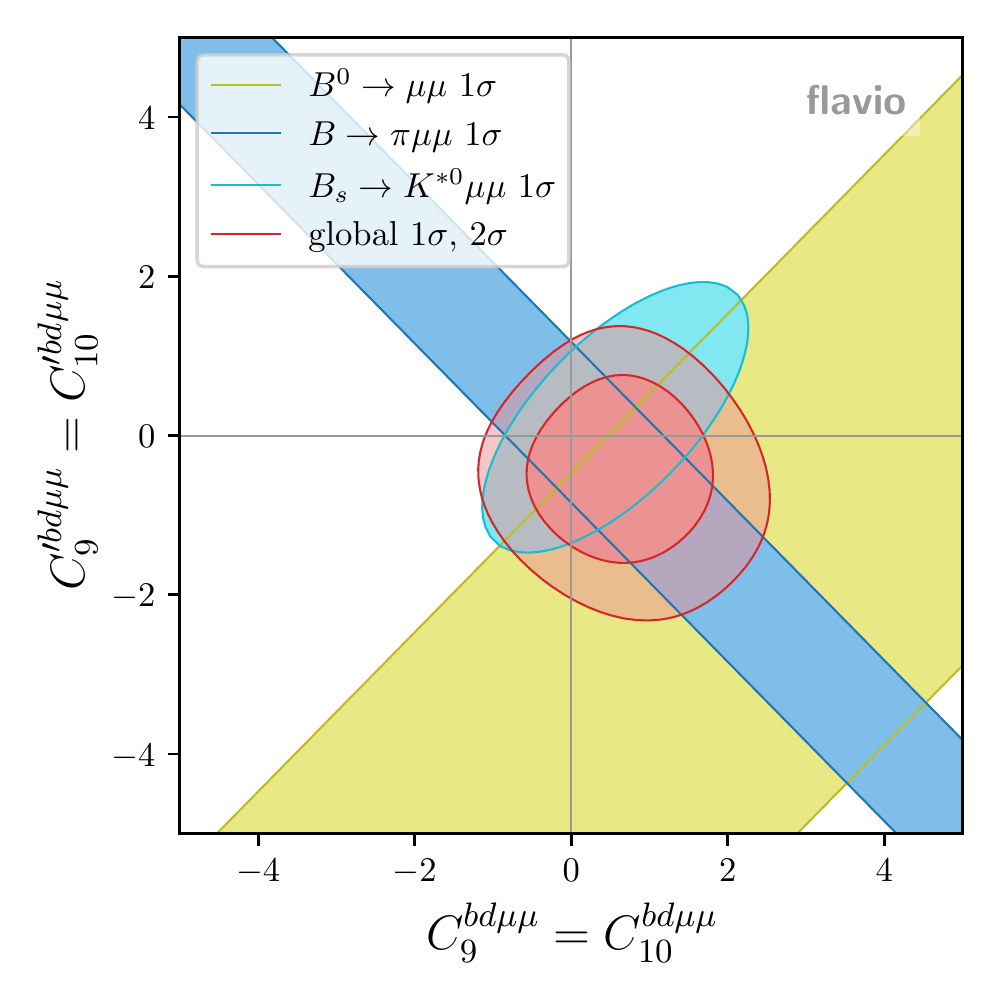}
     \end{subfigure}
        \caption{Constraints on the $C_{9,10}^{(\prime)bd\ell\ell}$ and $C_S^{bd\ell\ell}$ Wilson coefficients in various scenarios. See Section~\ref{sec:WETbdee} for details.}
        \label{fig:bdll}
\end{figure}

In this subsection, we show selected bounds on the WET Wilson coefficients from measurements of $b\to d \ell \ell$ transitions, emphasizing the key differences between electron and muon final states. We note that $b\to d \mu\mu$ transitions have been discussed to great extent in Ref.~\cite{Bause:2022rrs}.

Due to the limited number of experimental measurements available, the parameter space of $C_{9,10,S,P}^{(\prime) bdee}$ is only loosely constrained. The branching ratio of the purely leptonic decay of $B^0 \to e e$ is sensitive to $C_{10,S,P}^{(\prime)bdee}$ via~\cite{DeBruyn:2012wk}
\begin{equation}
    \frac{\br{B^0\to e e}}{\br{B^0\to e e}_\mathrm{SM}} = |P|^2 + |S|^2 \,,\\
\end{equation}
where
\begin{equation}
\begin{aligned}
    P &\equiv \frac{C_{10}^\mathrm{SM} + C_{10}^{bdee} - C_{10}^{\prime bdee}}{C_{10}^\mathrm{SM}} + \frac{M_{B^0}^2}{2 m_e} \left(\frac{m_b}{m_b+m_d}\right) \left(\frac{C_P^{bdee} - C_P^{\prime bdee}}{C_{10}^\mathrm{SM}}\right) \,,\\
    S &\equiv \sqrt{1-4\frac{m_e^2}{M_{B^0}^2}}\frac{M_{B^0}^2}{2 m_e}\left(\frac{m_b}{m_b+m_d}\right)\left(\frac{C_S^{bdee} - C_S^{\prime bdee}}{C_{10}^\mathrm{SM}}\right)\,.
\label{eq:Bee}
\end{aligned}
\end{equation}
These decays are particularly sensitive to $C_{S,P}^{\prime bdee}$ as their contributions lift the helicity suppression that the SM contributions suffer from. As for the branching ratio of $B\to \pi e e$, it is sensitive to NP in $C_{9,10}^{(\prime)bdee}$ through~\cite{Rusov:2019ixr}
\begin{equation}
\frac{d \br{B \to \pi e e}}{d q^2} = \tau_{B}
\frac{G_F^2 \alpha_{\rm em}^2 |V_{tb} V_{td}^\ast|^2}{1536 \pi^5 m_B^3}  \lambda^{3/2}(m_B^2,m_\pi^2,q^2) F_{B\pi}(q^2) \,,
\label{eq:Bpiee}
\end{equation}
with
\begin{equation}
\begin{split}
    F_{B\pi}(q^2) &= \left| (C_9^{\mathrm{eff}}(q^2) + C_9^{bdee}+C_9^{\prime bdee})f_+(q^2) +
\frac{2 m_b}{m_B + m_\pi} C_7^{\mathrm{eff}}(q^2) f_T (q^2)\right|^2
\\
&+ \left|( C_{10}^{\mathrm{eff}}(q^2) + C_{10}^{bdee}+C_{10}^{\prime bdee})f_+(q^2) \right|^2  \,,
\end{split}
\label{eq:BpieeF}
\end{equation}
where $f_+, f_T$ are the vector and tensor $B\to \pi$ form factors, $\lambda(m_B^2,m_\pi^2,q^2)$ is the Källén function \cite{ParticleDataGroup:2022pth}, and $\tau_{B}$ is the $B$ meson mean lifetime. We omit the dependence on $C_7^{(\prime)bdee}$ as we do not consider it in the following.

In the first column of Fig.~\ref{fig:bdll} we show the constraints from measurements of $B\to \pi e e$ and $B^0 \to e e$ in three NP scenarios: assuming NP in $(C_9^{bdee}, C_{10}^{bdee})$ (first row), assuming NP in $(C_9^{bdee}, C_{S}^{bdee})$ (second row), and assuming a right-handed NP scenario of $(C_9^{bdee}=C_{10}^{bdee}, C_9^{\prime bdee}=C_{10}^{\prime bdee})$ (third row). As expected, $B\to \pi e e$ plays an important role in constraining $C_9$ and $C_{10}$, whereas $B^0 \to e e$ is very sensitive to $C_S$. Let us point out here the flat direction in the last considered scenario. It can be understood by considering Eq.~\eqref{eq:BpieeF} --- inserting the scenario into the equation results in $F_{B\pi}(q^2)\sim |(C_9^\mathrm{eff}(q^2) + \delta C)f_+(q^2)|^2 + |(C_{10}^\mathrm{eff}(q^2) + \delta C)f_+(q^2)|^2$ where $\delta C_9 = C_9^{bdee}+C_9^{\prime bdee}$ and where we omit the $C_7$ dependence. Clearly, the branching ratio is independent of the direction orthogonal to $\delta C_9$, hence the flat direction. In Section~\ref{sec:pheno}, we will return to the feasibility of closing such flat directions in the context of SMEFT, where correlating with other measurements is possible.

For comparison purposes, we show in the second column of Fig.~\ref{fig:bdll} the bounds for the same WC combinations as in the first column, but now assuming NP in $b\to d \mu \mu$. Notice that all the considered WC, namely $C_{9}$, $C_{10}$ and $C_S$, are better constrained for muonic final states, thanks to both $\br{B^0\to \mu \mu}$ and $\br{B\to \pi \mu \mu}$ being significantly better measured compared to their electron counterparts. Moreover, in the right-handed scenario of $(C_9^{bd\mu\mu}=C_{10}^{bd\mu\mu}, C_9^{\prime bd\mu\mu}=C_{10}^{\prime bd\mu\mu})$ (third row), the aforementioned flat direction can still be seen in the constraint from $B\to \pi \mu \mu$, but it is now closed, firstly from $B^0 \to \mu \mu$, which in the muonic case can provide useful information on $C_{10}$ even with the accompanying helicity suppression, but secondly and more importantly from the measurement of $B_s \to K^{*0} \mu \mu$. Thanks to the vector meson in the final state, the dependence of this branching ratio on the various WC is rich in structure (see Ref.~\cite{Bause:2022rrs} for explicit expressions), and  the dependence is such that the flat direction does not appear. A future measurement of a $B\to V e e$ process, such as $B_s \to K^{*0} e e$, would play a crucial role in constraining flat directions such as the one shown on the bottom left plot of Fig.~\ref{fig:bdll}.

\section{Implementation of the high-mass Drell-Yan in {\tt flavio}}
\label{sec:implementation}

In Section~\ref{sec:low_energy} we have summarized and showcased the ability of {\tt flavio} to do phenomenological analyses in an EFT at the scale of a particular process, e.g.~in the WET at the scale of the $B$ meson decays, $\mu_b=4.8~\mathrm{GeV}$. However, {\tt flavio} is interfaced with {\tt wilson}~\cite{Aebischer:2018bkb}, a python package that takes care of running and matching Wilson coefficients below and above the electroweak scale. Below the electroweak scale {\tt flavio} operates within the WET (integrating out particular quark flavors as the scale decreases), whereas above the electroweak scale it operates in the SMEFT, for which we use the Warsaw basis~\cite{Grzadkowski:2010es}. In this paper, we use the following definition of the SMEFT effective Lagrangian at mass dimension~6,
\begin{equation}
\mathcal{L_{\mathrm{eff}}} = \mathcal{L_{\mathrm{SM}}} + \sum_{Q_i=Q_i^\dagger} \frac{C_i}{\Lambda^2} Q_i + \sum_{Q_i\neq Q_i^\dagger} \left( \frac{C_i}{\Lambda^2} Q_i + \frac{C_i^\ast}{\Lambda^2} Q_i^\dagger \right) \,,
\label{eq:SMEFT}
\end{equation}
which differs slightly from the one defined internally in {\tt flavio} as our $C_i$ are dimensionless. Thanks to the running and matching procedures, it is straightforward to do phenomenological analyses in {\tt flavio} with WC defined at arbitrarily high energies. As we will demonstrate in the following sections, one can constrain various SMEFT $C_i$ from low-energy processes, e.g. $B$ meson decays.

It has been demonstrated several times in the literature that high-mass Drell-Yan tails can act as powerful probes of NP effects at scales of $\mathcal{O}(\mathrm{TeV})$, for selected examples see Refs.~\cite{Greljo:2017vvb, Allwicher:2022gkm, Angelescu:2020uug, Fuentes-Martin:2020lea, Farina:2016rws}. In this section, we present the {\tt flavio} implementation of both theoretical predictions in the SMEFT as well as the latest experimental measurement of high-mass Drell-Yan tails by CMS and ATLAS, both in neutral-current (NC: $pp\to \ell \ell$) and charged-current (CC: $pp\to \ell \nu$) processes. Incorporating these datasets into the {\tt flavio} framework, together with aforementioned {\tt flavio} functionalities, enables examining the interplay between low- and high-energy processes within the parameter space of the SMEFT. For example, one can directly compare the limits on the high-energy Wilson coefficients obtained from the Drell-Yan tails to those from $B$ meson decays (Section~\ref{sec:pheno}) or exploit correlations when specific flavor symmetry and UV dynamics are assumed (Section~\ref{sec:models}).

Our high-mass Drell-Yan {\tt flavio} implementation includes the effects of semileptonic dimension-6 contact interactions with arbitrary flavor structure listed in Table~\ref{tab:SMEFToperators}. This is the complete set of dimension-6 operators in SMEFT contributing at tree level to $q \bar q \to \ell \bar \ell$  with the leading energy scaling $ \propto \hat s/\Lambda^2$, where $\sqrt{\hat s}$ is the invariant mass of the lepton pair. The effects of other dimension-6 operators contributing at the tree level are suppressed by powers of $v  / \sqrt{\hat s}$ in comparison with operators in Table~\ref{tab:SMEFToperators}. This additional power counting is possible thanks to the hierarchy between the relevant energy scale in the high-mass Drell-Yan tails and the electroweak scale. The subleading operators include dipoles $\psi^2 X \phi$ and Higgs-current operators $\psi^2 \phi^2 D$. Both classes enter by modifying the couplings of weak gauge bosons to fermions and are highly constrained from on-shell gauge boson processes. Neither of these classes will be further discussed here.\footnote{The dipole operators enter the high-mass tails at the next-to-leading order in energy scaling $\propto v \sqrt{\hat s}/\Lambda^2$. Indeed, the limits extracted in Ref.~\cite{Allwicher:2022gkm} (Figure 4.2) are relatively weak, questioning the validity of the EFT interpretation in perturbative UV completions (for the dipole operators, those start at the one-loop level). See Section~\ref{sec:validity} for more details on the validity of the EFT approach in the high-mass tails.}

The rest of this section is organized as follows. We begin with a discussion of the implementation of NC and CC Drell-Yan theoretical predictions at leading order in new physics effects, from parton-level to hadron-level cross-sections. Next, we discuss the inclusion of data from the latest experimental searches, define the experimental likelihood, including systematic uncertainties, and provide justification for implementing the predictions of cross-sections at leading order as an efficient approximation for scanning the SMEFT parameter space. The implementations discussed in this section can be found in the {\tt flavio} sub-module {\tt physics.dileptons}.

\begin{table}[t]
    \centering
    \begin{tabular}{c|c}
        $Q_{lq}^{(1)}$ & $(\bar l_p\gamma_\mu l_r)(\bar q_s \gamma^\mu q_t)$ \\
        $Q_{lq}^{(3)}$ & $(\bar l_p\gamma_\mu\sigma^i l_r)(\bar q_s \gamma^\mu\sigma^i q_t)$ \\ \hline
        $Q_{lu}$ & $(\bar l_p\gamma_\mu l_r)(\bar u_s \gamma^\mu u_t)$ \\
        $Q_{ld}$ & $(\bar l_p\gamma_\mu l_r)(\bar d_s \gamma^\mu d_t)$ \\
        $Q_{qe}$ & $(\bar q_p\gamma_\mu q_r)(\bar e_s \gamma^\mu e_t)$ \\ \hline
        $Q_{eu}$ & $(\bar e_p\gamma_\mu e_r)(\bar u_s \gamma^\mu u_t)$ \\
        $Q_{ed}$ & $(\bar e_p\gamma_\mu e_r)(\bar d_s \gamma^\mu d_t)$ \\ \hline
        $Q_{ledq}$ & $(\bar l_p^j e_r)(\bar d_s q_{tj})$ \\
        $Q_{lequ}^{(1)}$ & $(\bar l_p^j e_r)\varepsilon_{jk}(\bar q_s^k u_t)$ \\
        $Q_{lequ}^{(3)}$ & $(\bar l_p^j \sigma_{\mu\nu} e_r)\varepsilon_{jk}(\bar q_s^k \sigma^{\mu\nu} u_t)$ \\
    \end{tabular}
    \caption{Semileptonic four-fermion SMEFT operators at dimension 6 with $\Delta B =0$. The flavor indices $p,r,s,t$ are suppressed on the left-hand side, $\sigma^i$ are Pauli matrices and $\varepsilon$ is the totally anti-symmetric tensor in the $SU(2)_L$ space.}
    \label{tab:SMEFToperators}
\end{table}

\subsection{Predictions for semileptonic contact interactions}
\label{sec:xsec}
We begin by defining the effective Lagrangians for NC and CC Drell-Yan processes, in the mass basis of the fermions and separating the contributions from operators of different Lorentz structures:
\begin{equation}
\begin{split}
    \mathcal L_{\rm NC} \supset \sum_{X,Y = L,R}\sum_{q=u,d} & \left( \frac{c^{SXY}_{{eq}_{ijkl}}}{\Lambda^2} [\bar e_i\, P_X\, e_j][\bar q_i\, P_Y\, q_j] + \frac{c^{VXY}_{{eq}_{ijkl}}}{\Lambda^2} [\bar e_i\, \gamma_\mu P_X\, e_j][\bar q_i\, \gamma^\mu P_Y\, q_j] \right. \\
    &\left. \quad + \frac{c^{TXY}_{{eq}_{ijkl}}}{\Lambda^2} [\bar e_i\, \sigma_{\mu\nu} P_X\, e_j][\bar q_i\, \sigma^{\mu\nu} P_Y\, q_j]\right) + {\rm h.c.}\,,
\end{split}
\label{eq:lagNC}
\end{equation}
and
\begin{equation}
\begin{split}
    \mathcal L_{\rm CC} \supset \sum_{X = L,R} & \left( \frac{c^{SLX}_{{e\nu ud}_{ijkl}}}{\Lambda^2} [\bar e_i\, P_L\, \nu_j][\bar u_i\, P_X\, d_j] + \frac{c^{VLX}_{{e\nu ud}_{ijkl}}}{\Lambda^2} [\bar e_i\, \gamma_\mu P_L\, \nu_j][\bar u_i\, \gamma^\mu P_X\, d_j] \right.\\
    & \left. \quad + \frac{c^{TLX}_{{e\nu ud}_{ijkl}}}{\Lambda^2} [\bar e_i\, \sigma_{\mu\nu} P_L\, \nu_j][\bar d_i\, \sigma^{\mu\nu} P_X\, u_j] \right) + {\rm h.c.}\,.
\end{split}
\label{eq:lagCC}
\end{equation}
The sum over all flavor indices is implicitly assumed, while $P_{R,L}=\frac{1}{2}(1\pm\gamma_5)$ are the chirality projectors. The contributions of dimension-6 SMEFT operators from Table~\ref{tab:SMEFToperators} can be matched onto these Lagrangians. In Table~\ref{tab:SMEFTmatching} we give the coefficients $c$ in the Warsaw basis with up-alignment, in which the left-handed quark doublet reads $q = (u_L, ~ V d_L)^T$ where $u_L$ and $d_L$ are the mass eigenstates and $V$ is the CKM mixing matrix. Neutrinos are assumed to be massless and the PMNS mixing is a unit matrix. The up-aligned basis was chosen here for easier validation against \texttt{MadGraph} simulations (see Appendix~\ref{app:implementation}). Note, however, that the phenomenological studies in \texttt{flavio} can be done on either the up- or down-aligned basis, with automatic translation between the two taken care of by \texttt{wilson}. The studies presented later in this paper will be performed using the down-aligned basis.

\begin{table}[t]
    \centering
    \begin{tabular}[t]{c|c}
        Coefficient & Matching \\ \hline
        $c^{SRL}_{ed}$ & $C_{ledq} V$ \\
        $c^{SRR}_{eu}$ & $-C^{(1)}_{lequ}$ \\
        $c^{SLL}_{e\nu ud}$ & $C^\dagger_{ledq}$ \\
        $c^{SLR}_{e\nu ud}$ & $C^{(1)\dagger}_{lequ} V$ \\ \hline
        $c^{TRR}_{eu}$ & $-C^{(3)}_{lequ}$ \\
        $c^{TLR}_{e\nu ud} $ &  $C^{(3)\dagger}_{lequ} V$\\
    \end{tabular} \hspace{20pt}
    \begin{tabular}[t]{c|c}
        Coefficient & Matching \\ \hline
        $c^{VLL}_{eu}$ & $C^{(1)}_{lq} - C^{(3)}_{lq}$ \\
      $c^{VLR}_{eu}$ &  $C_{lu}$\\
    $c^{VRL}_{eu_{ijkl}}$ & $C_{qe_{klij}}$ \\
    $c^{VRR}_{eu}$ & $C_{eu}$ \\
    $c^{VLL}_{ed}$ & $V^\dagger \left(C^{(1)}_{lq}+C^{(3)}_{lq}\right) V$ \\
    $c^{VLR}_{ed}$ & $C_{ld}$\\
    $c^{VRL}_{ed_{ijkl}}$ & $V^*_{mk} C_{qe_{mnij}} V_{nl}$ \\
    $c^{VRR}_{ed}$ & $C_{ed}$ \\
    $c^{VLL}_{e\nu ud}$ & $2\, C^{(3)}_{lq} V$
    \end{tabular}
    \caption{The matching of the dimension 6 SMEFT operators defined in Table~\ref{tab:SMEFToperators} onto the effective Lagrangians defined in Eqs.~\eqref{eq:lagNC} and \eqref{eq:lagCC} in the up-diagonal mass basis for scalar and tensor operators (left) and vector operators (right). We suppress the flavor indices when it is straightforward to recover them, multiplication by the CKM matrix $V$ is blind to the lepton flavor indices, and $C_{ijkl}^\dagger = C^*_{jilk}$.}
    \label{tab:SMEFTmatching}
\end{table}

\subsubsection*{Partonic cross-sections}
\label{sec:parton_xsec}
The polarized scattering amplitudes corresponding to the neutral current process $\bar q^k\, q^l \to \bar e^j\, e^i$ and charged current process $\bar u^k\, d^l \to \bar \nu^j\, e^i$ can be written as
\begin{equation}
\begin{split}
    \mathcal A^{\Gamma XY}_{ijkl}(\bar q^k\, q^l \to \bar e^j\, e^i) &= i\, [\bar e_i\, \Gamma P_X\, e_j][\bar q_k\, \Gamma P_Y\, q_l]N^{\Gamma XY}_{q,ijkl}(\hat{s})\,,\\
    \mathcal A^{\Gamma XY}_{ijkl}(\bar u^k\, d^l \to \bar \nu^j\, e^i) &= i\, [\bar e_i\, \Gamma P_X\, \nu_j][\bar u_k\, \Gamma P_Y\, d_l]C^{\Gamma XY}_{ijkl}(\hat{s})\,,
\end{split}
\label{eq:amplitudes_NC_CC}
\end{equation}
where $q = u$ or $d$. The Lorentz structure is given by $\Gamma = S, V, T = 1, \gamma_\mu, \sigma_{\mu\nu}$, and $X,Y$ denote the chirality. We define the form factors $N$ and $C$ by taking into account the SM contributions as well as the contributions from effective Lagrangians defined in Eqs.~\eqref{eq:lagNC} and \eqref{eq:lagCC} as
\begin{equation}
\begin{split}
    N^{\Gamma XY}_{q,ijkl}(\hat{s}) &= \left(e^2\frac{Q_e Q_q}{\hat{s}} + \frac{g_Z^{e_X} g_Z^{q_Y}}{\hat{s}-m_Z^2+im_Z\Gamma_Z}\right)\delta^{ij}\delta^{kl}\delta^{V\Gamma} + \frac{c^{\Gamma XY}_{{eq}_{ijkl}}}{\Lambda^2}\,, \\
    C^{\Gamma XY}_{ijkl}(\hat{s}) &= \frac{g^2}{2}\frac{V_{kl}}{\hat{s}-m_W^2+im_W\Gamma_W}\delta^{ij}\delta^{LX}\delta^{LY}\delta^{V\Gamma} + \frac{c^{\Gamma XY}_{{e\nu ud}_{ijkl}}}{\Lambda^2}\,.
\end{split}
    \label{eq:FormFactors_NC_CC}
\end{equation}
Here $\hat{s} = (p_q + p_{\bar q})^2  = (p_e + p_{\bar e})^2$ is the partonic center-of-mass energy. Regarding the SM parts, in the photon contribution, we have $Q_e = -1$, $Q_u = 2/3$, $Q_d = -1/3$, whereas the couplings with the $Z$ boson are $g_Z^{f_{X}} = \frac{e}{s_W c_W}(T^3_{f_X}/2-s^2_W Q_f)$ and $e = gs_W$. Here $s_W$ and $c_W$ are the sine and cosine of the weak mixing angle, $g$ is the weak coupling constant and $T^3_{f_X}$ is the weak isospin of the fermion $f_X$, and $V_{kl}$ are the elements of the CKM matrix entering in the $W$ boson contribution to the CC process. With $m_{Z,W}$ and $\Gamma_{Z,W}$ we denote the mass and width of the massive gauge bosons. The SM gauge couplings are evaluated at the scale of $1~\mathrm{TeV}$.

With the amplitudes defined in Eq.~\eqref{eq:amplitudes_NC_CC} we can compute the unpolarized partonic differential cross-section for either an NC or CC process, still with free flavor indices $ijkl$, as
\begin{equation}
     \frac{d\hat\sigma^F_{ijkl}}{d\hat t}(\hat s, \hat t) = \sum_{\Gamma, \Gamma' = S,V,T}\sum_{X,Y=L,R}  \frac{1}{48\pi \hat s^2} f^{\Gamma\Gamma'}_{XY}(\hat s,\hat t) \left[F^{\Gamma XY}_{ijkl}(\hat s)\right]\left[F^{\Gamma' XY}_{ijkl}(\hat s)\right]^*\,,
\label{eq:dsigmadt_partonic}
\end{equation}
where $\hat t=(p_q - p_e)^2 $, the functions $f^{\Gamma\Gamma'}_{XY}(\hat s,\hat t)$ arise from the Dirac trace, and $F = N_u, N_d, C$ is a particular form factor. Assuming all initial and final state particles are massless
\begin{equation}
\begin{split}
    f^{SS}_{LL}(\hat s,\hat t) = f^{SS}_{LR}(\hat s,\hat t) &= f^{SS}_{RL}(\hat s,\hat t) = f^{SS}_{RR}(\hat s,\hat t)  = \frac{\hat s^2}{4}\,, \\
    f^{VV}_{LL}(\hat s,\hat t) = f^{VV}_{RR}(\hat s,\hat t) & = (\hat s+\hat t)^2\,, \\
    f^{VV}_{LR}(\hat s,\hat t) = f^{VV}_{RL}(\hat s,\hat t) & = \hat t^2\,, \\
    f^{TT}_{RL}(\hat s,\hat t) = f^{TT}_{LR}(\hat s,\hat t) & =  4(\hat s+2\hat t)^2\,, \\
    f^{TT}_{LL}(\hat s,\hat t) = f^{TT}_{RR}(\hat s,\hat t) & =  0\,, \\
    f^{ST}_{RL}(\hat s,\hat t) = f^{ST}_{LR}(\hat s,\hat t) & =  -\hat s(\hat s+2\hat t)\,, \\
    f^{ST}_{LL}(\hat s,\hat t) = f^{ST}_{RR}(\hat s,\hat t) & =  0\,. \\
\end{split}
\end{equation}
We note that only the scalar-tensor interference term is present at the differential level and no other Lorentz structures interfere with each other.

Integrating the differential cross-section given in Eq.~\eqref{eq:dsigmadt_partonic} over the whole range of $\hat{t}$ results in a total cross-section as a function of the partonic center-of-mass energy
\begin{equation}
    \hat\sigma^F_{ijkl}(\hat s) =  \int_{-\hat{s}}^0 d\hat t \frac{d\hat\sigma^F_{ijkl}}{d\hat t}(\hat s, \hat t).
\label{eq:sigma_partonic_total}
\end{equation}
In the case of NC Drell-Yan processes ($F=N_u,N_d$), $\hat{s}$ can be accessed in measurements through the invariant mass of the dilepton pair $m_{\ell\ell} \equiv \sqrt{\hat{s}}$. Moreover, integrating over $\hat t$ results in zero contribution of the scalar-tensor interference to the total partonic and hadronic cross-section.

However, in the case of CC Drell-Yan processes, $\hat{s}$ is difficult to access experimentally. Instead, the measurements are reported as differential distributions in transverse momentum $p_T$ of the charged lepton, or in transverse mass,
\begin{equation}
m_T = \sqrt{2p_T^lp_T^{miss}(1-\cos(\Delta \phi(\vec p_T^l,\vec p_T^{miss})))} ~.
\end{equation}
Note that theoretically, they are equivalent with $m_T=2 p_T$. In this case, we can perform a partial integration of Eq.~\eqref{eq:dsigmadt_partonic}, fully over the $z$ direction, but keeping the differential in the transverse direction. This results in the differential cross-section for CC Drell-Yan in terms of the transverse mass
\begin{equation}
    \frac{d\hat\sigma^C_{ijkl}}{d m_T}(\hat s, m_T) = \sum_{\Gamma,\Gamma' }\sum_{X,Y}\frac{1}{96\pi}\frac{m_T}{\sqrt{\hat s^2 - \hat s m_T^2}} f^{\Gamma\Gamma'}_{XY}(\hat s,\hat t^\pm(m_T))\left[C^{\Gamma XY}_{ijkl}(\hat s)\right]\left[C^{\Gamma' XY}_{ijkl}(\hat s)\right]^\ast\,,
\end{equation}
with $f^{\Gamma\Gamma'}_{XY}(\hat s, \hat t^\pm(m_T)) = f^{\Gamma\Gamma'}_{XY}(\hat s, -\frac{1}{2}(\hat s+\sqrt{\hat s(\hat s-m_T^2)})) + f^{\Gamma\Gamma'}_{XY}(\hat s, -\frac{1}{2}(\hat s-\sqrt{\hat s(\hat s-m_T^2)}))$. We note that the scalar-tensor interference again does not contribute to the differential cross-section due to $f^{ST}_{XY}(\hat s, \hat t^\pm(m_T)) = 0$.

\subsubsection*{Hadronic cross-sections}
\label{sec:hadron_xsec}
A differential hadronic cross-section is obtained by convoluting a partonic cross-section with parton luminosity functions
\begin{equation}
    \mathcal L_{\bar q q}(\tau,\mu_F) = \int_\tau^1 \frac{dx}{x} f_{\bar q}(x,\mu_F)f_{q}(\tau/x,\mu_F)\,,
\end{equation}
where $f_q$ are parton distribution functions (PDFs) and $\mu_F$ is the factorisation scale. We extended the functionality of \texttt{flavio} by interfacing it with the \texttt{parton} package, taking care of downloading and interpolating a chosen PDF set so that it is ready for efficient use. In this paper, we show results using the {\tt NNPDF40\_nnlo\_as\_0118} PDF set \cite{NNPDF:2021njg}, but it is straightforward to change to a different PDF set.

\begin{figure}[t]
    \centering
    \includegraphics[width=\textwidth]{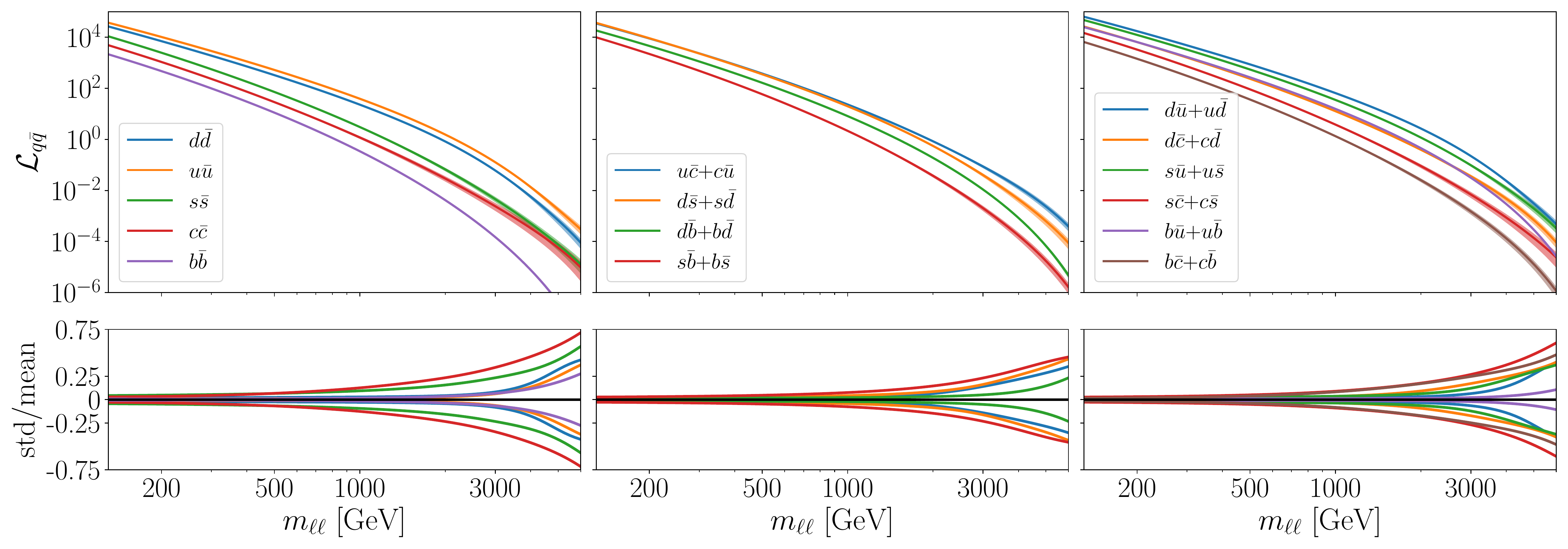}
    \caption{Parton luminosity functions using the {\tt NNPDF40\_nnlo\_as\_0118} PDF set for selected quark pairs. See Section~\ref{sec:hadron_xsec} for details.}
    \label{fig:luminostiy}
\end{figure}

Fig.~\ref{fig:luminostiy} shows a set of representative parton luminosity functions. The factorization scale for the PDFs is set to the center of the bin over which the integral is performed. By default the central PDF is used, however, we also made it straightforward to change to another PDF replica in order to study the uncertainties associated with PDFs. The relative uncertainty bands are shown in the bottom panel for various flavors. The uncertainties are most significant for the strange and charm quarks and explode for $m_{\ell \ell} \gtrsim 5$\,TeV. Luckily, the most sensitive bins are around $2$\,TeV, where the PDF knowledge is still adequate.\footnote{The Drell-Yan data from the LHC is used in global PDF fits. In principle, the correct procedure is to perform simultaneous fits of both the PDFs and the EFT parameters. However, as shown in Ref.~\cite{Greljo:2021kvv}, this approach might become necessary only at the HL-LHC, where the data will be much more precise. For the time being, it is a good approximation to use the SM determination of PDFs in the EFT fits.} More discussion on this point will follow at the end of Section~\ref{sec:validity}.

The differential cross section for the NC Drell-Yan process $p\, p \to \ell^+\, \ell^-$ can then be written as
\begin{equation}
    \frac{d\sigma(p\, p \to \ell^-_i\, \ell^+_j)}{dm_{\ell\ell}} = \sum_{q = u, d} \sum_{k,l=1}^3  2\frac{2 m_{\ell\ell}}{s}  \mathcal L_{\bar q^k q^l}\left(\frac{m_{\ell\ell}^2}{s},\mu_F\right) \hat\sigma^{N_q}_{ijkl}(m_{\ell\ell}^2)
    \label{eq:dsigmadmll}
\end{equation}
where $s$ is the hadronic center-of-mass energy. The first factor of $2$ is a symmetry factor due to the exchange of protons.

For the CC Drell-Yan process $p p \to \ell \nu$, the partonic cross-sections for the two charge-conjugated processes are equal, but when calculating the hadronic cross-section, they are convoluted with different parton luminosities. In particular,
\begin{equation}
\begin{split}
    \frac{d\sigma(p\, p \to \ell_i\, \nu)}{dm_T} &=  \int_{m_T^2/s}^1 d\tau\, \frac{d^2\sigma(p\, p \to \ell_i\, \nu)}{d\tau d m_T} \\
    &=\int_{m_T^2/s}^1 d\tau\,\sum_{j,k,l=1}^3 2 \mathcal L_{\bar u^k d^l + \bar d^l u^k}\left(\tau s,\mu_F\right) \frac{d\hat\sigma^C_{ijkl}}{dm_T}(\tau s, m_T)~,
    \label{eq:dsigmadmT}
\end{split}
\end{equation}
where
\begin{equation}
    \mathcal L_{\bar u^k d^l + \bar d^l u^k}\left(\tau s,\mu_F\right) = \mathcal L_{\bar u^k d^l}\left(\tau s,\mu_F\right) + \mathcal L_{\bar d^l u^k}\left(\tau s,\mu_F\right)\,.
\end{equation}
In this case, we also sum over the neutrino flavors in the final state as they cannot be determined experimentally.

Ultimately both the NC and CC hadronic differential cross-sections are numerically integrated in a particular $m_{\ell\ell}$ or $m_T$ bin. In the case of NC Drell-Yan, see Eq.~\eqref{eq:dsigmadmll}, this amounts to an evaluation of a single numerical integral
\begin{equation}
    \sigma^\mathrm{NC}_{\rm bin} = \int_{m_{\ell\ell}^{\rm min}}^{m_{\ell\ell}^{\rm max}} dm_{\ell\ell} \frac{d\sigma}{dm_{\ell\ell}}\,.
\end{equation}
As the luminosity functions are pre-computed and interpolated over $\tau$, and the partonic cross-section is analytically integrated, see Eq.~\eqref{eq:sigma_partonic_total}.
However, in the case of CC Drell-Yan, see Eq.~\eqref{eq:dsigmadmT}, there is an additional non-trivial integration over the $\tau$ variable
\begin{equation}\label{eq:3.16}
    \sigma^\mathrm{CC}_{\rm bin} = \int_{m_T^{\rm min}}^{m_T^{\rm max}}dm_T \int_{m_T^2/s}^1  d\tau\frac{d^2\sigma}{d\tau dm_T}\,.
\end{equation}
In an effort to decrease the computational complexity, we perform the following non-trivial change of variables, ultimately reducing also the CC Drell-Yan predictions to a single numerical integration. First, we substitute $u = \frac{m_T^2}{s}$, and get
\begin{equation}
    \sigma_{\rm bin}^\mathrm{CC} = \int_{u_{\rm min}}^{u_{\rm max}} du \int_u^{1} d\tau~ \frac{d^2\sigma}{d\tau du}(\tau,u) \,.
\end{equation}
In this expression, we always have $0 < u_{\rm min} < u < u_{\rm max} < 1$ and can therefore split the integral over $\tau$ as follows
\begin{equation}
    \sigma^{\rm CC}_{\rm bin} =
    \int_{u_{\text{min}}}^{u_{\text{max}}} d u
    \int_{u}^{u_{\text{max}}} d \tau
    \frac{d\sigma}{d u\, d\tau}(\tau,u)
    +
    \int_{u_{\text{min}}}^{u_{\text{max}}} d u
    \int_{u_{\text{max}}}^{1} d \tau
    \frac{d\sigma}{d u\, d\tau}(\tau,u)\,.
\end{equation}
In the second term, we can now simply exchange the $u$ and $\tau$ integrals since all integration boundaries are constants. The first term, on the other hand, corresponds to integration over a triangular area, for which we can use that
\begin{equation}
    \int_{u_{\rm min}}^{u^{\rm max}} du \int_u^{u_{\rm max}} d\tau = \int_{u_{\rm min}}^{u^{\rm max}} d\tau \int_{u_{\rm min}}^\tau du\,.
\end{equation}
Since the parton luminosities only depend on $\tau$, we can move them outside the $u$ integral
\begin{equation}
    \begin{aligned}
    \sigma_\mathrm{bin}^\mathrm{CC}(p\, p \to \ell_i\, \nu) &=
    \int_{u_{\text{min}}}^{u_{\text{max}}} d \tau\,
    \sum_{j,k,l=1}^3 2 \mathcal L_{\bar u^k d^l + \bar d^l u^k}\left(\tau s,\mu_F\right)
    \int_{u_{\text{min}}}^{\tau} d u
    \frac{d\hat\sigma^C_{ijkl}}{du}(\tau s, u)
    \\
    &+
    \int_{u_{\text{max}}}^{1} d \tau\,
    \sum_{j,k,l=1}^3 2 \mathcal L_{\bar u^k d^l + \bar d^l u^k}\left(\tau s,\mu_F\right)
    \int_{u_{\text{min}}}^{u_{\text{max}}} d u
    \frac{d\hat\sigma^C_{ijkl}}{du}(\tau s, u) \,,
    \end{aligned}
\end{equation}
where
\begin{equation}
    \frac{d\hat\sigma^C_{ijkl}}{du}(\tau s, u)=
    \sqrt{\frac{s}{4 u}}\frac{d\hat\sigma^C_{ijkl}}{dm_T}(\tau s, m_T=\sqrt{u s})\,.
\end{equation}
Let us now define
\begin{equation}
    \hat\sigma^C_{ijkl}(\tau s;a,b) \equiv
    \hat\sigma^C_{ijkl}(\tau s;b)-\hat\sigma^C_{ijkl}(\tau s;a)
    =
    \int_{a}^{b} d u
    \frac{d\hat\sigma^C_{ijkl}}{du}(\tau s,u)~,
\end{equation}
which can now be evaluated analytically. After doing so, there is only a single integration over $\tau$ left to be done numerically, and we arrive at the final expression
\begin{equation}
\begin{split}
    \sigma_\mathrm{bin}^\mathrm{CC}(p\, p \to \ell_i\, \nu) & =\int_{u_{\text{min}}}^{1} d \tau\,
    \sum_{j,k,l=1}^3 2
    \mathcal L_{\bar u^k d^l + \bar d^l u^k}\left(\tau s,\mu_F\right) \\
    & \quad
    \times\left(
        \Theta(\tau-u_{\text{max}})\,
        \hat\sigma^C_{ijkl}\Big(\tau s;u_{\text{max}}\Big)
        -
        \hat\sigma^C_{ijkl}\Big(\tau s;u_{\text{min}}\Big)
    \right)\,,
\end{split}\label{eq:3.23}
\end{equation}
where $\Theta$ is the Heaviside step function. The analytical simplification from Eq.~\eqref{eq:3.16} to Eq.~\eqref{eq:3.23} greatly reduces the computational resources needed for numerical integration.

To summarise, the predictions of both the NC and CC Drell-Yan hadronic cross sections at LO can be efficiently evaluated in \texttt{flavio} in a particular $m_{\ell\ell}$ or $m_T$ bin, effectively boiling down to a single $1\mathrm{D}$ numerical integration, taking into account both the SM contributions as well as the contributions from dimension-6 SMEFT operators from Table~\ref{tab:SMEFTmatching}.

\subsection{Measurements and Likelihoods}
\label{sec:data}
In order to contrast the NP predictions with experiments, we implement data from four recent experimental searches, corresponding to $\sim140~\mathrm{fb}^{-1}$, into the database of measurements in \texttt{flavio}. These are CMS \cite{CMS:2021ctt} and ATLAS \cite{ATLAS:2020yat} searches in the high-mass dilepton final states, and CMS \cite{CMS:2022yjm} and ATLAS \cite{ATLAS:2019lsy} searches in charged lepton and missing transverse momentum final states. We summarise the information about quantities extracted from the available measurements in Table~\ref{tab:SearchSummary} and in the following paragraphs.

\begin{table}[t]
\renewcommand*{\arraystretch}{1.2}
\centering
\caption{Summary information on the implemented searches. The asterisk indicates a special treatment of systematic uncertainties. See the text for more details.}
\label{tab:SearchSummary}
\begin{tabular}{cc|ccc|cc}
Search                 & Ref.                         & Channel & Luminosity        & Figure          & HepData & Digitized \\ \hline
\multirow{2}{*}{ATLAS} & \multirow{2}{*}{\cite{ATLAS:2020yat}} & $pp\to ee$ & 139 fb$^{-1}$    & Aux.Fig.~1a     & \multirow{2}{*}{\thead{$\Delta^*$}}        & \multirow{2}{*}{\thead{$N_{\rm obs}, N_{\rm SM}$\\$N_{\rm DY}$\\}}        \\
               &                                   & $pp\to \mu\mu$ & 139 fb$^{-1}$ & Aux.Fig.~1b     &         &         \\ \hline
\multirow{2}{*}{CMS}   & \multirow{2}{*}{\cite{CMS:2021ctt}} & $pp\to ee$ & 137 fb$^{-1}$    & Fig.~2 (left)    & \multirow{6}{*}{\thead{$N_{\rm obs}$\\ $N_{\rm SM}$\\ $\Delta$}}        & \multirow{6}{*}{\thead{$N_{\rm DY}$}}           \\
                       &                                   & $pp\to \mu\mu$ & 140 fb$^{-1}$ & Fig.~2 (right)  &         &           \\ \cline{1-5}
\multirow{2}{*}{ATLAS} & \multirow{2}{*}{\cite{ATLAS:2019lsy}} & $pp\to e\nu$ & 139 fb$^{-1}$  & Fig.~1 (top)    &        &          \\
                       &                                   & $pp\to \mu\nu$ & 139 fb$^{-1}$ & Fig.~1 (bottom) &         &     \\ \cline{1-5}
\multirow{2}{*}{CMS}   & \multirow{2}{*}{\cite{CMS:2022yjm}} & $pp\to e\nu$ & 138 fb$^{-1}$  & Fig.~4 (left)   &        &      \\
                       &                                   & $pp\to \mu\nu$ & 138 fb$^{-1}$ & Fig.~4 (right)  &    &
\end{tabular}
\end{table}

\paragraph{Observed number of events:} The CMS and ATLAS searches of the NC and CC Drell-Yan processes report the observed number of events $N_\mathrm{obs}$ as binned distributions in the dilepton invariant mass $m_{\ell\ell}$ and transverse mass $m_T$, respectively. Moreover, they report results for the electron and muon channels separately. The events were subjected to a standard set of $p_T$, $\eta$, and isolation cuts, see Refs.~for details. We extract the information on $N_\mathrm{obs}$ either from HepData (if available) or by digitizing the Figures provided in the search publications (see Table~\ref{tab:SearchSummary}).

\paragraph{Expected number of events:} All the searches provide state-of-the-art binned predictions of the expected number of signal ($N_{\rm DY}$) events at NNLO order in QCD with NLO EW corrections, and background ($N_\mathrm{bkg}$) events in the SM, including detector and cut effects. The main sources of background are $t\bar t$, $tW$, $VV$, $\tau\tau$ and jet misidentification. The total number of expected events in each bin $N_\mathrm{SM} = N_\mathrm{DY} + N_\mathrm{bkg}$ is reported with a combined systematic uncertainty $N_\mathrm{SM} \pm \Delta$. The information on $N_\mathrm{SM}, N_\mathrm{DY}$ and $\Delta$ is extracted either from HepData (if available) or digitized from Figures in search publications (see Table~\ref{tab:SearchSummary}), with $N_\mathrm{bkg}$ determined as $N_\mathrm{bkg}=N_\mathrm{SM}-N_\mathrm{DY}$. The ATLAS search in neutral currents required special treatment, as the systematic uncertainty was not reported. We resorted to extracting the relative systematic uncertainty from an older search by the same collaboration, with $36~\mathrm{fb}^{-1}$ of data~\cite{ATLAS:2017fih}. Under a reasonable assumption that the relative systematic uncertainty did not significantly change with increasing luminosity, we used these numbers to calculate the absolute uncertainty at the current luminosity. As a by-product, we leave the older search implemented in the database of measurements, but in the following, we report results only with the latest data.

\paragraph{Likelihood:} Next, we discuss how to contrast the predictions implemented in \texttt{flavio} with the available experimental data. We rely on the reported expected number of SM binned events $N_\mathrm{SM} = N_\mathrm{DY} + N_\mathrm{bkg}$. In each bin, we then reweigh the expected number of SM Drell-Yan events with the ratio of cross-sections, predicted using the implementation discussed in the previous subsection, in the following way~\cite{Greljo:2017vvb}
\begin{equation}
\label{eq:R-ratio}
    R = \frac{\sigma^{\rm SM + NP}_{\rm bin}}{\sigma^{\rm SM}_{\rm bin}}\,, \qquad N_\mathrm{DY}^\mathrm{NP} = R \, N_\mathrm{DY} \,,
\end{equation}
where $\sigma^{\rm SM}_{\rm bin}$ includes only the SM contributions, whereas $\sigma^{\rm SM + NP}_{\rm bin}$ contains both SM and the NP contributions under a certain NP hypothesis.\footnote{For muons and electrons the detector level $m_{\ell \ell}$ and $m_T$ variables are highly correlated with the corresponding truth level variables as illustrated in Fig.~2.1 of Ref.~\cite{Allwicher:2022mcg} with the detector response matrix. This, however, is not the case for taus for which the migration of the events across bins is more pronounced. Thus, the approximation of the NP weights with the truth level $R$ ratio might be questionable. Instead, one should perform full-fledged simulations, see  e.g.~\cite{Greljo:2018tzh,Fuentes-Martin:2020lea,Faroughy:2016osc}. We urge experimental collaborations to report unfolded distributions as those could be straightforwardly included in the {\tt flavio} framework.} Notice that in the limit of $\mathrm{NP}\to0$ we recover the state-of-the-art SM predictions for the expected number of DY events provided by the experimental searches.\footnote{For recent precision calculations of high-mass Drell-Yan in the SM, see Refs.~\cite{Duhr:2021vwj,Bonciani:2021zzf,Armadillo:2022bgm}. The SM prediction of the $m_{\ell \ell}$ spectrum in the TeV region is known to $\lesssim 1\%$ modulo PDF uncertainties, while the statistical uncertainty in the tails is $\mathcal{O}(10\%)$.} By considering a certain NP hypothesis, we are only smoothly deforming the distribution of the expected number of DY events. Moreover, in the $R$ ratio, we expect the higher order corrections to the cross sections to factorize and largely cancel. The detector and cut effects are expected to be sufficiently captured by the approach.  The validation of these assumptions is detailed in Appendix \ref{app:implementation}. Since the $R$ ratio does not contribute significantly to the systematic uncertainty, we assume that the reported $\Delta$ remains unchanged for the newly calculated expected number of events under a NP hypothesis $N_\mathrm{tot} = N_\mathrm{DY}^\mathrm{NP} + N_\mathrm{bkg}$.

\begin{figure}[t]
    \centering
    \includegraphics[width=0.5\textwidth]{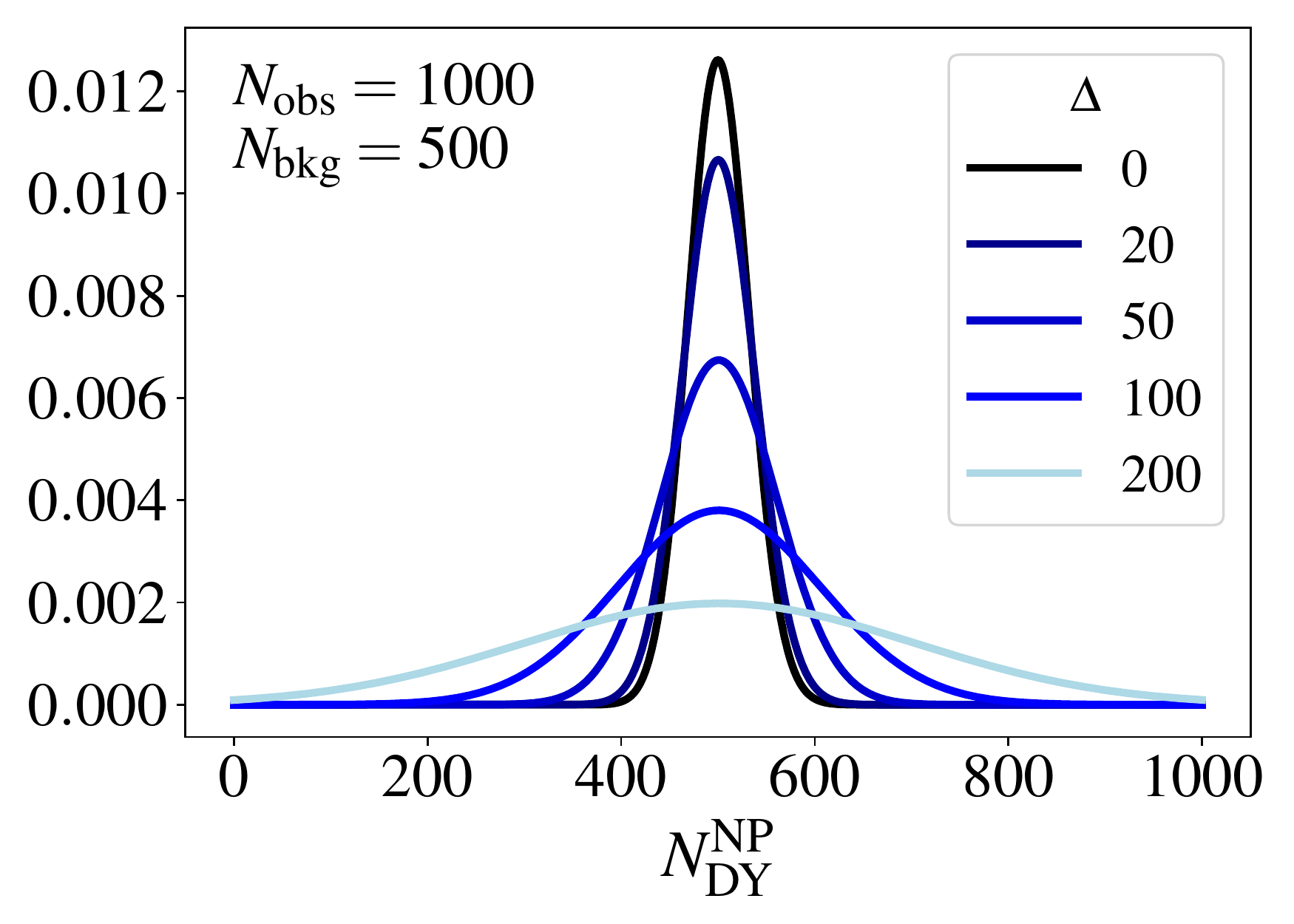}~\includegraphics[width=0.5\textwidth]{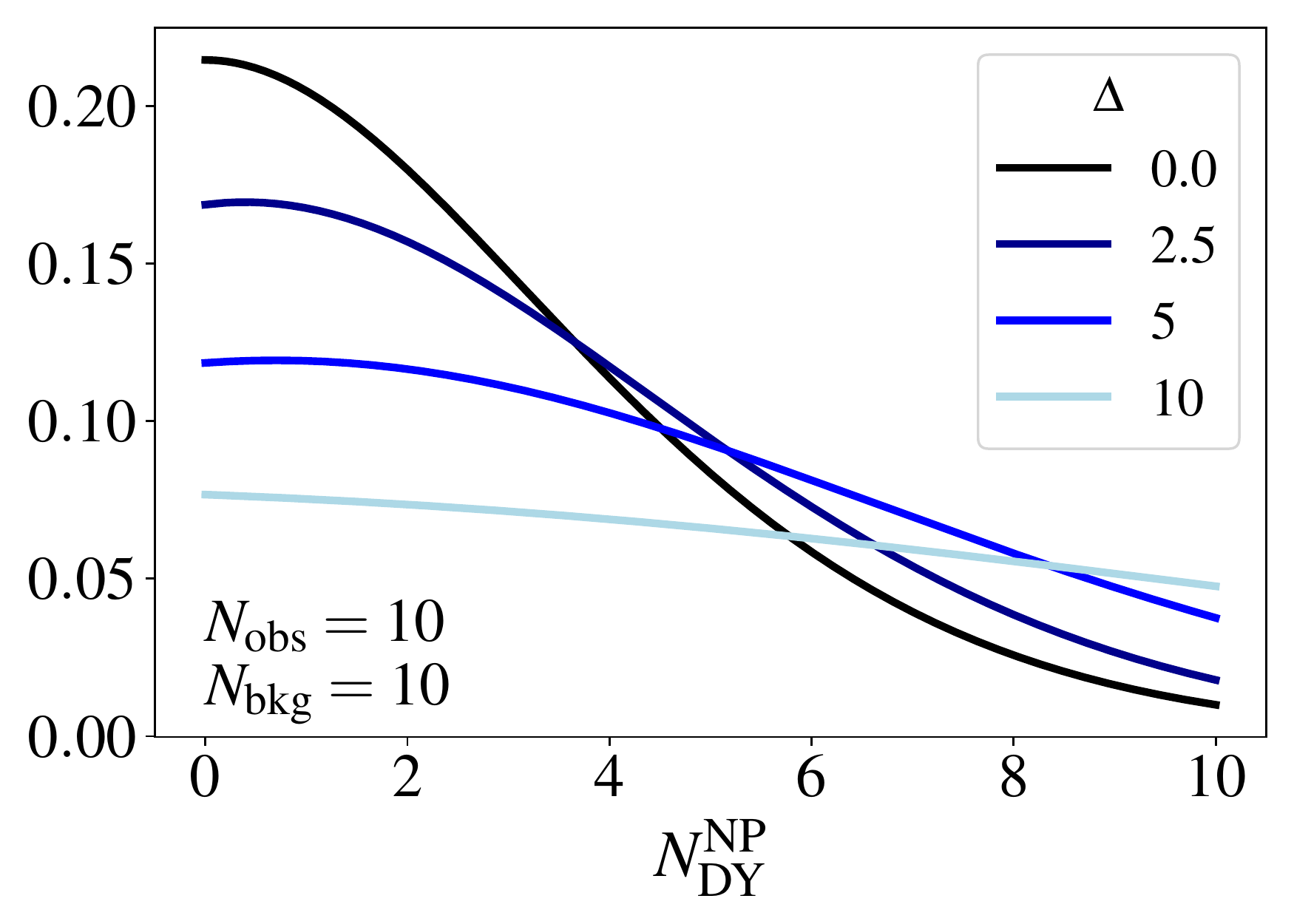}~
    \caption{Illustrative examples of the likelihood defined in Eq.~\eqref{eq:likelihood}. See text for details.}
    \label{fig:likelihood}
\end{figure}

With both the expected $N_\mathrm{tot}$ and observed $N_\mathrm{obs}$ number of events at hand, we construct a likelihood by considering the number of events in each bin as an independent Poisson variable. Moreover, we account for the systematic uncertainty discussed in the previous paragraphs by convolving the Poisson distribution with the Normal distribution centered at $0$ with standard deviation $\Delta$, so that the resulting likelihood is
\begin{equation}
f(N_\mathrm{DY}^\mathrm{NP}|N_\mathrm{obs}) = \int d\tau \frac{(N_\mathrm{DY}^\mathrm{NP} + N_\mathrm{bkg} + \tau)^{N_\mathrm{obs}} e^{-(N_\mathrm{DY}^\mathrm{NP} + N_\mathrm{bkg}+\tau)}}{N_\mathrm{obs}!} \mathcal{N}(0, \Delta)(\tau).
\label{eq:likelihood}
\end{equation}
We show the effect of the convolution with the Normal distribution in Fig.~\ref{fig:likelihood} on two illustrative examples. The left plot is assuming $N_\mathrm{obs}=1000$ and $N_\mathrm{bkg}=500$, whereas the right plot is assuming $N_\mathrm{obs}=10$ and $N_\mathrm{bkg}=10$. In both cases, we show the likelihood of having $N_\mathrm{DY}^\mathrm{NP}$ Drell-Yan events given the observed number of events and expected number of background events, with various values of the uncertainty $\Delta$. Notice that increasing $\Delta$ results in smoothly broadening the Poisson likelihood, corresponding to our imperfect knowledge of the expected number of events.

Finally, as experimental correlations between different bins are not reported, they are assumed to be uncorrelated. The total likelihood is constructed as a product of $f(N_\mathrm{DY}^\mathrm{NP}|N_\mathrm{obs})$ over all bins $L = \Pi_{\rm bin} f_{\rm bin}$. The log-likelihood is then expected to follow a chi-square distribution (in the Gaussian limit), $\chi^2 = -2\log(L)$.
The best-fit point corresponds to the global minimum of this function while the $n\sigma$ confidence level (CL) region satisfies $\chi^2 - \chi^2_{min} < \Delta_{n\sigma}$. Explicitly, $\Delta_{\{1,2\}\sigma} = \{1,4\}$ for one degree of freedom and $\Delta_{\{1,2\}\sigma} = \{2.3,6.18\}$ for two degrees of freedom.

The likelihood was validated by comparing the bounds on Wilson coefficients reported by the experimental collaborations in the references given in Table~\ref{tab:SearchSummary} to the bounds derived by our method. Despite a difference in the data treatment, different treatment of uncertainties, and the overall difference in the statistical analysis, our bounds are in excellent agreement with the experimental reports, with up to $\mathcal{O}(10\%)$ agreement or better.

\paragraph{Numerical efficiency}
In order to be able to efficiently scan the multidimensional SMEFT parameter space and at the same time to include hundreds of observables in the likelihood, we use the numerical method of~\cite{Altmannshofer:2021qrr} to implement our likelihood function.
In particular, we make use of the fact that each observable is given in terms of amplitudes that are linear functions of the NP Wilson coefficients.
Consequently, we can express each of the observables~$O_k$ as a function~$f_k$ of $n$ polynomials $p_i$ that is of second order in the NP Wilson coefficients,
\begin{equation}
 O_k = f_k(p_1,p_2, ..., p_n)\,.
\end{equation}
These functions can be trivial, like in the case of cross sections and branching ratios that are themselves second-order polynomials, i.e., $f_k(p_1)=p_1$.
But the function of a single observable can also depend on several polynomials and might involve ratios and square roots of polynomials.
In any case, given the numerical values of the polynomials~$p_i$, the observables can be computed very efficiently, while still keeping their full, potentially non-polynomial dependence on the Wilson coefficients.

The problem of computing the observables $O_k$ in a numerically efficient way thus reduces the problem of efficiently computing the polynomials $p_i$.
Following~\cite{Altmannshofer:2021qrr}, they are written as the scalar vector product
\begin{equation} \label{eq:polynomial}
 p_i
 = \vec{p}_i \cdot \vec{V}
 = \begin{pmatrix}
    a_i\\
    \vec b_i\\
    \vec c_i
   \end{pmatrix}
 \cdot \begin{pmatrix}
         1\\
         \vec{C}\\
         \vec{D}
         \end{pmatrix}
 = a_i +  \vec{b}_i \cdot \vec{C} + \vec{c}_i\cdot \vec{D}\,,
\end{equation}
where $\vec{C}=(C_1,C_2,...,C_M)^T$ is a vector containing the NP Wilson coefficients and $\vec{D}=\text{vec}(\vec{C}\otimes\vec{C})$ is a vector containing their products.
The NP dependence only enters in the vector $\vec V$, while the vectors $\vec p_i$ are independent of the values of the Wilson coefficients.
Consequently, we can precompute the values of the $\vec p_i$ and then the NP predictions can be computed by evaluating the scalar vector product of the precomputed $\vec p_i$ and the NP-dependent $\vec V$, which is efficiently performed using the \texttt{NumPy}~\cite{harris2020array} library.
Since especially the computationally expensive numerical integrations can be precomputed and stored in $\vec p_i$, the overall computation time for the theoretical predictions of the observables entering our likelihood is reduced by a factor of $\mathcal{O}(10^{-3})$ compared to computing the predictions without the precomputed $\vec p_i$.

To further improve the computational efficiency, as done in Ref.~\cite{Aebischer:2018iyb}, we separate the observables into two classes: the first class consists of observables with negligible theory uncertainties, whereas the second class consists of observables with sizable correlated theory uncertainties, for which we precompute the theoretical covariance matrix including NP effects, as described in more detail in Ref.~\cite{Altmannshofer:2021qrr}.

\section{EFT validity and tree-level completions}
\label{sec:validity}

When interpreting effective field theory likelihoods in terms of explicit models, one has to be careful about the validity of the approach. As a rule of thumb, the EFT is valid when $M_{{\rm NP}}^2 > q^2$, where $|q|$ is the relevant scale of the process and $M_{{\rm NP}}$ is the new physics mass threshold. In this respect, the EFT analysis of $b$ hadron decays has a more extensive range of validity than the high-$p_T$ tails analysis. In this section, we discuss the correctness of the EFT description in the high-mass Drell-Yan, identifying classes of perturbative ultraviolet completions which admit the interpretation. We also discuss cases for which the Drell-Yan study should be performed in the explicit model and exemplify how to do that with {\tt flavio}.

Our first task is to identify the energy scale $|q| \equiv m_{\ell^+ \ell^-}$ that provides the most sensitive probe of the EFT effects in $p p \to \ell^+ \ell^-$. To this purpose, we construct a jack-knife likelihood for each bin defined by removing a given bin from the full expected likelihood, see the supplemental material of Ref.~\cite{Greljo:2018tzh}. Extracting the expected 95\% CL bound on a coefficient $C$ based on the full ($C_{\rm full}$) and jack-knife ($C_{\rm jack}$) likelihoods, we calculate the quantity $R_{\rm jack}=C_{\rm jack}/C_{\rm full}$. The bins with the higher $R_{\rm jack}$ contribute more to the overall bound and are, therefore, more sensitive.

\begin{figure}[t]
    \centering
    \includegraphics[width=0.48\textwidth]{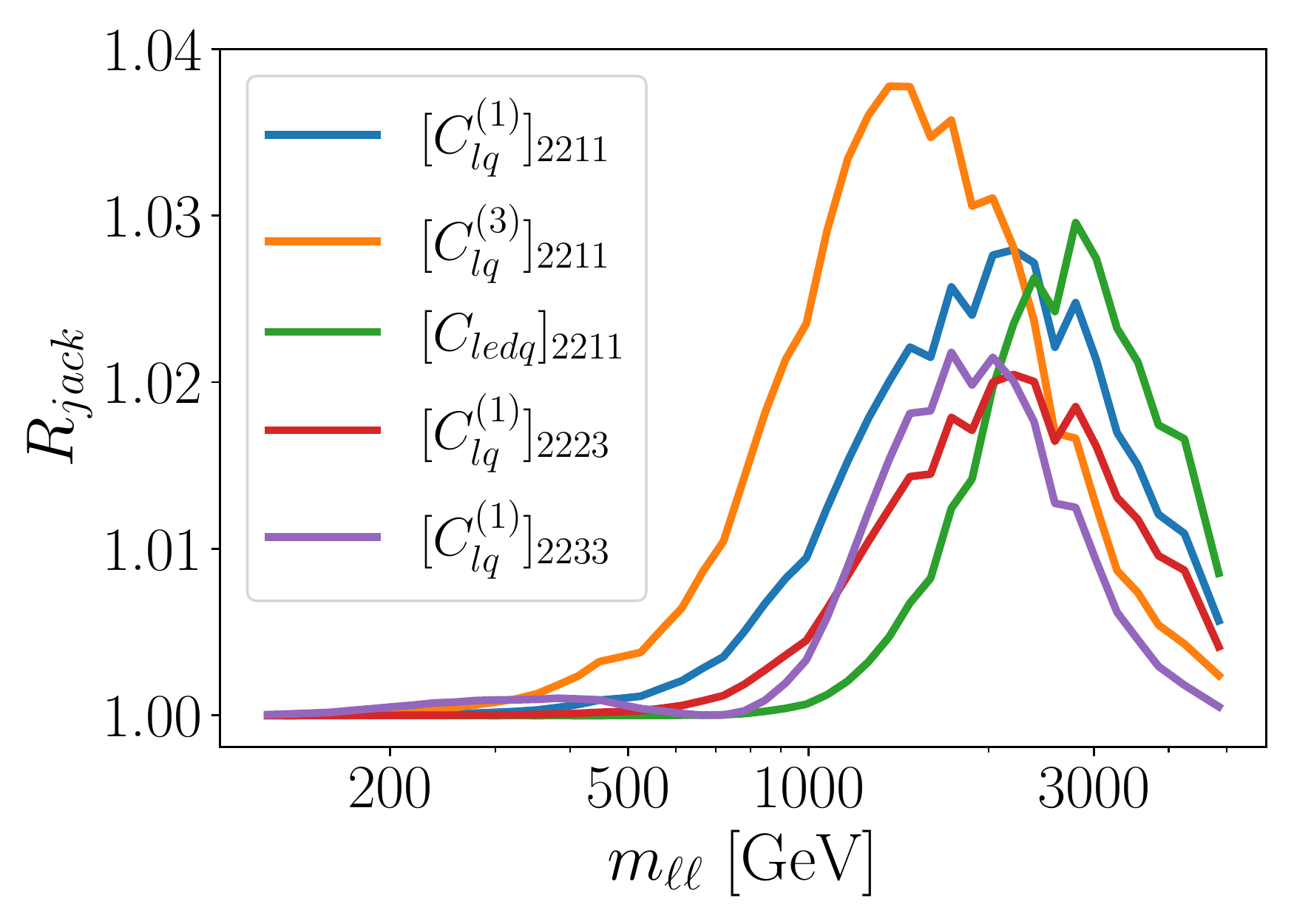}
    \includegraphics[width=0.48\textwidth]{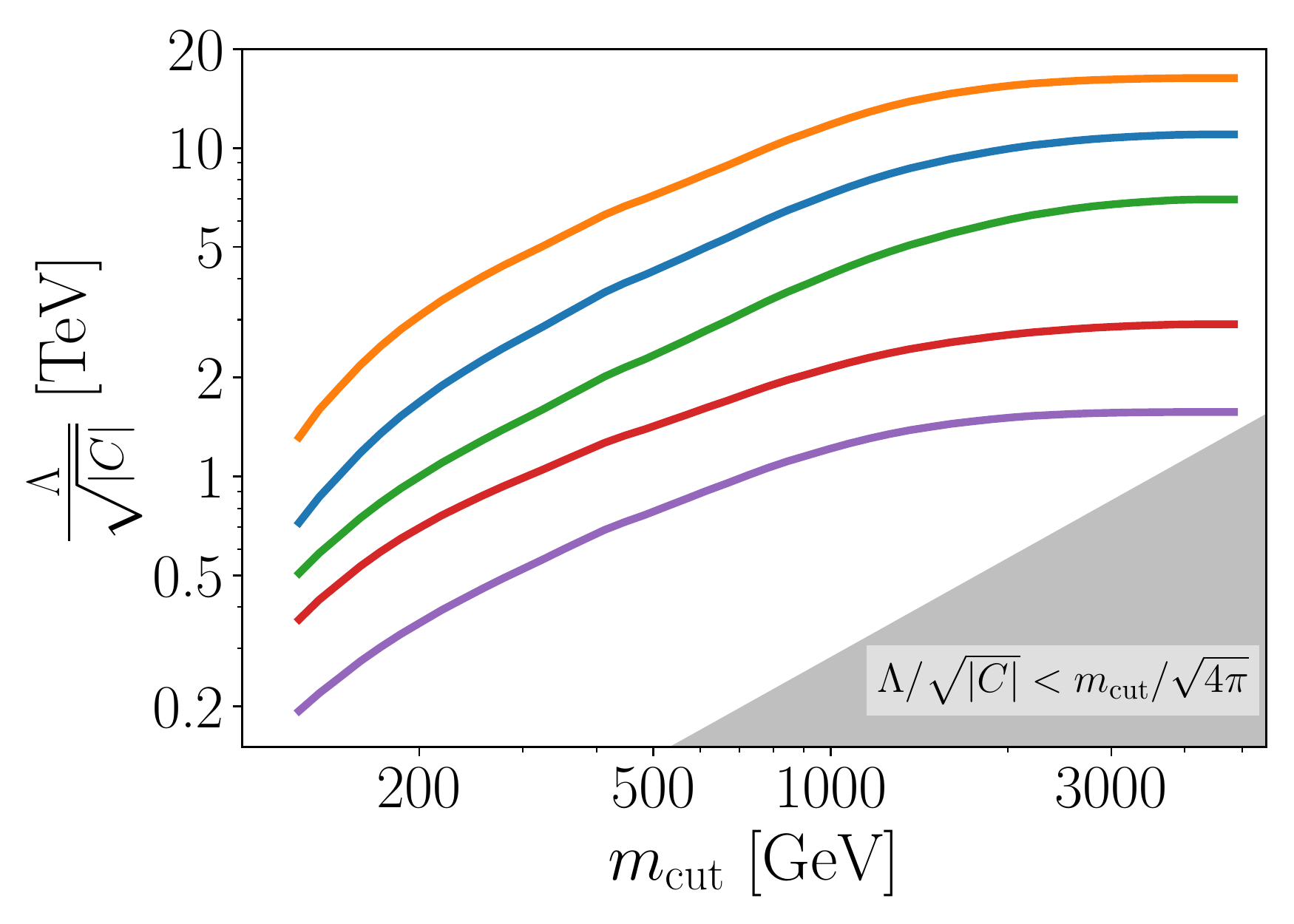}
    \caption{Sensitivity to selected Wilson coefficients from the invariant mass spectrum of $pp \to \mu^+ \mu^-$ search~\cite{CMS:2021ctt}. The plot assumes negative real values of the Wilson coefficients in Eq.~\eqref{eq:SMEFT}. This affects the sign of the interference with the SM which is, in this case, positive for up-type quarks. \textit{Left panel:} Values of $R_{\rm jack}$ for five different Wilson coefficients. \textit{Right panel:} The bound on the effective NP scale $\Lambda/\sqrt{|C|}$ as a function of the upper cut on the invariant mass of the lepton pair for a given Wilson coefficient. The shaded region can not be consistently achieved in a perturbative UV model. See Section~\ref{sec:validity} for details.}
    \label{fig:sensitivity}
\end{figure}

In the left panel of Fig.~\ref{fig:sensitivity} we plot $R_{\rm jack}$ for selected Wilson coefficients analyzing the data from the CMS search in the $pp \to \mu^+ \mu^-$ channel~\cite{CMS:2021ctt}. Broadly speaking, the most sensitive bins are found between 1\,TeV and 3\,TeV. However, the sensitivity depends on the size and the type of new physics contribution.  For example, in the case of $[Q_{lq}^{(3)}]_{2211}$, where the interference term largely dominates over the NP squared term, the sensitivity is shifted towards lower bins. On the other hand, the operator $[Q_{lq}^{(1)}]_{2211}$ receives partial interference cancellation due to the opposite sign for up- and down-type quarks. Here, the bound comes from an interplay of the interference and the NP squared term, which has a sizable effect only at higher $m_{\ell\ell}$. Thus, the sensitivity is shifted towards higher bins. Similarly, for operators that do not interfere with the SM when neglecting fermion masses (e.g. $[Q_{ledq}]_{2211}$), or for operators where the SM channel is loop-suppressed (e.g. $[Q_{lq}^{(1)}]_{2223}$), or in general when no valence quarks are involved (e.g. $[Q_{lq}^{(1)}]_{2233}$), the bound is (mostly) driven by the NP squared term and the most sensitive bins are between 2\,TeV and 3\,TeV.

When the sensitivity is dominated by the dimension-6 squared term, there is a potential issue with the missing contributions in the EFT. The dimension-8 operator interference with the SM diagram is of the same order in the EFT power counting as the dimension-6 squared contribution. However, the importance of those effects depends very much on the partonic channel involved. For flavor-violating channels generated by operators in Table~\ref{tab:1dbounds}, or chirality-flipping channels, such as those induced by $[Q_{ledq}]_{2211}$, the effect of dimension-8 operators interfering with the SM can be neglected since the SM part itself is negligible. The worrisome cases are flavor-conserving operators with heavy quark flavors such as $b \bar b$, where the dimension-6 squared term dominates the bound and the SM contribution is tree-level. The explicit model analyses have shown that even for such cases, there are valid model interpretations with tree-level mediators and large(ish) couplings in comparison with the electroweak gauge couplings. See Ref.~\cite{Allwicher:2022gkm} for explicit examples with dimension-8 operators included.

The above analysis can be improved by restricting the bins to be below some designated cutoff and extracting the EFT coefficient as a function of the cutoff. In Fig.~\ref{fig:sensitivity} (right), we show the expected bounds on the effective NP scale $\Lambda/\sqrt{|C|}$ for the same benchmarks as in the left plot, by increasingly including more and more bins in the likelihood (the region below the curves is excluded). The plot assumes negative real values of the Wilson coefficients in Eq.~\eqref{eq:SMEFT}. A different sensitivity for different cases can be understood in terms of the interplay of the linear and quadratic dimension-6 contributions and (or) the valence quark versus sea quark comparison.  We observe that for all presented operators, the bound saturates for $m_{{\rm cut}}$ between 1\,TeV and 3\,TeV as expected, and the inclusion (or exclusion) of bins at higher masses has a negligible effect on the bound.

Shown in gray in Fig.~\ref{fig:sensitivity}~(right) is the region which does not admit interpretation in a weakly coupled model. As a crude estimate, the lightest tree-level mediator consistent with the EFT assumption has a mass $M_{{\rm NP}} \sim m_{{\rm cut}}$ and the largest coupling to SM quarks and leptons consistent with the perturbativity limit $g_{NP} \sim \sqrt{4 \pi}$, implies $\frac{|C|}{\Lambda^2} \lesssim \frac{4\pi}{m_{{\rm cut}}^2}$ in a perturbative model. The plot shows that for all operator examples there is an interpretation in a perturbative tree-level model with $|C| \sim 1$. At the one-loop level, $|C| \sim 1/16\pi^2$, the EFT bounds are useful for much fewer operators, typically those involving valence quarks.

The catalog of all tree-level mediators with spins 0, $1/2$, or 1, which can be matched to dimension-6 semileptonic four-fermion operators in the SMEFT can be found in Refs.~\cite{deBlas:2017xtg,Allwicher:2022gkm}. These include scalar and vector leptoquarks, $Z'$ and $W'$ bosons, and extra Higgs bosons. The EFT approach approximates the $t$ (and $u$) channel resonances (leptoquarks) better than the colorless mediators exhibiting a resonance enhancement. For resonances with masses $|q| \gtrsim m_X \gtrsim v_{EW}$, the EFT analysis gives only qualitative bounds, while the true limits are driven by the on-shell production. For an $s$-channel mediator, the actual bounds are much more constraining due to the resonant enhancement, while for a $t$-channel, the EFT analysis is a good approximation (slightly aggressive, see, e.g., Fig.~3 of Ref.~\cite{Greljo:2018tzh}). The interplay between the on-shell $s$-channel production and the off-shell contribution to the tails depends on the mass of the resonance, as illustrated in Fig.~5 of Ref.~\cite{Greljo:2017vvb}. For heavy enough resonances, the limits from the tails always dominate due to the rapid kinematical suppression of the on-shell production.

\begin{figure}[t]
    \centering
    \includegraphics[width=\textwidth]{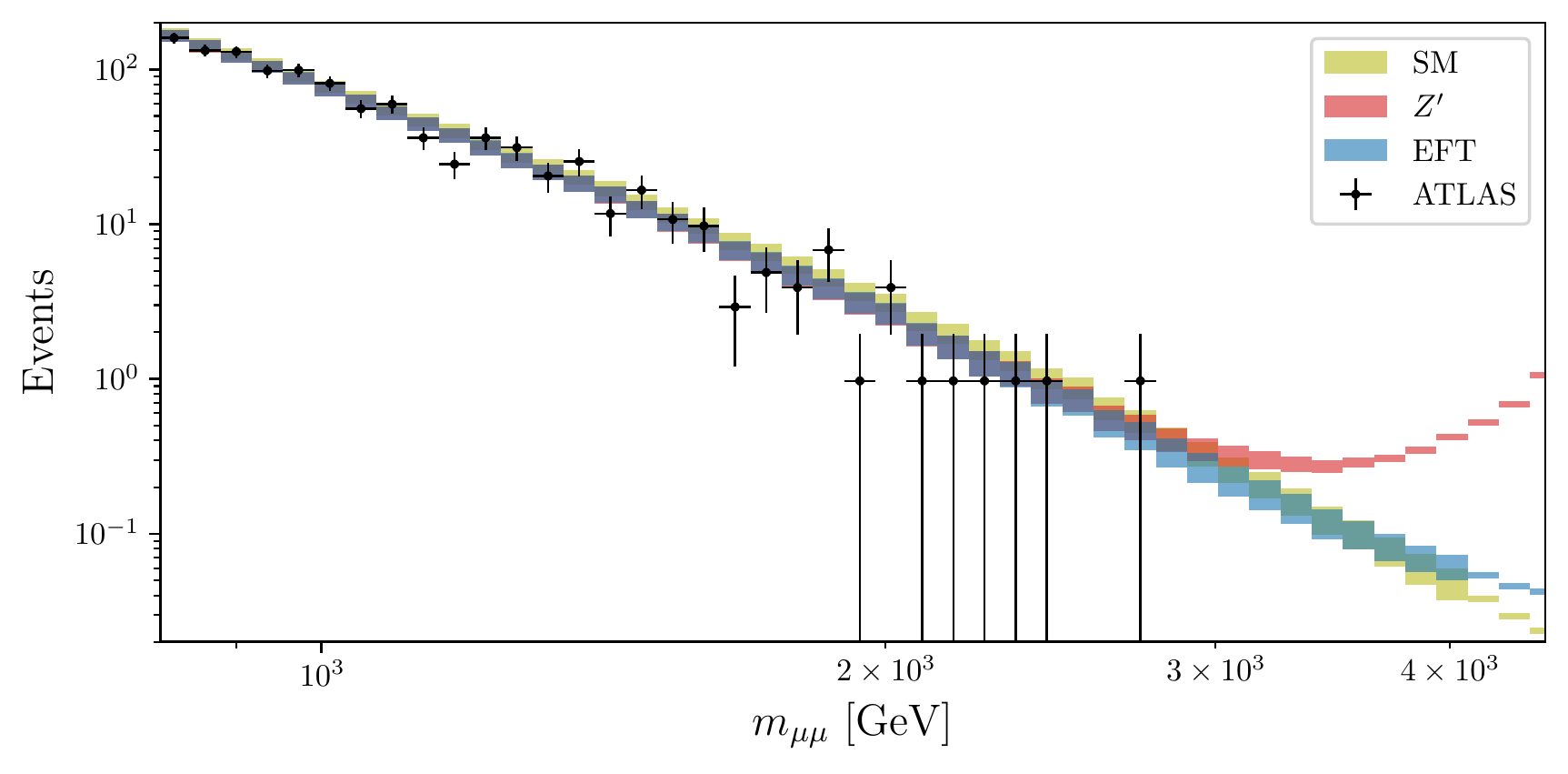}
    \caption{Comparison between experimental data and theory prediction in the $U(1)_{B-L}$ model with $m_X = 6\,$TeV, $g_X = 0.765$ both in the EFT and with a dynamical $Z'$ gauge bosons. The SM prediction (in yellow) and data (black markers) are taken from ATLAS search \cite{ATLAS:2020yat} in $pp\to\mu\mu$ channel. Both the $Z'$ (red) and the EFT (blue) predictions are obtained by reweighing the SM values as detailed in Section \ref{sec:data}.}
    \label{fig:tails_B-L}
\end{figure}

To illustrate this, we consider an explicit model example which we will closely study in Sec.~\ref{sec:modelI}. The heavy mediator $X_\mu$ is a gauge boson of the $U(1)_{B-L}$ symmetry. It is an $s$-channel mediator, $\bar q q \to X^* \to \ell^+ \ell^-$, see Eq.~\eqref{eq:B-Lcurrent}. The prediction for $p p \to \ell^+ \ell^-$ in the full model can easily be recovered in {\tt flavio} by introducing kinematics-dependent Wilson coefficients. In the concrete example, one replaces the mass $m_X$ with the full propagator in the matching expressions Eqs.~\eqref{eq:607} and~\eqref{eq:608}, $m_X^2 \to m_X^2 - \hat s - i\, m_X\, \Gamma_X$, where $\hat s \equiv m_{\ell^+ \ell^-}^2$ and $\Gamma_X$ is the total decay width of $X_\mu$ given by Eq.~\eqref{eq:Width}. The benchmark point used in the following discussion, $m_X = 6\,$TeV, $g_X = 0.765$ and $\epsilon_{ij}=0$, predicts $\Gamma_X / m_X = 10\%$. The search for an on-shell narrow resonance from Ref.~\cite{CMS:2021ctt} (Fig.~6) is insensitive to this benchmark point. However, as we will see, this benchmark is in tension with the high-mass Drell-Yan tails.\footnote{In passing, it is worth noting that the interpretation of the cross-section limits from the on-shell production on the high-mass resonances is polluted by the significant PDF uncertainties; see Fig.~\ref{fig:luminostiy}. Instead, the limits from the tails are more robust since those come from the lower energy bins where the PDFs are under control.}

Shown in Fig.~\ref{fig:tails_B-L} is the observed number of events (black markers with the error bars representing the statistical uncertainty) together with the SM predictions (yellow bands) taken from the ATLAS search \cite{ATLAS:2020yat} in the $pp\to\mu\mu$ channel.
The $Z'$ and EFT predictions were obtained by reweighing the SM expected number of events as detailed in Section \ref{sec:data}.
The EFT and the full model prediction agree very well up to $\sim2.5$ TeV, in the region of the most sensitive bins.
The exclusion of this benchmark in the EFT is driven by the negative interference in the region of the spectrum, where the EFT and the full dynamical model agree very well.
Furthermore, above $\sim2.5$ TeV, the full $Z'$ model modifies the SM prediction much more drastically compared to the EFT.
Including the full propagation effects would therefore only result in strengthening the bound.

\section{Model-independent global fits and complementarity}
\label{sec:pheno}
With the well-established low-energy flavor phenomenology in \texttt{flavio} (discussed in Sec.~\ref{sec:low_energy}), and the newly implemented high-mass Drell-Yan phenomenology (discussed in Sec.~\ref{sec:implementation}) at hand, we study in this section the interplay between the two in the context of the SMEFT. We consider minimalistic flavor scenarios, where we only turn on certain SMEFT Wilson coefficients of a chosen flavor (Sec.~\ref{sec:minsce}), as well as a more realistic flavor scenario (Sec.~\ref{sec:MFV}), where we assume a particular flavor pattern in the SMEFT parameter space, namely Minimal Flavor Violation (MFV). All results are presented assuming the down-diagonal quark mass basis.

\subsection{Minimalistic flavor scenarios}
\label{sec:minsce}

\begin{table}[t]
\centering
\caption{The $2\sigma$ bounds on different flavor structures of single Wilson coefficients at $\Lambda=1~\mathrm{TeV}$. See Sec.~\ref{sec:minsce} for details.}
\label{tab:1dbounds}
\begin{tabular}{cc|cc|cc}
 &  & \multicolumn{2}{c|}{Drell-Yan tails} & \multicolumn{2}{c}{$B$ decays} \\
Operator & Flavor & NC & CC & $b\to q\ell\ell$ &  $b\to q\nu\nu$ \\
\hline \multirow{4}{*}{$\mathcal O^{(1)}_{lq}$} & 1113 & [-0.068, 0.068] & - & [-0.005, 0.002] & [-0.035, 0.039] \\
 & 2213 & [-0.031, 0.032] & - & [-4.96, 0.78]$\times 10^{-4}$ &[-0.035, 0.039] \\
 & 1123 & [-0.145, 0.152] & - & [-4.26, 0.98]$\times 10^{-4}$ & [-0.038, 0.017] \\
 & 2223 & [-0.066, 0.071] & - & [7.71, 51.86]$\times 10^{-5}$ & [-0.038, 0.017] \\
\hline \multirow{4}{*}{$\mathcal O^{(3)}_{lq}$} & 1113 & [-0.068, 0.068] & [-0.017, 0.017] & [-0.005, 0.002] & [-0.037, 0.033] \\
 & 2213 & [-0.032, 0.031] & [-0.029, 0.029] & [-4.85, 0.7]$\times 10^{-4}$ & [-0.037, 0.033] \\
 & 1123 & [-0.152, 0.145] & [-0.054, 0.051] & [-4.26, 0.98]$\times 10^{-4}$ & [-0.015, 0.035] \\
 & 2223 & [-0.071, 0.066] & [-0.089, 0.089] & [7.71, 51.86]$\times 10^{-5}$ & [-0.015, 0.035] \\
\hline \multirow{4}{*}{$\mathcal O_{ld}$} & 1113 & [-0.068, 0.068] & - & [-0.005, 0.002] & [-0.038, 0.038] \\
 & 2213 & [-0.032, 0.032] & - & [-2.79, 2.43]$\times 10^{-4}$ & [-0.038, 0.038] \\
 & 1123 & [-0.149, 0.149] & - & [-4.04, 1.09]$\times 10^{-4}$ & [-0.007, 0.023] \\
 & 2223 & [-0.069, 0.069] & - & [-1.68, 2.14]$\times 10^{-4}$ & [-0.007, 0.023] \\
\hline \multirow{4}{*}{$\mathcal O_{qe}$} & 1311 & [-0.068, 0.068] & - & [-0.003, 0.004] & - \\
 & 1322 & [-0.032, 0.032] & - & [-3.35, 7.56]$\times 10^{-4}$ & - \\
 & 2311 & [-0.148, 0.149] & - & [-0.003, 0.001] & - \\
 & 2322 & [-0.068, 0.069] & - & [-2.39, 4.97]$\times 10^{-4}$ & - \\
\hline \multirow{4}{*}{$\mathcal O_{ed}$} & 1113 & [-0.068, 0.068] & - & [-0.003, 0.004] & - \\
 & 2213 & [-0.032, 0.032] & - & [-7.03, 3.76]$\times 10^{-4}$ & - \\
 & 1123 & [-0.149, 0.149] & - & [-0.002, 0.002] & - \\
 & 2223 & [-0.069, 0.069] & - & [-4.05, 4.37]$\times 10^{-4}$ & - \\
\hline \multirow{8}{*}{$\mathcal O_{ledq}$} & 1113 & [-0.079, 0.079] & - & [-1.19, 1.18]$\times 10^{-4}$ & - \\
 & 1131 & [-0.079, 0.079] & [-0.037, 0.037] & [-1.18, 1.18]$\times 10^{-4}$ & - \\
 & 2213 & [-0.037, 0.037] & - & [-3.48, 0.67]$\times 10^{-5}$ & - \\
 & 2231 & [-0.037, 0.037] & [-0.061, 0.061] & [-3.49, 0.68]$\times 10^{-5}$ & - \\
 & 1123 & [-0.173, 0.173] & - & [-1.78, 1.79]$\times 10^{-4}$ & - \\
 & 1132 & [-0.173, 0.173] & [-0.113, 0.113] & [-1.77, 1.78]$\times 10^{-4}$ & - \\
 & 2223 & [-0.08, 0.08] & - & [-6.82, 16.57]$\times 10^{-6}$ & - \\
 & 2232 & [-0.08, 0.08] & [-0.194, 0.194] & [-6.8, 16.48]$\times 10^{-6}$ & - \\

\end{tabular}
\end{table}

We first study the semi-leptonic contact interactions from Table~\ref{tab:SMEFToperators} that can impact (semi)leptonic rare $B$-decays at low energies one by one. These are $Q_{lq}^{(1,3)}, Q_{ld}, Q_{qe}, Q_{ed}$ and $Q_{ledq}$. We fix their flavor indices so that at the scale $\Lambda=1~\mathrm{TeV}$ we only activate those that correspond to the $bd\ell\ell$ or $bs\ell\ell$ flavors with $\ell=e,\mu$. We constrain each coefficient separately from both NC and CC (if possible) high-mass Drell-Yan tails at the scale of $\Lambda=1~\mathrm{TeV}$, whereas we use \texttt{wilson} to run and match the chosen coefficient down to the scale of $\mu=4.8~\mathrm{GeV}$ in order to constrain it also from various $B$-decays with the underlying $b\to q \ell \ell$ transitions, with $q=d,s$ and $\ell=e,\mu$. Due to the $SU(2)_L$ gauge invariance, the same coefficient might enter also processes with the underlying $b \to q \nu \nu$ transition with $\nu=\nu_e,\nu_\mu$. We consider various $B\to M \nu \nu$ with $M=K, \pi, \rho$ already implemented in \texttt{flavio}, with the upper limits of their branching ratios measured by Belle~\cite{Belle:2017oht}. We collect the results in Table~\ref{tab:1dbounds}, showing the $2\sigma$ bounds on the chosen set of SMEFT Wilson coefficients, separately from NC and CC Drell-Yan, as well as $b\to q \ell \ell$ and $b\to q \nu \nu$ processes. We summarise the main conclusions for each operator here:
\begin{itemize}
    \item $Q_{lq}^{(1,3)}$: Regarding high-mass Drell-Yan, the singlet operator is only constrained by NC, whereas the triplet is also constrained by CC. Notice the $\mu$ channel constraints are comparable between NC and CC, while the $e$ channel constraints from CC are more stringent than those from NC. This is due to the anomalous events in the CMS $pp\to e e$ data~\cite{CMS:2021ctt}. The bounds from $b\to q \nu \nu$ are comparable, if slightly more stringent with respect to the bounds from high-mass DY. The $b\to q \ell \ell$ bounds however are about two orders of magnitude stronger. We point out the $2223$ flavor indices, with which we are solving the various $B$-anomalies with the well-known low-energy scenario of $C_9=-C_{10}$, albeit with slight tension with the new $R_{K^{(\ast)}}$ measurement. In electrons ($1123$) the fits are consistent with the SM.

    \item $Q_{ld}, Q_{qe}, Q_{ed}$: The CC high-mass DY is not sensitive to the effects of these operators. They are constrained from NC DY, $b\to q \ell \ell$, and in the case of $Q_{ld}$ also from $b\to q \nu \nu$. The latter constraints are again slightly more stringent than those coming from tails. The NC Drell-Yan bounds are very similar between all these operators (and also $Q_{lq}^{(1,3)}$) as all of them are vector operators, only differing in chiralities of the fermions. We point out that the operators with the $13$ quark flavor structure are in general more constrained than those with the $23$ structure, due to the presence of valence quarks in the first case. Nevertheless, the $b\to q \ell \ell$ processes are again significantly more constraining. The operators $Q_{ld}, Q_{qe}, Q_{ed}$ generate at low energies the scenarios $C_9^\prime=-C_{10}^\prime$, $C_9=C_{10}$ and $C_9^\prime=C_{10}^\prime$ respectively. All of the scenarios are consistent with the SM.

    \item $Q_{ledq}$: This non-hermitian scalar operator is constrained from NC high-mass DY, $b\to q \ell \ell$, and for some flavor indices also from CC DY --- the $13$ and $23$ quark flavor indices can not be constrained from CC DY due to the top quark not being present in the proton. The $b\to q \ell \ell$ are dominating here as the leptonic $B$ decays are highly sensitive to scalar operators (see Sec.~\ref{sec:low_energy}). As the leptonic $B$ decays to muons are better measured, the coefficients with the $22$ lepton flavor are better constrained.
\end{itemize}

\begin{figure}[h]
     \centering
     \begin{subfigure}[b]{0.45\textwidth}
         \centering
         \includegraphics[width=\textwidth]{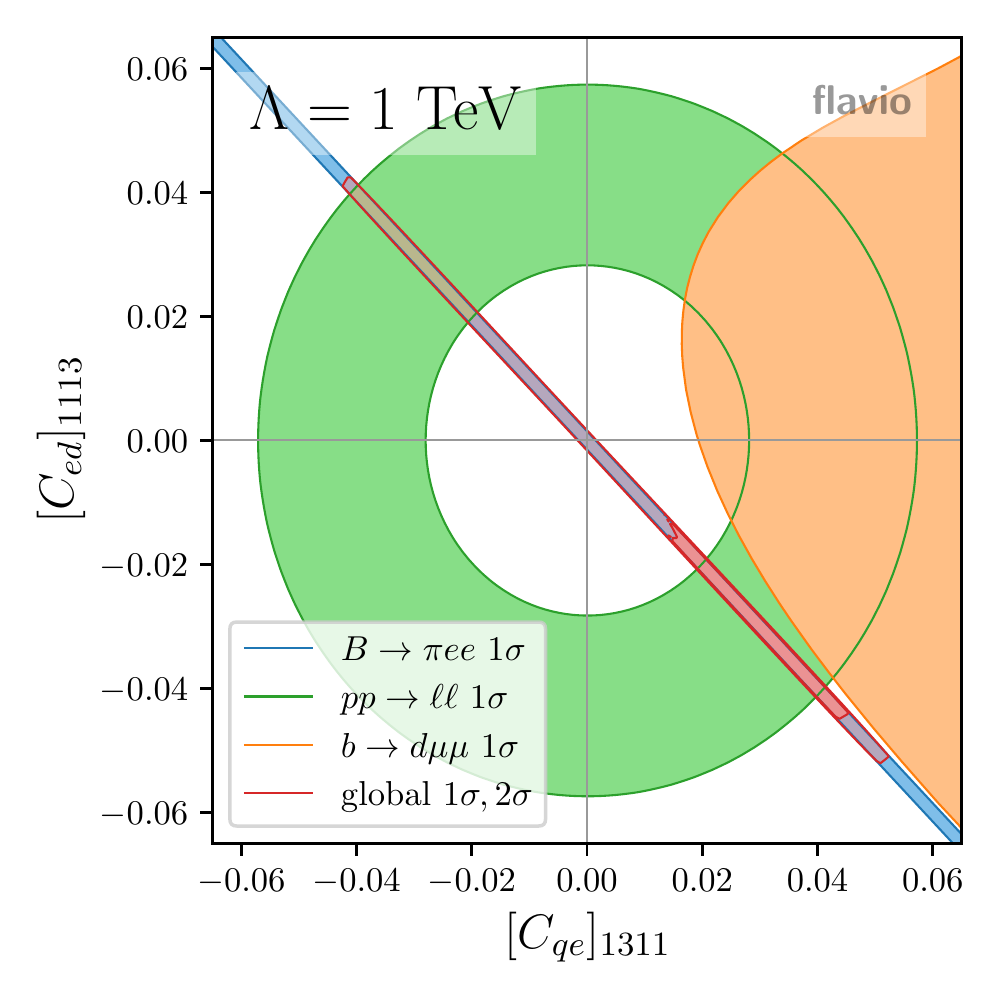}
     \end{subfigure}~
     \begin{subfigure}[b]{0.45\textwidth}
         \centering
         \includegraphics[width=\textwidth]{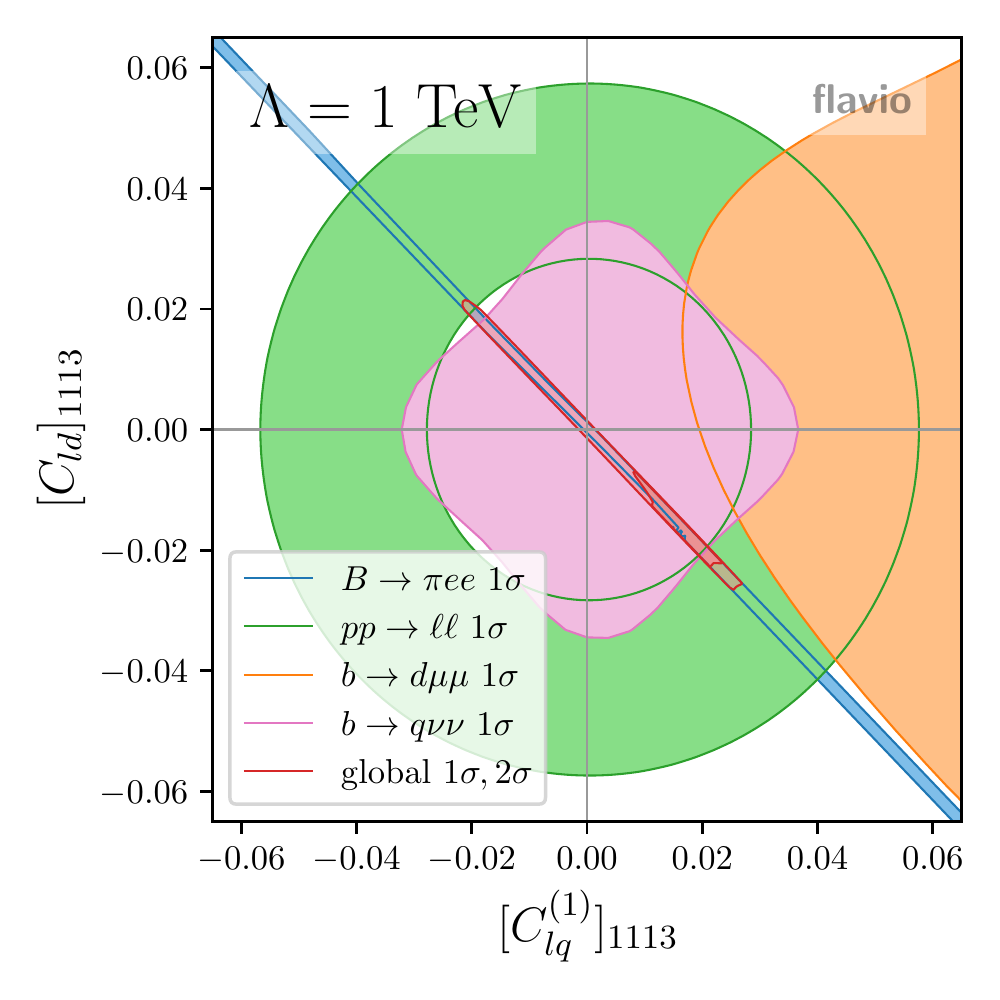}
     \end{subfigure}
     \\

     \begin{subfigure}[b]{0.45\textwidth}
         \centering
         \includegraphics[width=\textwidth]{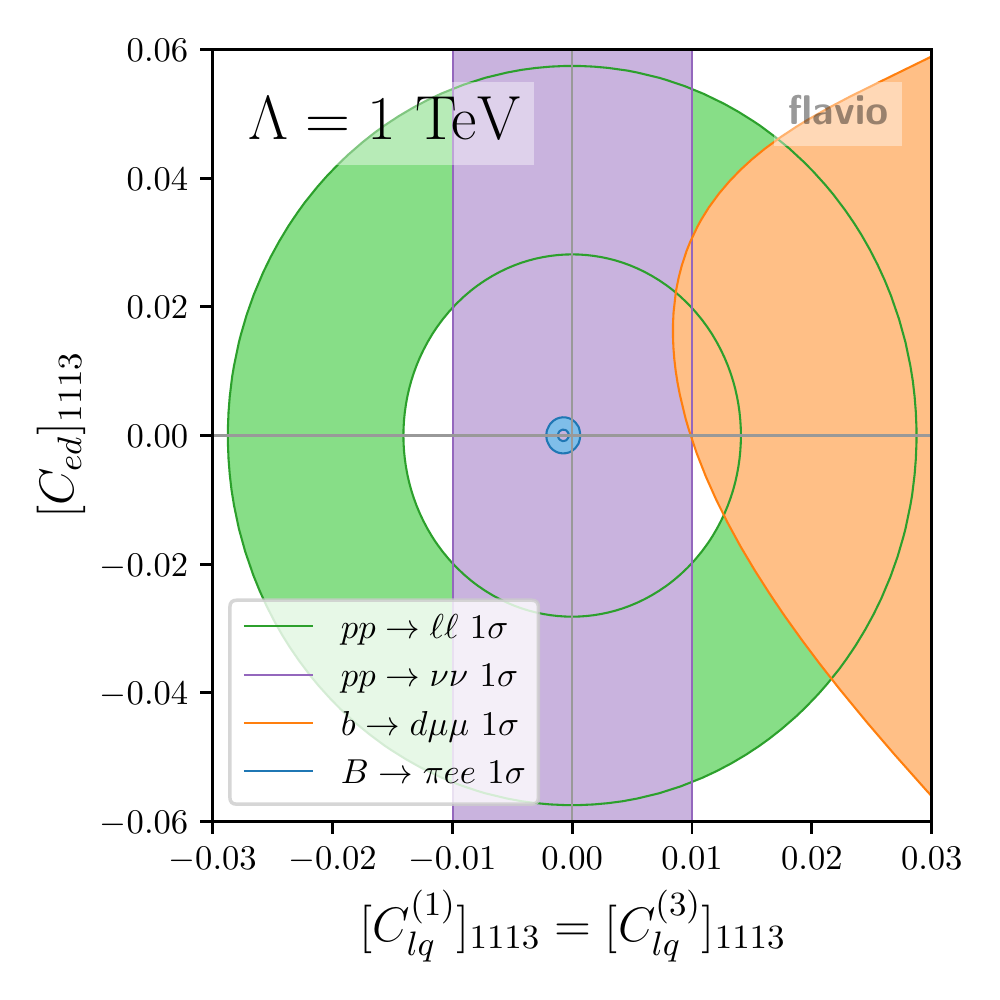}
     \end{subfigure}
        \caption{Selected representative $2$D minimalistic flavor scenarios. See discussion in Sec.~\ref{sec:minsce}.}
        \label{fig:minim-bdee-smeft}
\end{figure}

Next, we consider selected $2$D scenarios, related to the model-independent bounds from $b\to d e e$ at low-energies presented in Sec.~\ref{sec:low_energy}. The upper two scenarios presented on Fig.~\ref{fig:minim-bdee-smeft}, namely the $({[C_{qe}]}_{1311}, {[C_{ed}]}_{1113})$ and  $([C_{lq}^{(1)}]_{1113}, {[C_{ld}]}_{1113})$ showcase the flat direction in the bounds from $B\to \pi e e$ already discussed in Sec.~\ref{sec:low_energy}. Note, however, that the global fits shown in the first two figures on Fig.~\ref{fig:minim-bdee-smeft} are closed, as we can now constrain the same parameter space also from other processes. In the upper-left scenario, the flat direction is closed firstly by the NC high-mass Drell-Yan constraint and secondly due to the RGE running effects --- the coefficient with the $1113$ flavor structure at high-energies generates through RGE mixing also $2213$ and $3313$ (universally) at low-energies through an electroweak penguin. This in turn means we can constrain the same parameter space from measurements of $B\to\pi \mu \mu$ and $B_s\to K^{\ast 0} \mu \mu$ with the underlying $b\to d \mu \mu$ transition, discussed in Sec.~\ref{sec:low_energy}. The upper-right scenario is however mostly closed already with constraints from $b\to q \nu \nu$. The constraint on the bottom scenario on Fig.~\ref{fig:minim-bdee-smeft} is already dominated by low-energy $b\to d e e$ measurements. Nonetheless, we overlay the constraints from $b\to d \mu \mu$, NC, and also CC high-mass Drell-Yan, as now the $Q_{lq}^{(3)}$ operator is active.

\subsection{Minimal Flavor Violation}
\label{sec:MFV}
In the limit of vanishing Yukawa couplings, the SM Lagrangian is invariant under a large flavor group $G_Q = U(3)_Q \times U(3)_u \times U(3)_d$ (here we focus on the quark sector). The non-vanishing Yukawa coupling matrices $Y_{u,d}$ in the Yukawa Lagrangian:
\begin{equation}
    \mathcal{L} \supset -\bar{Q}_L Y_d d_R H - \bar{Q}_L Y_u u_R \tilde{H} + \mathrm{h.c.}\,,
\label{eq:SMYukawa}
\end{equation}
break the $G_Q$ symmetry down to $U(1)_B$. In MFV, we assume all the flavor structure is contained in $Y_{u,d}$ also beyond the SM~\cite{DAmbrosio:2002vsn}. Formally, we promote $Y_u$ and $Y_d$ to be spurions, transforming as $Y_u\sim ({\bf 3}, \bar{{\bf 3}}, 1)$ and $Y_d\sim ({\bf 3}, 1, \bar{{\bf 3}})$, rendering the whole SM Lagrangian formally invariant under $G_Q$. Furthermore, under the MFV assumption also the BSM Lagrangian should be invariant under $G_Q$, which correlates various flavors of a particular operator~\cite{Greljo:2022cah}. The results here are presented in the down-diagonal quark mass basis, such that $Y_d=\mathrm{diag}(y_d, y_s, y_b)$ and $Y_u = V_\mathrm{CKM}^\dagger \mathrm{diag}(y_u, y_c, y_t)$.

Consider an operator with a $\bar{Q}Q$ bilinear, such as $Q_{lq}^{(1)}$. One can decompose its Wilson coefficient in the following way
\begin{equation}
    [C_{lq}^{(1)}]_{st}^{(l)} \bar{L}_l \gamma_\mu L_l \bar{Q}_s \gamma^\mu Q_t \rightarrow [C_{lq}^{(1)}]_{st}^{(l)} = \delta_{st} [C_{lq}^{(1)}]_{\delta}^{(l)} + (Y_u Y_u^\dagger)_{st} [C_{lq}^{(1)}]_{Y_u Y_u^\dagger}^{(l)}\,,
\label{eq:MFV:lq1}
\end{equation}
where we fix the lepton flavor $l$ by assuming NP only in electrons ($l=e$), only in muons ($l=\mu$), or universal NP ($l=\ell$ with $[C]^{(\ell)} \equiv [C]^{(e)} = [C]^{(\mu)}$). The first term here is flavor-diagonal and universal, with an overall coefficient $[C_{lq}^{(1)}]_{\delta}^{(l)}$. The second term is flavor-violating, in particular, we have
\begin{equation}
     Y_u Y_u^\dagger\sim y_t^2\begin{pmatrix}
 V_{td} V_{td}^* & V_{ts} V_{td}^* & V_{tb} V_{td}^* \\
 V_{td} V_{ts}^* & V_{ts} V_{ts}^* & V_{tb} V_{ts}^* \\
 V_{td} V_{tb}^* & V_{ts} V_{tb}^* & V_{tb} V_{tb}^* \\
 \end{pmatrix} \,,
 \label{eq:MFV:YuYuflavor}
\end{equation}
where we neglect $y_{u,c}$. This sets the flavor-violating structure of the operator, with only an overall coefficient $[C_{lq}^{(1)}]_{Y_u Y_u^\dagger}^{(l)}$ remaining free. The rest of the semileptonic operators containing the $\bar{Q}Q$ bilinear decompose in the same way.

Similarly, we can decompose the scalar operator $Q_{ledq}$ in the following way
\begin{equation}
    [C_{ledq}]_{st}^{(l)} (\bar{L}_l e_l) (\bar{d}_s Q_{t}) \to [C_{ledq}]_{st}^{(l)} = (Y_d^\dagger)_{st} [C_{ledq}]_{Y_d^\dagger}^{(l)} + (Y_d^\dagger Y_u Y_u^\dagger)_{st}[C_{ledq}]_{Y_d^\dagger Y_u Y_u^\dagger}^{(l)}\,,
\label{eq:MFV:ledq}
\end{equation}
where again the lepton flavor $l$ is fixed. In this case, the leading term is flavor-diagonal but not universal, with an overall coefficient $[C_{ledq}]_{Y_d}^{(l)}$. The flavor-violating part again has only a single overall free parameter $[C_{ledq}]_{Y_d^\dagger Y_u Y_u^\dagger}^{(l)}$, and has the following flavor structure
\begin{equation}
    Y_d^\dagger Y_u Y_u^\dagger \sim y_b y_t^2\begin{pmatrix}
 0 & 0& 0\\
 0 & 0 & 0 \\
 V_{td} V_{tb}^* & V_{ts} V_{tb}^* & V_{tb} V_{tb}^* \\
 \end{pmatrix} \,,
\label{eq:MFV:YdYuYuflavor}
\end{equation}
where we neglect all Yukawas but $y_{b,t}$.

\begin{figure}[t]
     \centering
     \begin{subfigure}[b]{0.43\textwidth}
         \centering
         \includegraphics[width=\textwidth]{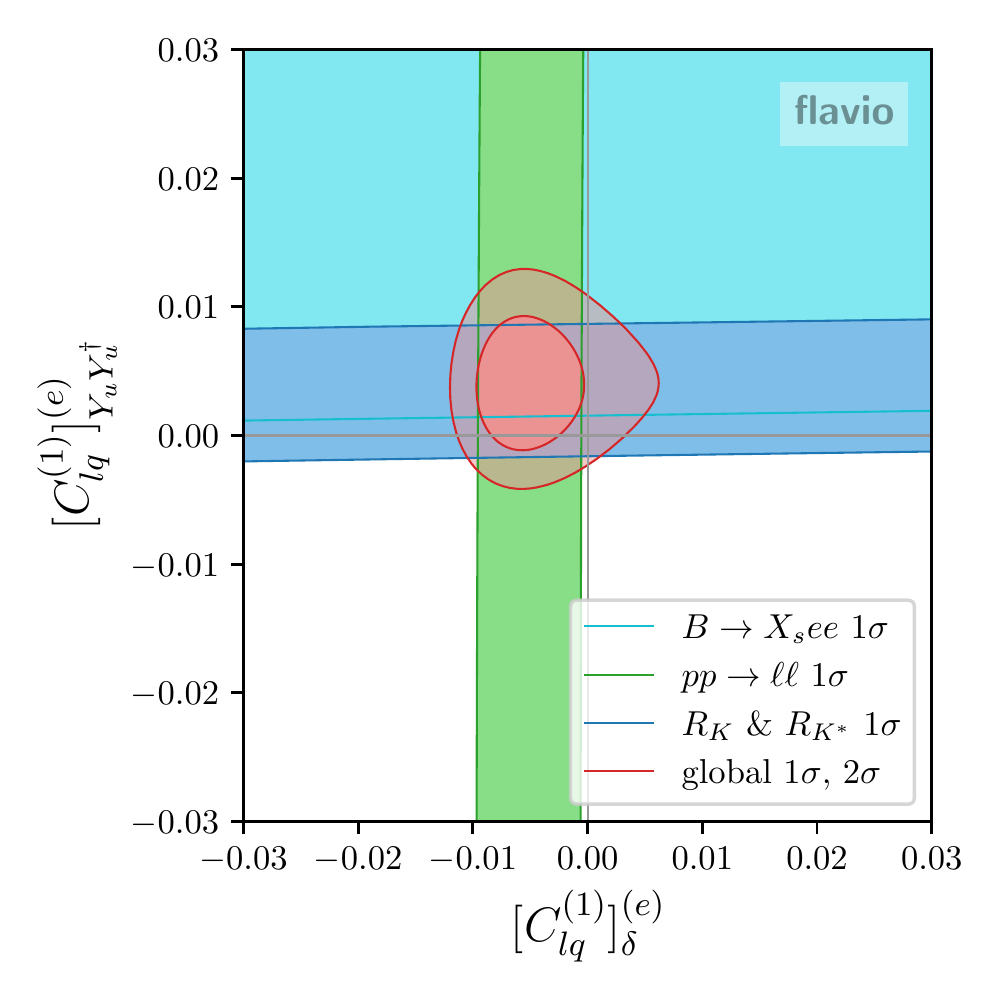}
     \end{subfigure}~
          \begin{subfigure}[b]{0.43\textwidth}
         \centering
         \includegraphics[width=\textwidth]{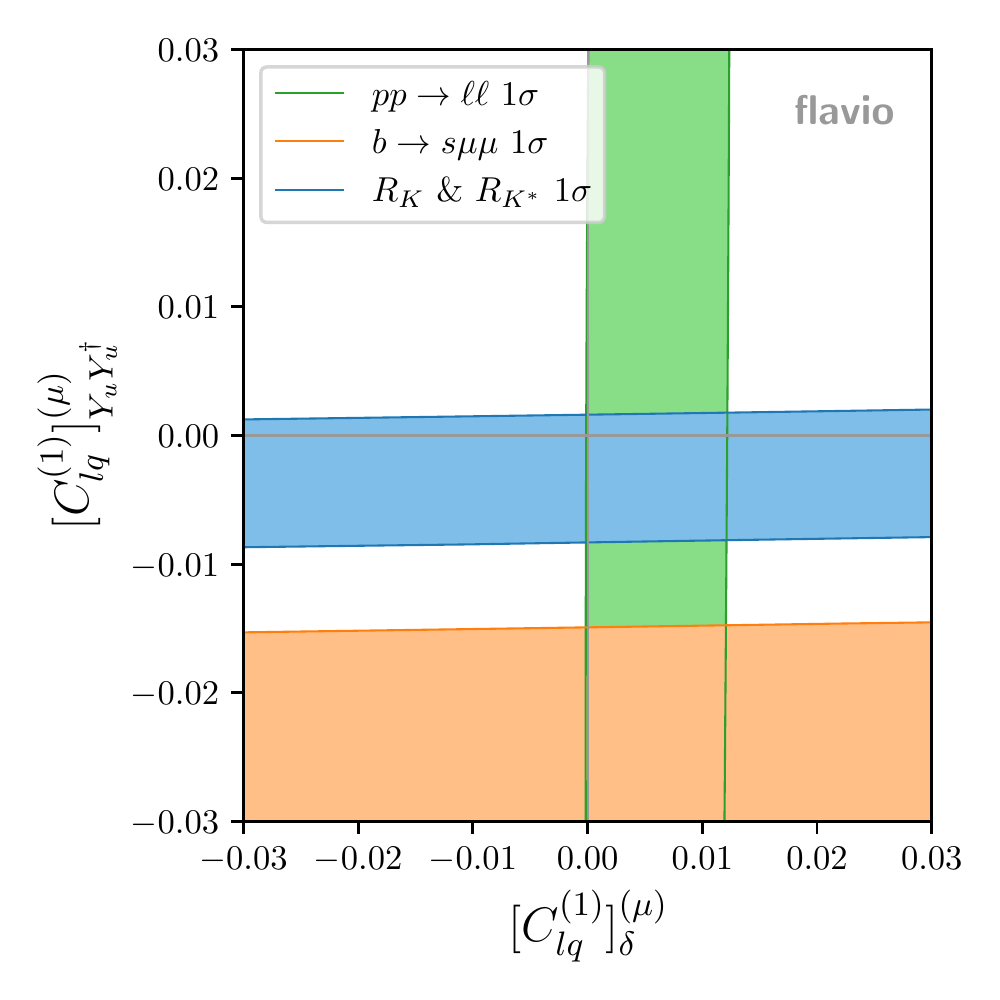}
     \end{subfigure}
     \\
     \begin{subfigure}[b]{0.43\textwidth}
         \centering
         \includegraphics[width=\textwidth]{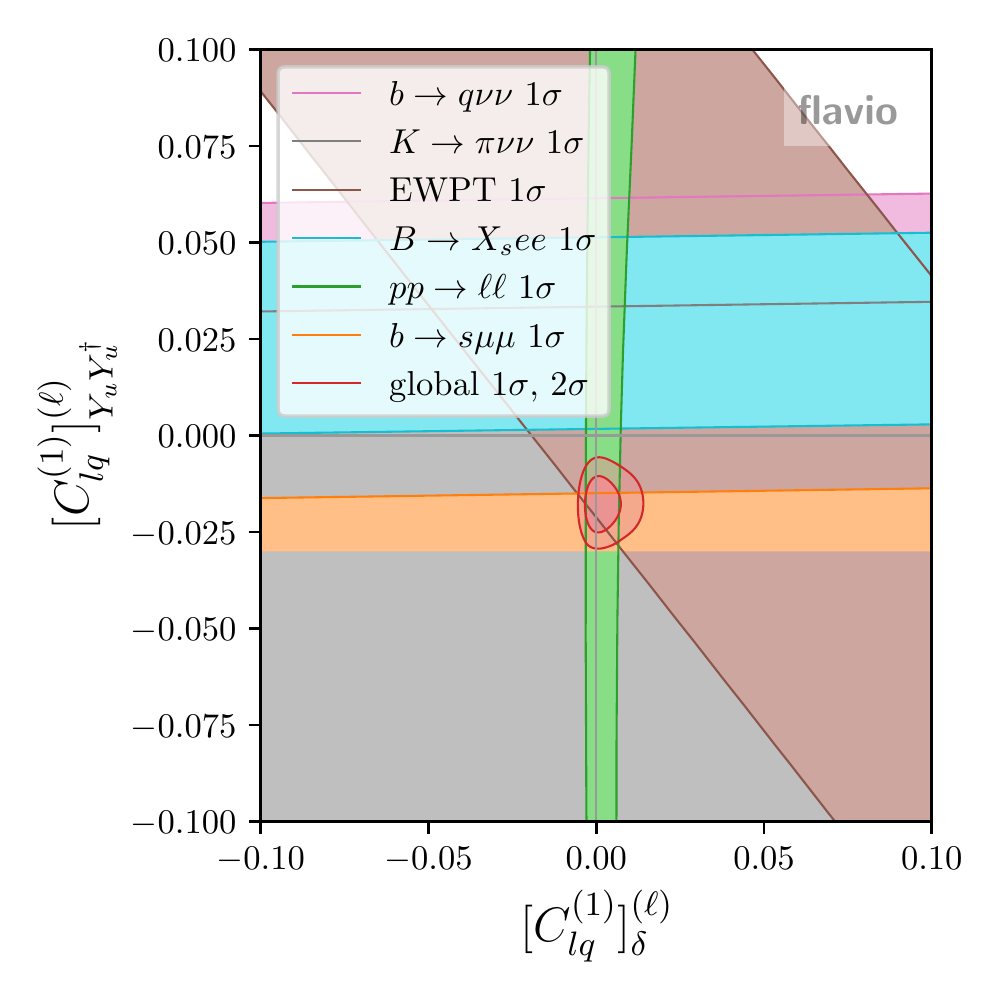}
     \end{subfigure}~
     \begin{subfigure}[b]{0.43\textwidth}
         \centering
         \includegraphics[width=\textwidth]{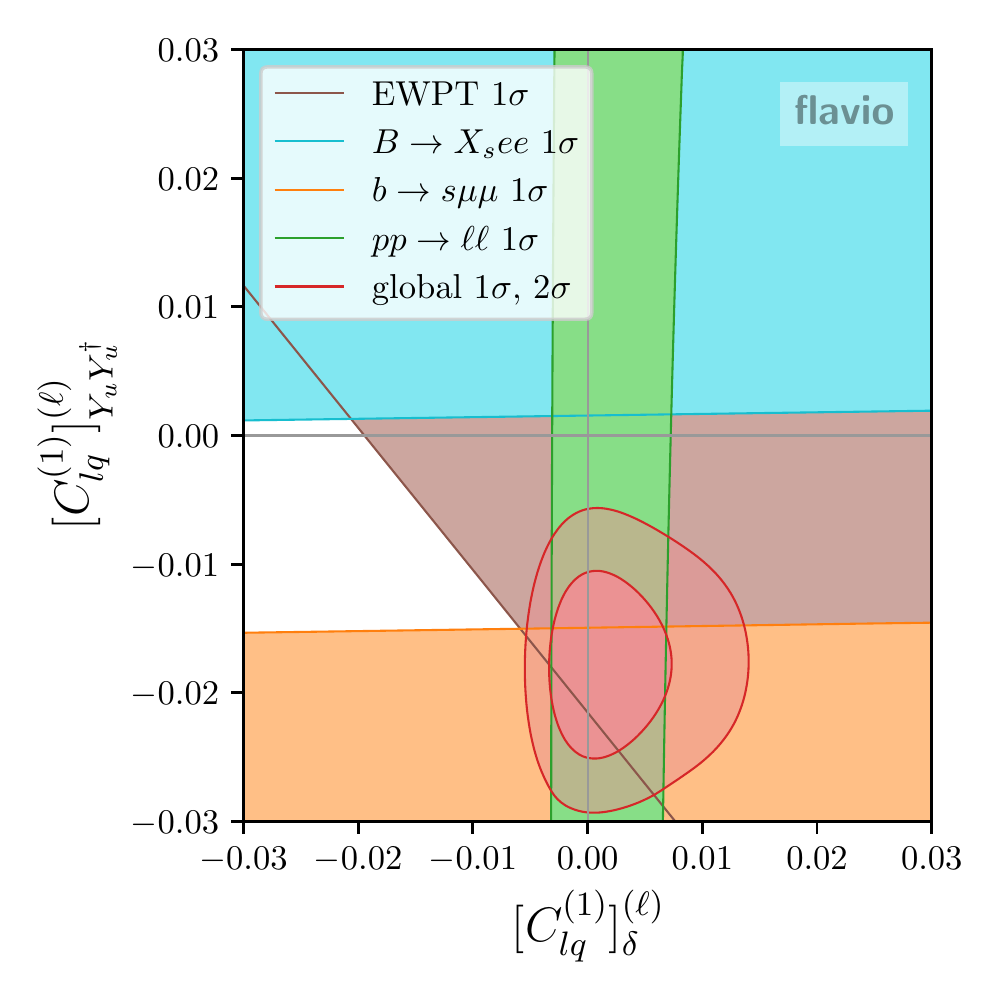}
     \end{subfigure}
        \caption{$[C_{lq}^{(1)}]_{\delta}^{(l)}$ versus $[C_{lq}^{(1)}]_{Y_u Y_u^\dagger}^{(l)}$ separate for assuming NP only in electrons ($l=e$ in Eq.~\eqref{eq:MFV:lq1}), only in muons ($l=\mu$), and assuming universal NP ($l=\ell$). See Sec.~\ref{sec:MFV} for discussion.}
        \label{fig:MFV_lq1C0vslq1Cu}
\end{figure}

Under the assumption of MFV, flavor violation in right-handed quark currents is highly suppressed. Consider the $\bar{d}{d}$ bilinear. Decomposing a Wilson coefficient belonging to an operator containing this bilinear would result in an unsuppressed flavor diagonal and universal term, a $y_d^2$ suppressed flavor diagonal and non-universal term, and only at $\mathcal{O}(y_d^2 y_u^2)$ a flavor violating term (the spurion insertion is $Y_d^\dagger Y_u Y_u^\dagger Y_d$). Due to this suppression, we do not consider such operators in the following.

\begin{figure}[t]
     \centering
     \begin{subfigure}[b]{0.43\textwidth}
         \centering
         \includegraphics[width=\textwidth]{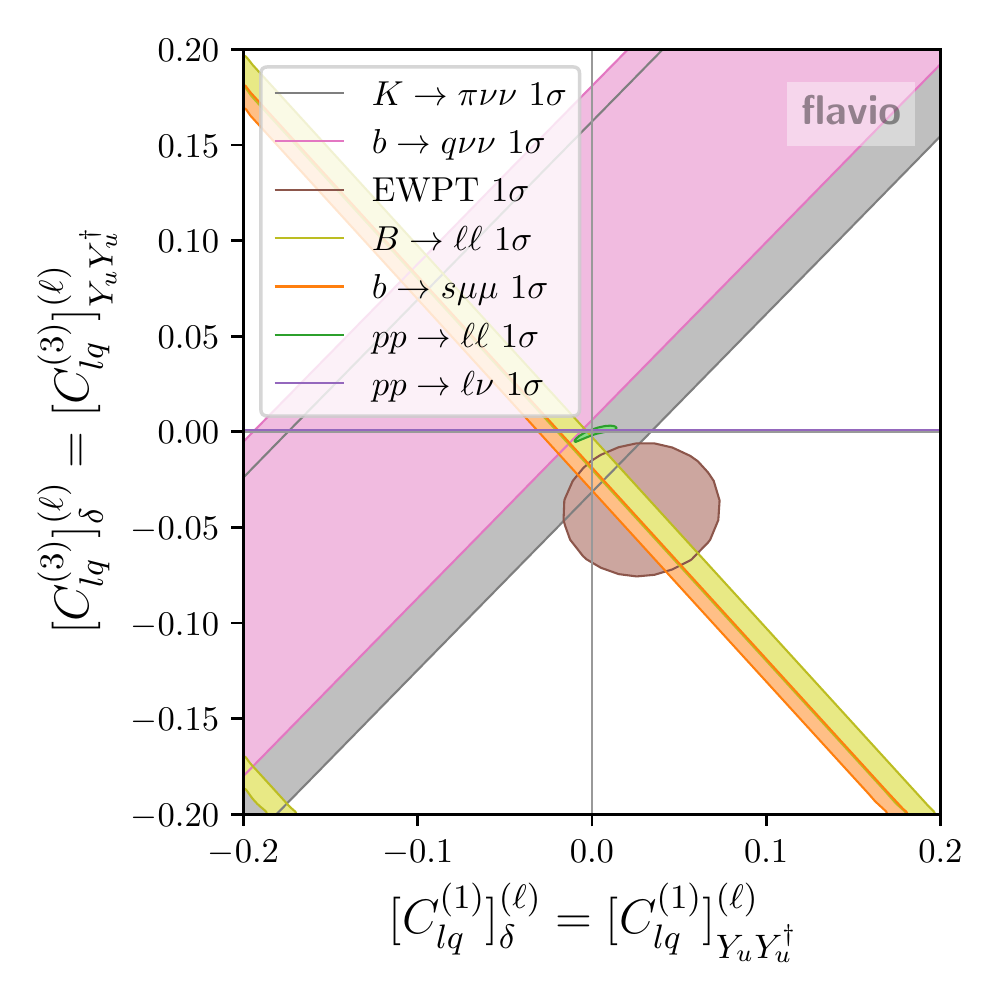}
     \end{subfigure}~
     \begin{subfigure}[b]{0.43\textwidth}
         \centering
         \includegraphics[width=\textwidth]{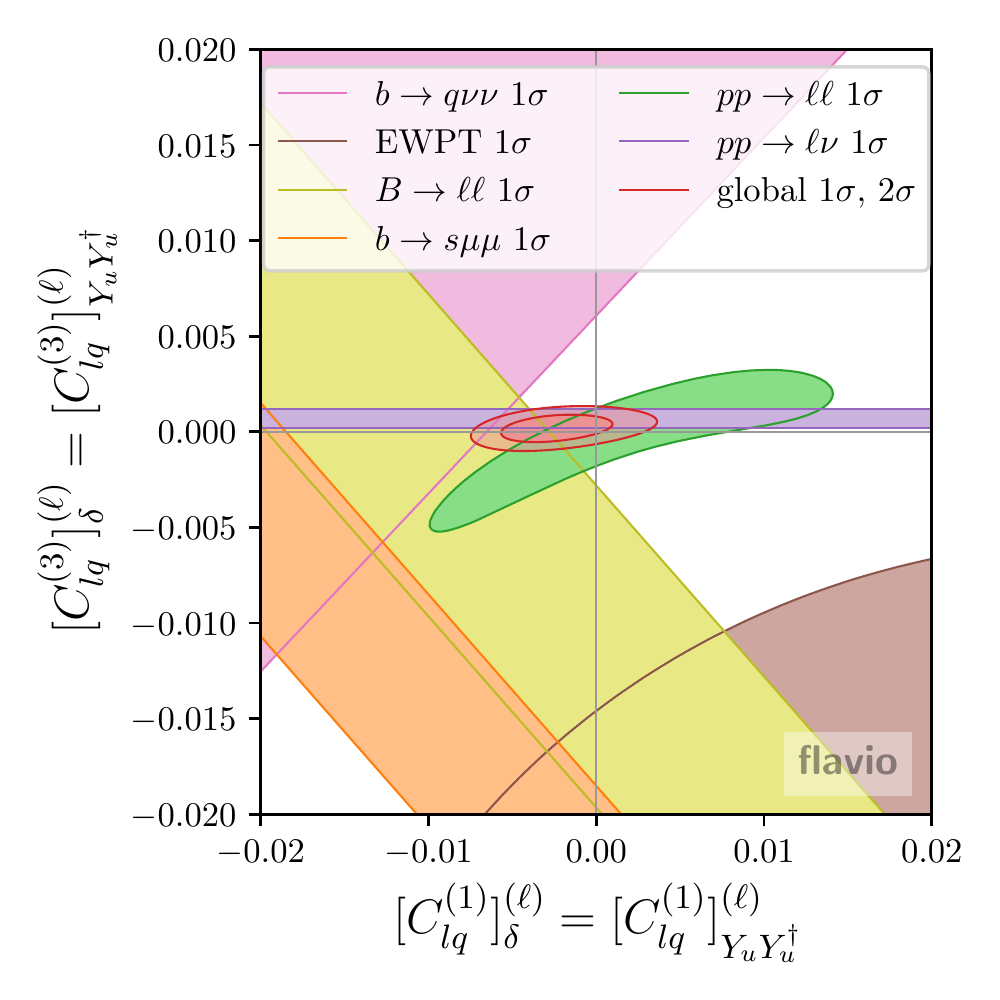}
     \end{subfigure}
    \\
     \begin{subfigure}[b]{0.43\textwidth}
         \centering
         \includegraphics[width=\textwidth]{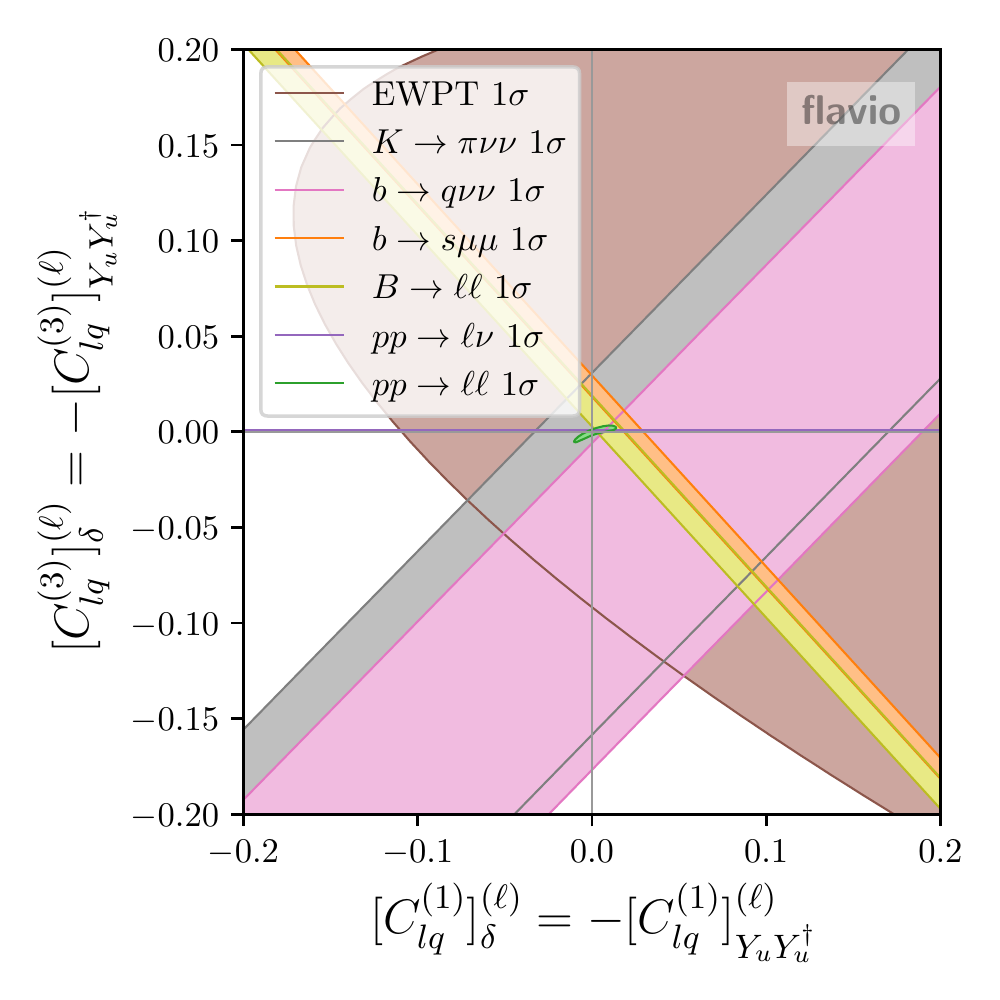}
     \end{subfigure}~
     \begin{subfigure}[b]{0.43\textwidth}
         \centering
         \includegraphics[width=\textwidth]{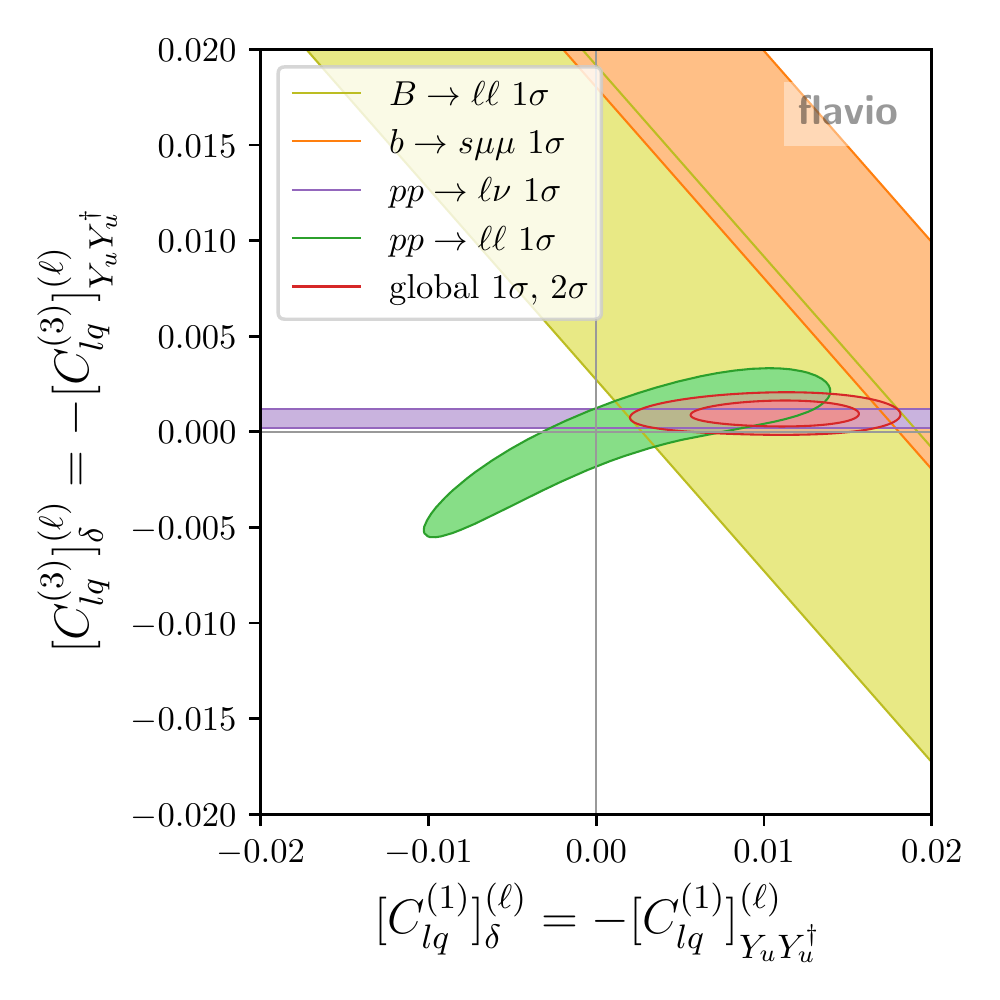}
     \end{subfigure}
        \caption{$[C_{lq}^{(1)}]^{(l)}_{\delta} = \pm[C_{lq}^{(1)}]^{(l)}_{Y_u Y_u^\dagger}$ versus $[C_{lq}^{(3)}]^{(l)}_{\delta} = \pm[C_{lq}^{(3)}]^{(l)}_{Y_u Y_u^\dagger}$ for lepton flavor universal case $l=\ell$ in Eq.~\eqref{eq:MFV:lq1}. See discussion in Sec.~\ref{sec:MFV}.}
        \label{fig:MFV_lq1C0=+lq1Cuvslq3C0=+lq3Cu_universal}
\end{figure}

In each of the plots on Fig.~\ref{fig:MFV_lq1C0vslq1Cu}, we keep the ranges of both axis the same, showcasing that the flavor-conserving and flavor-violating coefficients are constrained to values of similar absolute value. The flavor-conserving coefficients $[C_{lq}^{(1)}]_{\delta}^{(l)}$ are very much constrained from $pp\to \ell \ell$ due to their contribution to underlying valence quark transitions, whereas the low energy $b\to q \ell \ell$ processes mostly constrain the flavor-violating coefficients $[C_{lq}^{(1)}]_{Y_u Y_u^\dagger}^{(l)}$. In the electron case, the fit is consistent with the SM, whereas in the muon case, a non-zero negative value of $[C_{lq}^{(1)}]_{Y_u Y_u^\dagger}^{(\mu)}$ is preferred by the various $b\to s \mu \mu$ anomalies, solved by the negative interference with the SM in the $C_9=-C_{10}$ low-energy scenario, decreasing the muon branching ratios. This is, however, in slight tension with the latest constraints from $R_{K^{(\ast)}}$. Finally, in the last row of Fig.~\ref{fig:MFV_lq1C0vslq1Cu}, the universal NP scenario between electrons and muons is presented, so that we only impact the $b\to s \mu \mu$ anomalies, while predicting $R_{K^{(\ast)}}=1$. Here the $b\to s \mu \mu$ anomalies prefer negative values of $[C_{lq}^{(1)}]_{Y_u Y_u^\dagger}^{(\ell)}$ and a global fit can be performed which is found to be incompatible with the SM at the level of $2-3\sigma$. (Note that there is a slight tension with the $B\to X_s e e$ data.) In the left plot of the last row of Fig.~\ref{fig:MFV_lq1C0vslq1Cu} we show a zoomed-out version of the same scenario, showing the global fit is consistent also with $b\to q \nu \nu$, $K\to \pi \nu \nu$ and EWPT constraints.

On Fig.~\ref{fig:MFV_lq1C0=+lq1Cuvslq3C0=+lq3Cu_universal} we again consider the MFV expansion from Eq.~\eqref{eq:MFV:lq1}, however now we assume the flavor conserving and flavor violating coefficients are either equal (first row plots) or opposite to each-other (second row plots). Moreover, we show this in the triplet versus singlet operator planes. In both rows on Fig.~\ref{fig:MFV_lq1C0=+lq1Cuvslq3C0=+lq3Cu_universal} we show the zoomed-out versions in the first column and zoomed-in versions in the second column.
Considering the scenario in the first row of Fig.~\ref{fig:MFV_lq1C0=+lq1Cuvslq3C0=+lq3Cu_universal}, there is a slight tension between $b\to q \nu \nu$, EWPT and $b\to s \ell \ell$ data themselves, all of which are incompatible at the $1\sigma$ level with high-mass Drell-Yan tails, which dominate the global fit. A compatible combined fit between low-energy and high-energy data can be obtained by assuming a negative sign between the flavor-violating and flavor-conserving coefficients, as demonstrated in the second row of Fig.~\ref{fig:MFV_lq1C0=+lq1Cuvslq3C0=+lq3Cu_universal}. In the zoomed-in plot, we do not show the rest of the low energy constraints, which are consistent with the whole regions shown in the plot. A combined fit to $b\to s \ell \ell$ and high-mass Drell-Yan data can be performed, showing a tension with the SM at the level of $2-3~\sigma$.

\begin{figure}[t]
     \centering
     \begin{subfigure}[b]{0.45\textwidth}
         \centering
         \includegraphics[width=\textwidth]{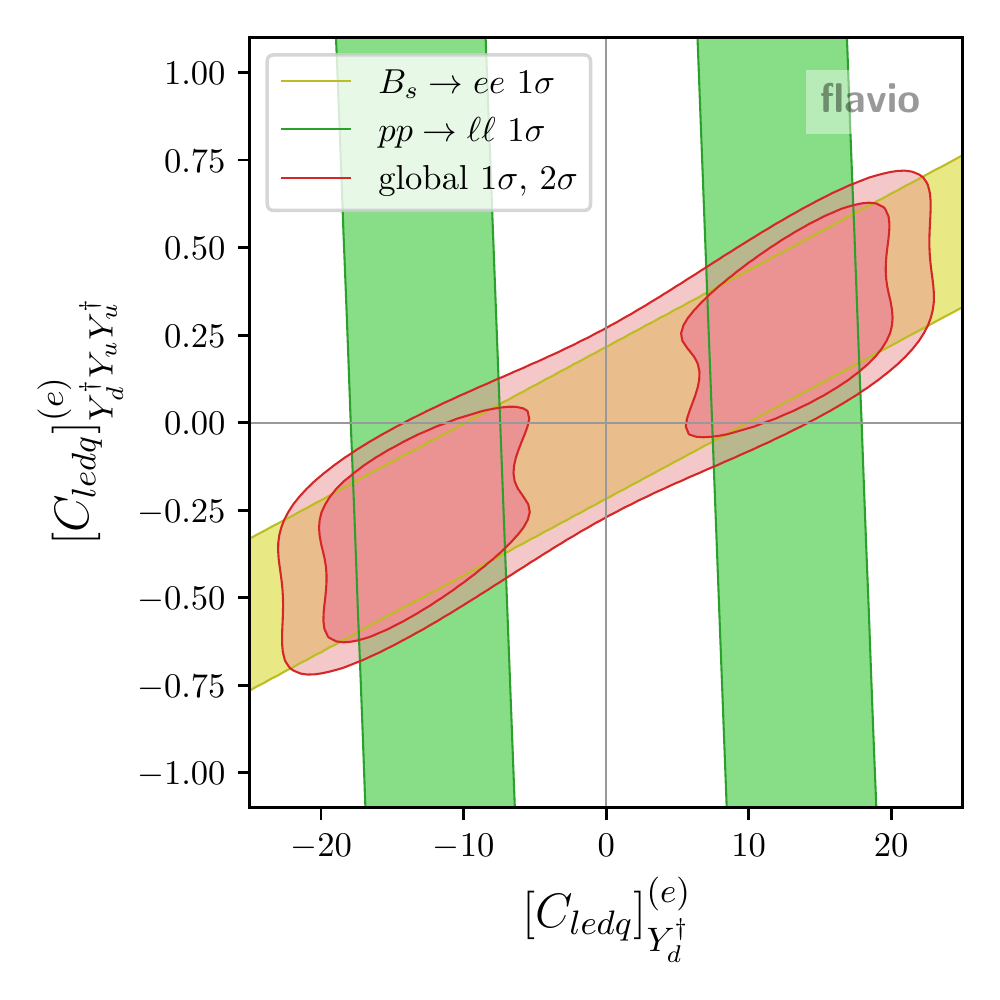}
     \end{subfigure}~
     \begin{subfigure}[b]{0.45\textwidth}
         \centering
         \includegraphics[width=\textwidth]{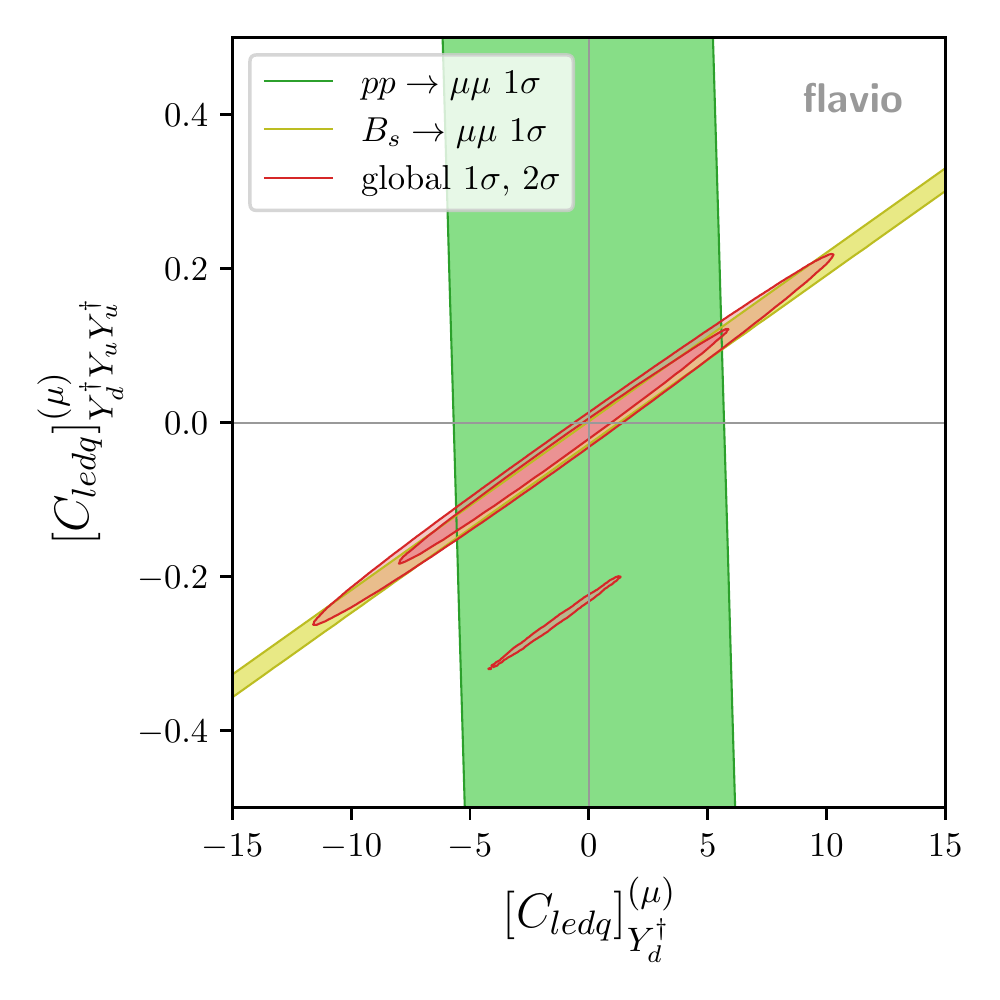}
     \end{subfigure}

        \caption{$[C_{ledq}]_{Y_d^\dagger}^{(l)}$ versus $[C_{ledq}]_{Y_d^\dagger Y_u Y_u^\dagger}^{(l)}$ separate for assuming NP in electrons ($l=e$ in Eq.~\eqref{eq:MFV:ledq}) and muons ($l=\mu$). See Sec.~\ref{sec:MFV} for details.}
        \label{fig:MFV_ledqCdvsledqCduu}
\end{figure}

Finally, in Fig.~\ref{fig:MFV_ledqCdvsledqCduu} we show the results for the scalar operator $Q_{ledq}$, with the flavor decomposition defined in Eqs.~\eqref{eq:MFV:ledq} and~\eqref{eq:MFV:YdYuYuflavor}. The leading term is now flavor-diagonal but not universal, with the largest coupling for the $33$ quark flavors. This makes the NC high-mass DY bound much stronger than the one coming from CC high-mass DY, as the leading constraint will come from the $22$ quark flavor combination. In fact, the NC high-mass DY constraint is the leading constraint in the flavor conserving $[C_{ledq}]_{Y_d^\dagger}^{(l)}$ direction. The flavor violating direction of $[C_{ledq}]_{Y_d^\dagger Y_u Y_u^\dagger}^{(l)}$ is significantly constrained from $B$ decays at low energies, particularly the fully leptonic decay modes $B\to \ell \ell$. These remarks hold true for both the assumption of NP in electrons(Fig.~\ref{fig:MFV_ledqCdvsledqCduu}, left) and the assumption of NP in muons (Fig.~\ref{fig:MFV_ledqCdvsledqCduu}, right). The $B\to \mu \mu$ measurements are more precise compared to the electron mode, hence a better constraint in the flavor-violating direction for the muon case. Moreover, in the muon case at the $2\sigma$ level, there are two allowed bands from low-energy $B$ decays, where in one of them NP is a small correction to the SM contribution, whereas the other one is due to a fine-tuned cancellation of the SM contribution. Lastly, the bound from NC high-mass DY is also better for the muon case, due to the anomalous events in the CMS $pp\to ee$ data~\cite{CMS:2021ctt}--- these are also the reason for SM being excluded to $1\sigma$ in the electron case, with two minima forming at negative and positive values of $[C_{ledq}]_{Y_d^\dagger}^{(e)}$.

\section{Model examples}
\label{sec:models}

In this section, we study four explicit model examples. Our goal is:
\begin{itemize}

    \item to generate large lepton flavor universal effects in $b \to q \ell^+ \ell^-$ transitions while predicting $R_{K^{(*)}} \approx R_{K^{(*)}}^{{\rm SM}}$, and

    \item to contrast several different phenomenological cases: when a new physics mediator, either a $Z'$ or a leptoquark, dominantly interacts with valence quarks versus when it interacts with sea quarks.

\end{itemize}
Common $Z'$ models rather naturally predict both lepton and quark universality, e.g. $U(1)_{B-L}$, while only in recent years there has been an increased interest in the non-universal $Z'$ models, see e.g.~\cite{Alonso:2017uky,Allanach:2018vjg, Bonilla:2017lsq, Allanach:2020kss, Bause:2021prv, Calibbi:2019lvs, Bian:2017xzg, Greljo:2022dwn,Greljo:2021npi}. On the contrary, common leptoquark models have hierarchical couplings in both quark and lepton flavor spaces, typically favoring heavier generations. Charging leptoquarks under flavor symmetries, here we construct LFU (and MFV) leptoquark models.

This section is organized as follows. The first and the second model examples, Sections~\ref{sec:modelI} and~\ref{sec:modelIb3}, extend the SM by TeV-scale $U(1)_{B-L}$ and $U(1)_{3 B_3 - L }$ gauge bosons, respectively, where the quark flavor violation occurs due to some heavy new dynamics (e.g., vector-like quarks). In Section~\ref{sec:modelIIb}, the SM is augmented by a triplet of scalar leptoquarks realizing LFU. Finally, in Section~\ref{sec:modelII}, the scalar leptoquark forms a bi-triplet under the quark and lepton flavor symmetry and predicts MFV.

All models respect lepton flavor universality but can affect angular distributions and branching ratios in $b \to s \ell^+ \ell^-$ decays. While the mediators in the first and the fourth model couple to valence quarks, they dominantly interact with the $b$ quark in the second and the third model. Thus, the importance of the high-mass Drell-Yan tails is very different in the two cases.

\subsection{Gauged $U(1)_{B-L}$}
\label{sec:modelI}

\begin{figure}[t]
     \centering
     \begin{subfigure}[b]{0.5\textwidth}
         \centering
         \includegraphics[width=\textwidth]{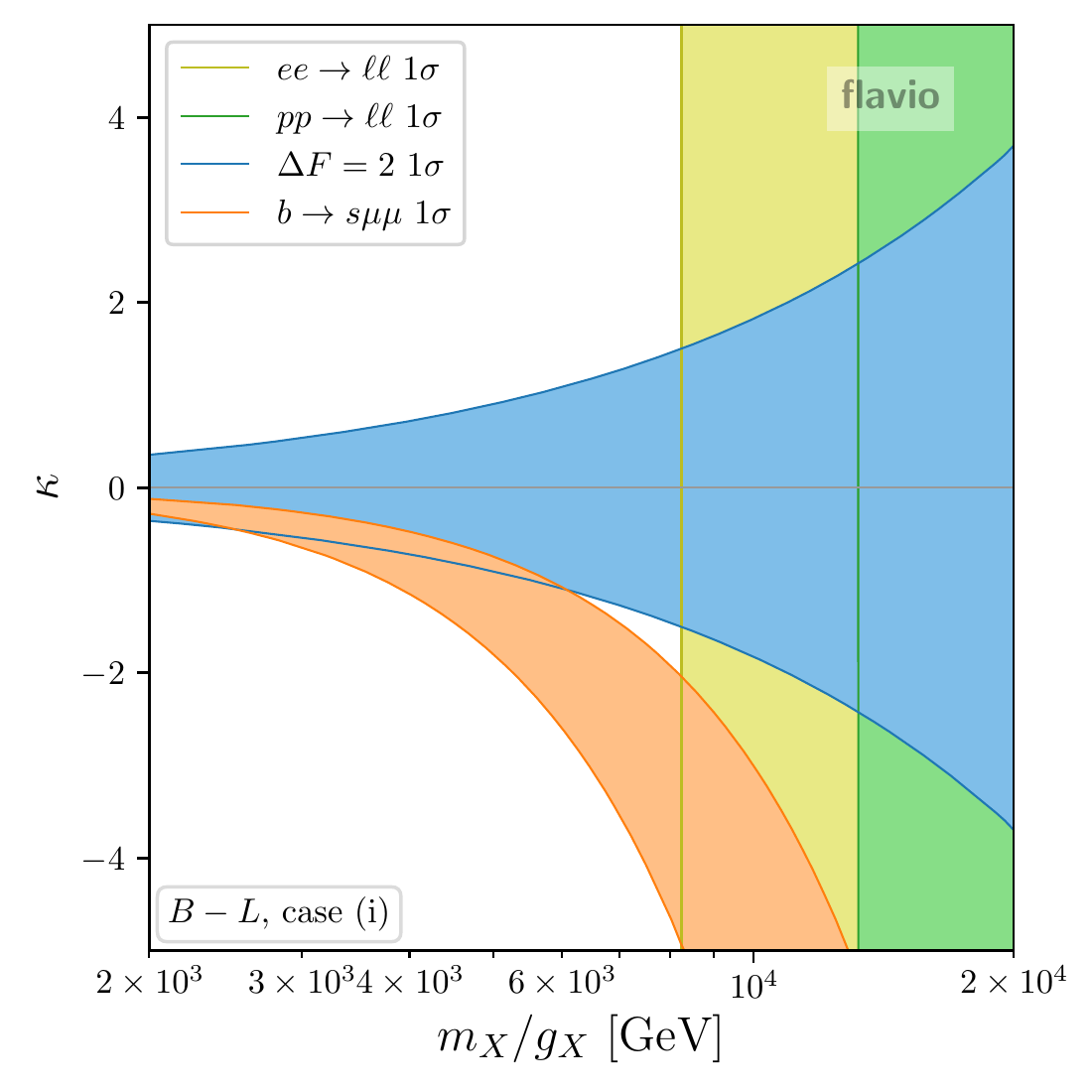}
     \end{subfigure}~
     \begin{subfigure}[b]{0.5\textwidth}
         \centering
         \includegraphics[width=\textwidth]{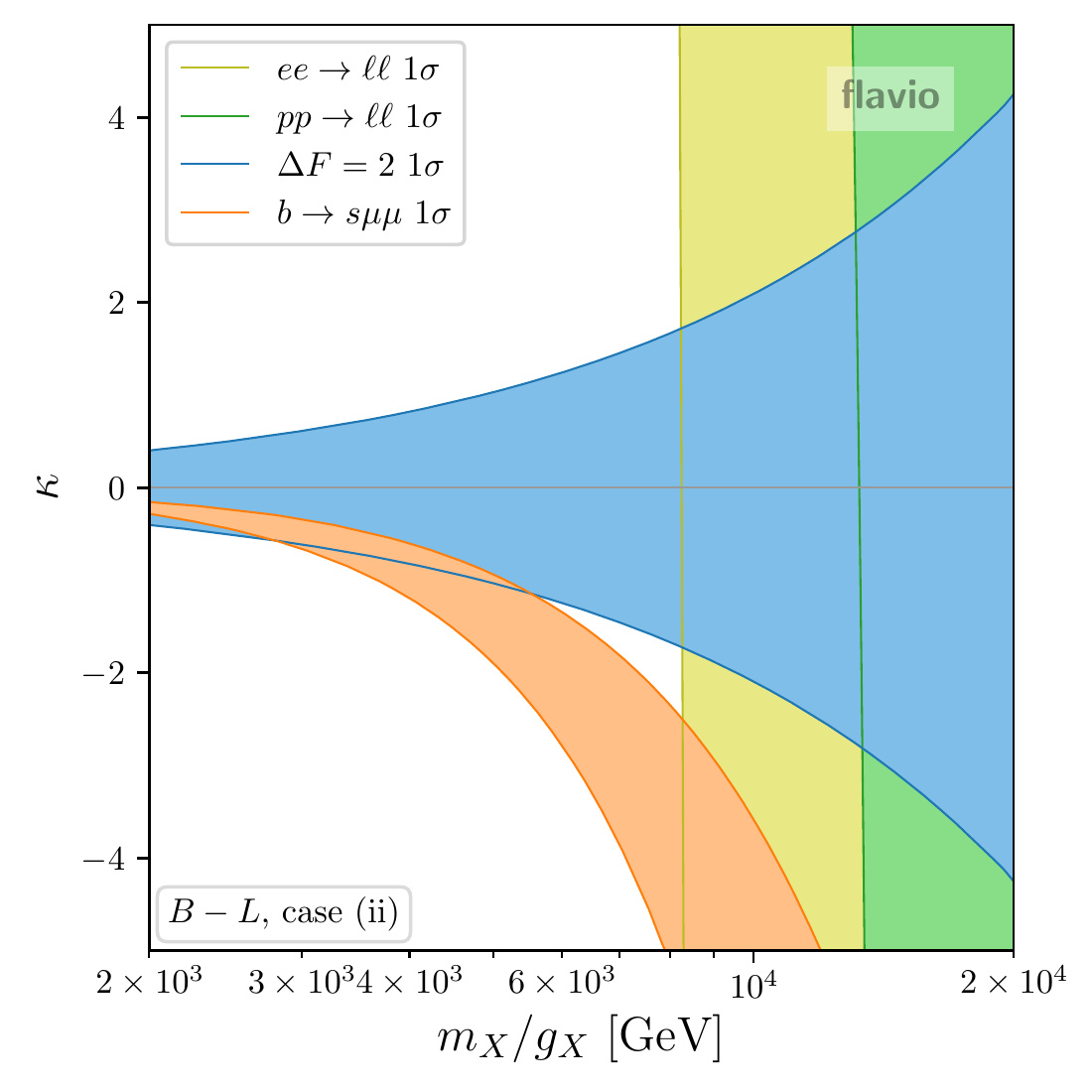}
     \end{subfigure}
    \caption{The best-fit regions at $1\sigma$ for all relevant constraints for the $U(1)_{B-L}$ model, see Section~\ref{sec:modelI}. We show the two cases described in Eq.~\eqref{eq:modItwoCases}, case $(i)$ on the left-hand side and case $(ii)$ on the right-hand side.}
        \label{fig:B-L}
\end{figure}

The $U(1)_{B-L}$ is the most celebrated $U(1)$ gauge extension of the SM; it fits into the Pati-Salam quark-lepton unification~\cite{Pati:1974yy} and $SO(10)$ grand unification~\cite{Fritzsch:1974nn}. The $U(1)_{B-L}$ is the exact global symmetry of the SM free of all gauge anomalies. Under this symmetry, all quark fields $q_L^i$, $u_R^i$, and $d_R^i$ have the same charge $+1/3$, while all lepton fields $l^i_L$ and $e_R^i$ are charged with $-1$. Adding three right-handed neutrinos, $\nu_R^i$ ($i=1,2,3$), which are the SM gauge singlets, allows for gauging the $B-L$ symmetry.\footnote{The $U(1)_{B-L}^3$ and the mixed $U(1)_{B-L} \times $Gravity$^2$ anomalies are absent when the right-handed neutrinos carry the universal lepton charge.} Spontaneous breaking of the $U(1)_{B-L}$ by an SM gauge singlet scalar field $\phi$, makes the associated gauge boson $X_\mu$ heavy. The phenomenological decoupling limit is achieved when the scalar condensate $\langle \phi \rangle \to \infty$ (or when the gauge coupling $g_{X} \to 0$). The high-energy breaking scenario is very attractive as an explanation of the small neutrino masses through the seesaw mechanism~\cite{Minkowski:1977sc,Mohapatra:1979ia,Yanagida:1979as,Gell-Mann:1979vob}. The Majorana mass comes from the term $\mathcal{L}\supset x_{ij}\, \nu^i_R \nu^j_R \phi$ with $\phi$ charge +2, when $\langle \phi \rangle \neq 0$.

At the renormalizable level, $X_\mu$ interacts with light fermions, $\mathcal{L} \supset g_{X}\, X_\mu J^\mu_{B-L}$,
where
\begin{equation} \label{eq:B-Lcurrent}
   J^\mu_{B-L} = \frac{1}{3} \left (\bar q_i \gamma^\mu q_i + \bar u_i \gamma^\mu u_i + \bar d_i \gamma^\mu d_i \right) - \left( \bar l_i \gamma^\mu l_i + \bar e_i \gamma^\mu e_i \right)~.
\end{equation}
Right-handed neutrinos are omitted assuming they become heavy enough after the $U(1)_{B-L}$ breaking. We also neglect the kinetic mixing contribution (typically loop suppressed), $\mathcal{L}  \supset \epsilon_{BX} X_{\mu\nu} B^{\mu \nu}$, and consider $\langle \phi \rangle \gg v_{{\rm EW}}$. The current in Eq.~\eqref{eq:B-Lcurrent} is flavor-universal, i.e. the summation over flavor index $i$ is assumed.

To get interesting flavor violation, let us imagine new states at some scale $\Lambda$ heavier than $X_\mu$ which integrate out to produce
\begin{equation}\label{eq:B-L:eft}
   \mathcal{L}_{eff} \supset \frac{c_{ij}}{\Lambda^2} (\phi^\dagger D_\mu \phi) (\bar q_i \gamma^\mu q_j)~.
\end{equation}
For example, this can be achieved by introducing new vector-like quarks $Q_{L,R}$ in the same SM representation as $q_L^i$ while having the same $U(1)_{B-L}$ charge as $\phi$, such that
\begin{equation} \label{eq:63}
    - \mathcal{L} \supset \lambda\,  \bar Q_R q_L \phi + \Lambda \,  \bar Q_R Q_L + {\rm h.c.}~.
\end{equation}
The details of this sector are not relevant to the rest of the discussion. The important effect of Eq.~\eqref{eq:B-L:eft} is that
\begin{equation}
    J^\mu_{X} = J^\mu_{B-L} + \frac{1}{3} \epsilon_{ij} \, \bar q_i \gamma^\mu q_j ~,
\end{equation}
where $|\epsilon_{ij}| \ll 1$. Otherwise, $\epsilon_{ij}$ is an arbitrary hermitian matrix. For convenience, we are working in the down-aligned basis, $q^i = ( V^*_{ji} u^j_L, d^i_L)^T$, such that the off-diagonal entries in $\epsilon_{ij}$ induce FCNCs in the down-quark sector. In the following analysis, we will consider two cases:
\begin{equation}\label{eq:modItwoCases}
  (i)~ \epsilon_{i j} = - \kappa \, |V_{ts}| (\delta_{i 2} \delta_{j 3} + \delta_{i 3} \delta_{j 2})  \quad {\rm and} \quad  (ii)~ \epsilon_{i j} = \kappa\, Y_u Y_u^\dagger~,
\end{equation}
where $\kappa$ is a real parameter and $Y_u$ is the up-quark Yukawa matrix. The second case corresponds to the MFV.\footnote{This can be generated, for example, by integrating out a vector-like quark triplet under $SU(3)_q$, $Q^i$ where $i=1,2,3$. The $SU(3)_q$ flavor symmetry is softly broken by $\Lambda \to \Lambda + \tilde \Lambda Y_u Y_u^\dagger$ in Eq.~\eqref{eq:63}. The condition $\epsilon \ll 1$ implies $\tilde \Lambda \ll \Lambda$.}

The total decay width of the $X_\mu$ boson is
\begin{equation}\label{eq:Width}
    \frac{\Gamma_{X}}{m_{X}} \approx \frac{13 g_X^2}{24 \pi}~,
\end{equation}
where we assumed $m_f \ll m_X$ for all SM fermions and set $\kappa = 0 $. For the perturbativity criteria, we will assume $\Gamma_{X} / m_{X} < 0.25 $ which implies $g_X \lesssim 1.2$.

Direct resonance searches in the Drell-Yan spectrum set stringent limits on the $X_\mu$, see e.g.~\cite{CMS:2021ctt}. Let us now consider $X_\mu$ boson with mass above the limits from direct resonance searches. See Section~\ref{sec:validity} for a benchmark example.

With this assumption, we can safely integrate out the $X_\mu$ field and match the model to the SMEFT at the tree level. We get
\begin{equation}\label{eq:ZpEFT}
    \mathcal{L} \supset -\frac{g_{X}^2}{2 m_X^2} J^\mu_{X} J_{X \mu}~.
\end{equation}
Expanding this expression, we find the Wilson coefficients for the dimension-6 SMEFT operators in the Warsaw basis in terms of $ m_X / g_X $ and $\epsilon_{ij}$. We get the following two-quark--two-lepton, four-lepton, and  four-quark operators:
\begin{align}
    [C_{lu}]_{\alpha \beta i j} &=[C_{ld}]_{\alpha \beta i j} =[C_{eu}]_{\alpha \beta i j} =[C_{ed}]_{\alpha \beta i j} = \frac{g_X^2}{3 m_X^2} \delta_{\alpha \beta} \delta_{i j} ~,\label{eq:607}\\
    [C_{lq}^{(1)}]_{\alpha \beta i j} &=[C_{qe}]_{i j \alpha \beta} = \frac{g_X^2}{3 m_X^2} \delta_{\alpha \beta} (\delta_{i j}+ \epsilon_{i j})~,\label{eq:608}\\
    [C_{le}]_{\alpha \beta \gamma \delta} &=2 [C_{ee}]_{\alpha \beta \gamma \delta} =2 [C_{ll}]_{\alpha \beta \gamma \delta} = - \frac{g_X^2}{m_X^2} \delta_{\alpha \beta} \delta_{\gamma \delta}~,\label{eq:609}\\
    [C_{qq}^{(1)}]_{i j k l} &= - \frac{g_X^2}{18 m_X^2} (\delta_{i j}+ \epsilon_{i j})\,(\delta_{k l}+ \epsilon_{k l})~,\label{eq:610}\\
    [C_{ud}^{(1)}]_{i j k l} &= 2 [C_{dd}]_{i j k l} = 2 [C_{u u}]_{i j k l} = - \frac{g_X^2}{9 m_X^2} \delta_{i j} \delta_{k l}~,\\
    [C_{qu}^{(1)}]_{i j k l} &=[C_{qd}^{(1)}]_{i j k l} = -\frac{g_X^2}{9 m_X^2} (\delta_{i j}+ \epsilon_{i j}) \delta_{k l}~.\label{eq:612}
\end{align}
The most important observables include rare meson decays from Eq.~\eqref{eq:608}, neutral meson mixings from Eq.~\eqref{eq:610},  high-mass Drell-Yan tails from Eqs.~\eqref{eq:607} and \eqref{eq:608}, and $e^+ e^- \to \ell^+ \ell^-$ from Eq.~\eqref{eq:609}. The last one was searched for at the LEP-II collider~\cite{Babich:2002jb,Falkowski:2015krw}.\footnote{The LEP-II experiment has also searched for contact interactions in $e^+ e^- \to j j$ which can be most directly compared with the high mass Drell-Yan $p p \to e^+ e^-$. However, those provide a subleading constraint on this model.}
We implement in {\tt flavio} the SMEFT $\chi^2$ for four-lepton contact interactions reported in Ref.~\cite{Falkowski:2015krw}. Other operators correct dijet and multijet tails at high-$p_T$, as well as flavor-violating hadronic decays, which are expected to give subleading bounds. For neutral $B$ meson mixings, we use values of the bag parameters resulting from a combination of their determination by the FNAL/MILC~\cite{FermilabLattice:2016ipl} and HPQCD~\cite{Dowdall:2019bea} collaborations, as well as a recent sum rules determination~\cite{King:2019lal}. We provide details on the combination, including correlations, in Appendix~\ref{appendix:BdBsBag}.

We can efficiently scan the global likelihood with our {\tt flavio} framework. The interesting case is a 2D scan in $(m_X / g_X, \kappa)$ where the tension in $P_5'$ and other observables in $b \to s \mu \mu$ sector is confronted against complementary constraints. At the same time, the LFU ratios are predicted to be SM-like. The results are shown in Fig.~\ref{fig:B-L} for the two cases for $\epsilon_{ij}$, see Eq.~\eqref{eq:modItwoCases}. The colored regions show preferred parameter space by different data sets. In both cases, there is no parameter space in which all constraints overlap. The second case (MFV) is very similar to the first case, showing that the $bs$ system dominates the mixing bounds. To sum up, the combination of $\Delta F =2 $ and high-mass dilepton tails at the LHC (and LEP-II) excludes the explanation of the $b \to s \mu^+ \mu^-$ anomalies. The LHC tails here are more effective than the LEP-II tails since the $X_\mu$ couples to valance quarks. This will not be the case in the following example.

\subsection{Gauged $U(1)_{3 B_3-L}$}
\label{sec:modelIb3}

\begin{figure}[t]
     \centering
     \begin{subfigure}[b]{0.5\textwidth}
         \centering
         \includegraphics[width=\textwidth]{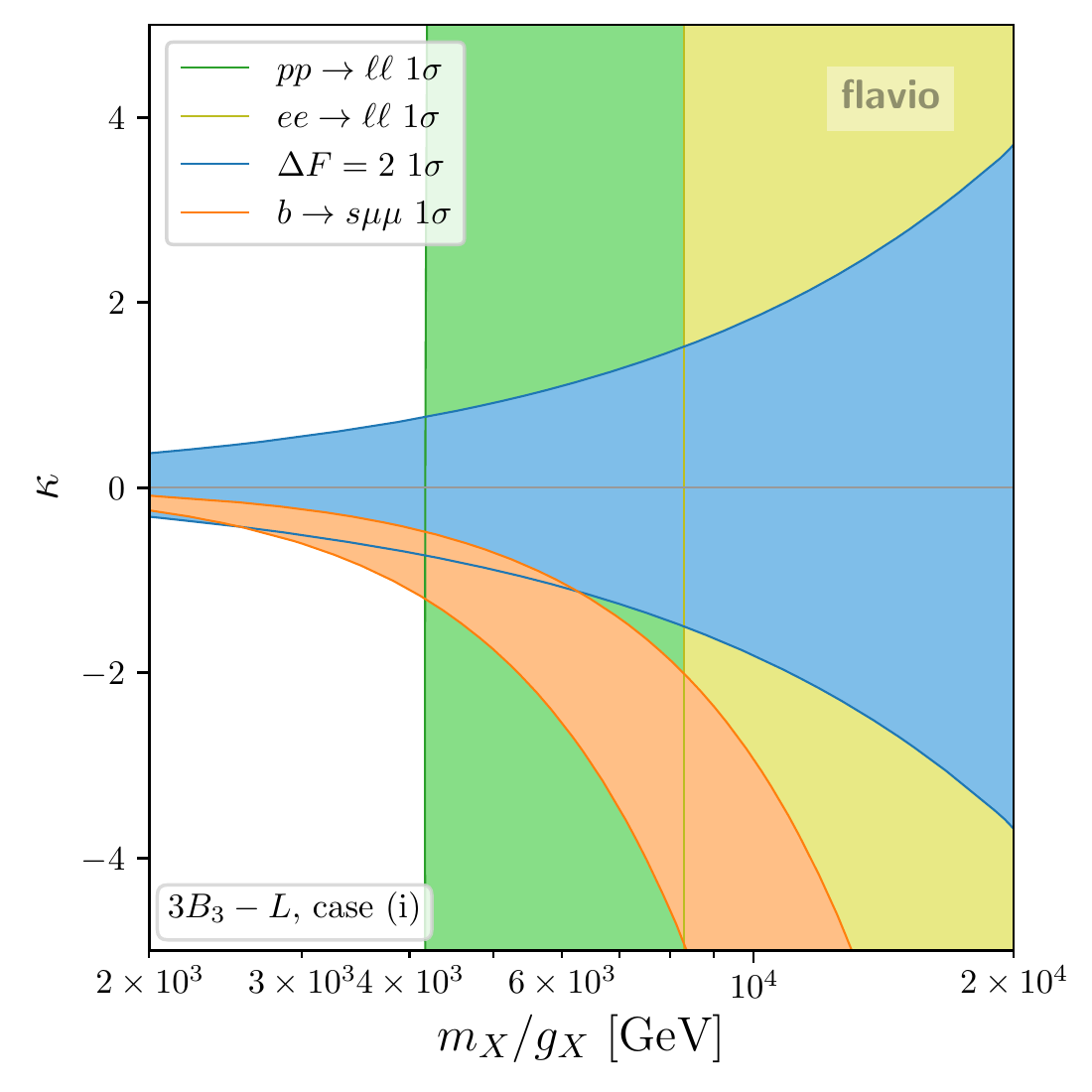}
     \end{subfigure}~
     \begin{subfigure}[b]{0.5\textwidth}
         \centering
         \includegraphics[width=\textwidth]{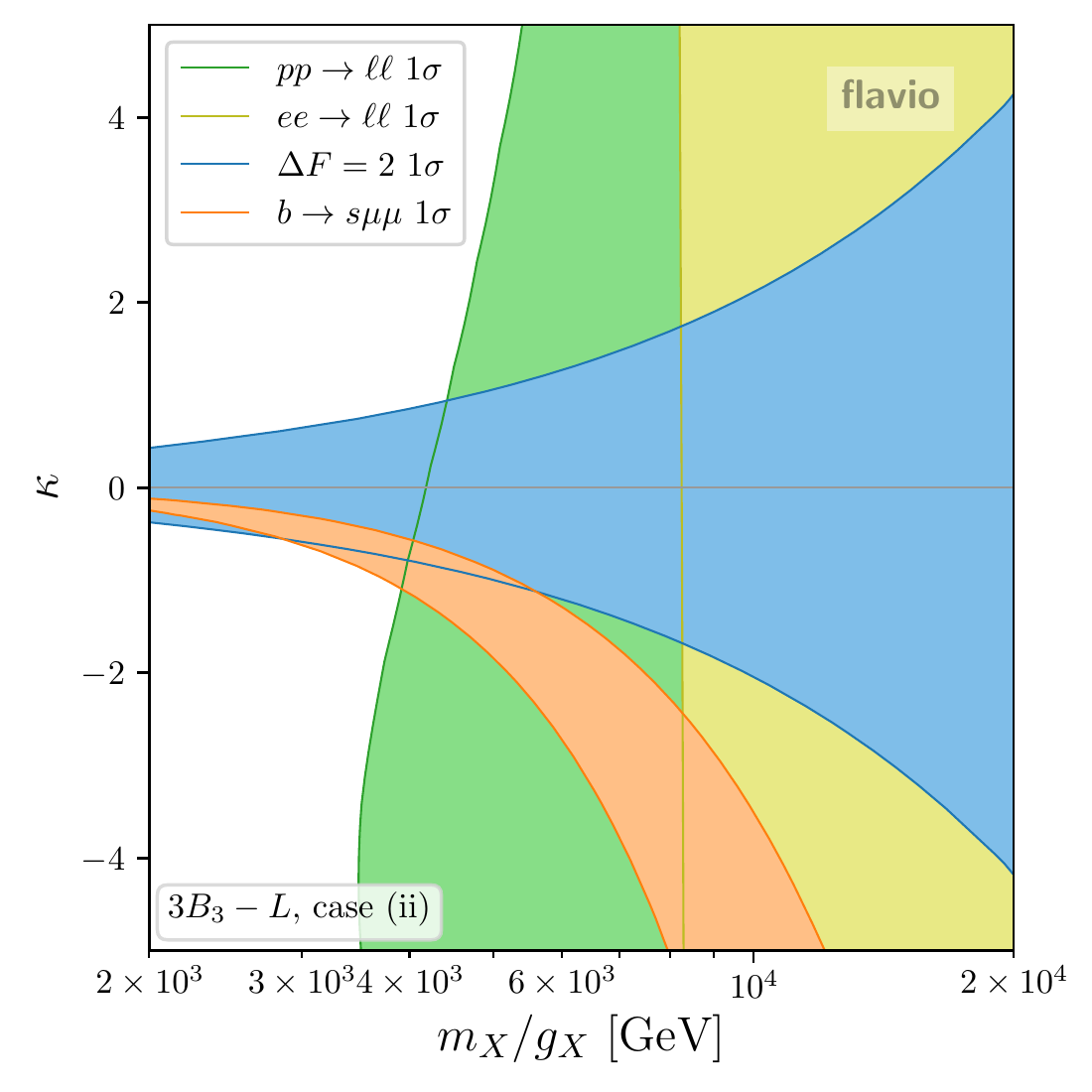}
     \end{subfigure}
        \caption{The best-fit regions at $1\sigma$ for all relevant constraints for the $U(1)_{3 B_3-L}$ model, see Section~\ref{sec:modelIb3}. We show the two cases described in Eq.~\eqref{eq:modItwoCases}, case $(i)$ on the left-hand side and case $(ii)$ on the right-hand side.}
        \label{fig:3B3-L}
\end{figure}

Let us consider a variation of the previous model in which only a single generation of quarks is charged under the additional $U(1)$ gauge group, while the other two generations carry zero charges. The anomaly cancellation conditions (Eqs.~(2.2)--(2.7) in Ref.~\cite{Greljo:2021npi}) are fulfilled for the charge assignment $3 B_3 - L$, where $B_3$ is the baryon number for the third family. While the generation of leptonic masses and mixings proceeds as before, the quark Yukawa matrices decompose into a direct sum of $2\times 2$ (light generations) and $1\times1$ (the third generation). Thus, the CKM mixing elements between the light and the third generation are absent at the renormalizable level. Those can be generated by dimension-5 operators, for example $\frac{1}{\Lambda}\bar q_i H \phi b_R$ and $\frac{1}{\Lambda}\bar q_i \tilde H \phi t_R$, where $i=1,2$ and the symmetry-breaking SM-singlet scalar $\phi$ has a charge $-1$ to annihilate the charge of $b_R / t_R$. The $U(1)$ symmetry breaking slightly above the TeV scale explains the smallness of the CKM elements $V_{td}$ and $V_{ts}$ if the associated scale is $\Lambda \sim 100$\,TeV. Thus, the TeV-scale model could be the first layer of a UV structure addressing the rest of the flavor puzzle.

The phenomenological advantage of this setup is that the associated $X_\mu$ boson couples dominantly with the third generation of quarks and its production is therefore suppressed at the LHC,
\begin{equation} \label{eq:3B3-Lcurrent}
   J^\mu_{3 B_3-L} = \left (\bar q_3 \gamma^\mu q_3 + \bar u_3 \gamma^\mu u_3 + \bar d_3 \gamma^\mu d_3 \right) - \left( \bar l_i \gamma^\mu l_i + \bar e_i \gamma^\mu e_i \right)~.
\end{equation}
In this model, we expect the high-mass Drell-Yan bounds to be dominated by the $b \bar b$ channel and therefore suppressed. Instead, the tree-level $X_\mu$ contribution to FCNC transitions is automatically generated together with the CKM matrix when rotating the quark fields from the interaction to mass eigenstate basis.

Interestingly, the model can be made compatible with the minimally broken $U(2)^3$ flavor symmetry~\cite{Barbieri:2011ci} (see also~\cite{Kagan:2009bn}). A part of the global symmetry of the quark kinetic term is $U(2)^3$ under which light generations form doublets. The symmetry is minimally broken by a doublet under $U(2)_q$ denoted as $V_q=(V_{td}, V_{ts})$, and two bidoublets for the light quark masses. In this model, the former is realized by the aforementioned dimension-5 operator while the latter is present already at dimension 4. The minimally broken $U(2)^3$ predicts the left-handed rotations to dominate over the right-handed ones.\footnote{For the explicit realization of the left-handed dominance with vector-like quarks, see a closely related model in Section~2.3 in Ref.~\cite{Greljo:2021npi}. By choosing appropriate representations, operators of the type $\frac{1}{\Lambda}\bar q_3 H \phi^\dagger d_i $ are absent, while $\frac{1}{\Lambda}\bar q_i H \phi d_3 $ is present.} The important effect is that the leading $X_\mu$ interactions in the mass basis are associated with the current
\begin{equation}
    J^\mu_{X} = J^\mu_{3 B_3-L} + \frac{1}{3} \epsilon_{ij} \, \bar q_i \gamma^\mu q_j ~,
\end{equation}
where $|\epsilon_{ij}| \ll 1$. The rest of the matching calculation proceeds as in Section~\ref{sec:modelI}. The matching results in Eqs.~\eqref{eq:607}--\eqref{eq:612} stay the same after the replacement of $\delta_{ij} \to 3 \,\delta_{i 3} \delta_{j 3}$ for all quark indices. We again choose the down-aligned basis.

In Fig.~\ref{fig:3B3-L}, we show the best-fit regions for different data sets assuming the two cases for $\epsilon_{ij}$ as in the previous section, see Eq.~\eqref{eq:modItwoCases}. The only difference with respect to the $U(1)_{B-L}$ case is that the high-mass Drell-Yan bound is less stringent (dominant couplings are with $b$ quarks) and now compatible with the intersection of $\Delta F = 2$ and $b \to s \ell^+ \ell^-$. However, the four-lepton contact interactions are inconsistent with this parameter space at the $1\sigma$ level. This is a general feature of the LFU $Z'$ models --- the $e^+ e^- \to \ell^+ \ell^-$ becomes a critical constraint. This is in contrast to the LFU violating models such as $L_\mu - L_\tau$, where the analogous bound was a neutrino trident production, and thus much weaker~\cite{Altmannshofer:2014pba}.

\subsection{LFU leptoquark}
\label{sec:modelIIb}

\begin{figure}[t]
         \centering
         \includegraphics[width=0.5\textwidth]{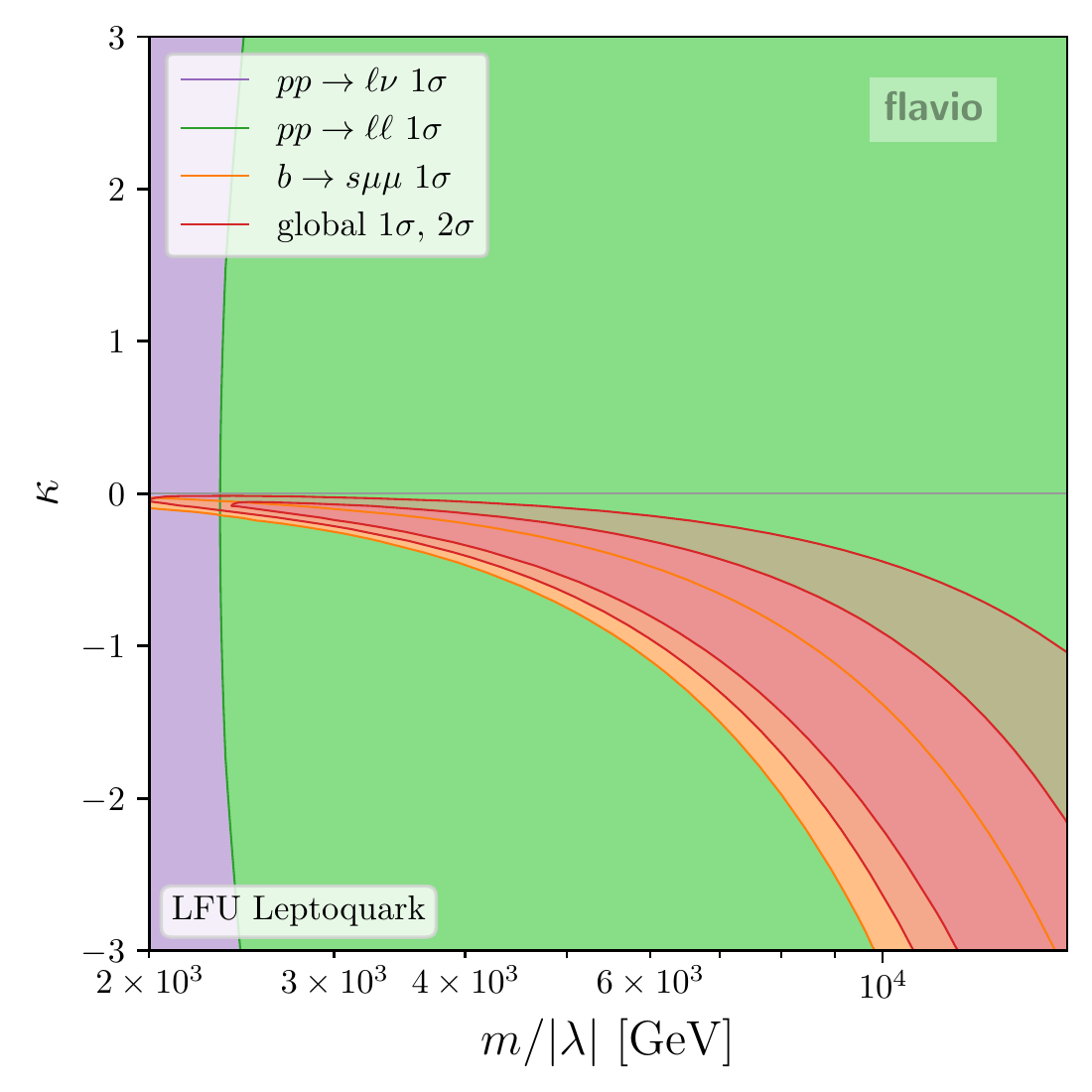}
        \caption{The best-fit regions at $1\sigma$ for all relevant constraints for the LFU leptoquark model. The global fit is shown at $1$ and $2\sigma$. See Section~\ref{sec:modelIIb} for details.}
        \label{fig:LQ_LFU}
\end{figure}

Let us consider a triplet of scalar leptoquarks $S^\alpha \, (\alpha=1,2,3)$ in the same SM gauge representation $(\bar {\bf 3}, {\bf 3}, 1/3)$.\footnote{Adding $n$ copies of scalars in the (anti)fundamental representation of $SU(3)_c$ and the adjoint representation of $SU(2)_L$ modifies the
SM beta functions of both $g_S$ and $g_2$. While the strong coupling $g_S$ stays asymptotically free for $n \leq 13$, the $SU(2)_L$ coupling acquires a Landau pole for any $n > 1$ \cite{Bandyopadhyay:2021kue}. However, this Landau pole stays way above the EW scale. Using the results from Ref.~\cite{Bandyopadhyay:2021kue}, we found the pole at one loop to be above $10^{31},10^{12}$ and $10^{8}$ GeV for $n=3,6,9$, respectively. The two-loop corrections push the pole to lower values. However, it still stays above $\sim10^{7}$ GeV even for $n=9$.} The flavor index $\alpha$ refers to the lepton flavor, and leptoquarks form a ${\bf \bar 3}$ under $U(3)_l$ symmetry.\footnote{An alternative would be to consider a leptoquark doublet of a flavored $U(2)_l$. Under this symmetry, the first two lepton generations form a doublet while the third generation forms a singlet. Even more minimal global symmetry assumption would be introducing two scalar leptoquarks $S_e$ and $S_\mu$ which carry (anti)electron and (anti)muon numbers, respectively, and a $Z_2$ parity under which $S_e \leftrightarrow S_\mu$ while at the same time $l_e \leftrightarrow l_\mu$. Thus, the symmetry is $U(1)_e  \times U(1)_\mu \times Z_2$ where $Z_2$ is generated by the group element $\begin{pmatrix}
    0 & 1\\
    1 & 0
\end{pmatrix}$ of the $O(2)$ subgroup of $U(2)_l$. The first two factors ensure lepton flavor conservation, while the third factor implies lepton flavor universality. The smaller symmetry could act in the IR as a remnant of a larger flavor symmetry breaking in the UV.} The Lagrangian is
\begin{equation}\label{eq:LagLQLFU}
    \mathcal{L} \supset (D_\mu S^\alpha)^\dagger (D^\mu S^\alpha) - m^2 S^{\alpha \dagger} S^\alpha - ( \lambda_i \, \bar q^c_i l_\alpha S^\alpha + {\rm h.c.})~,
\end{equation}
where the sum over repeated indices is assumed. The $SU(2)_L$ contraction in the last term is $\bar q^c i \sigma^2 \sigma^a l S^a$, where $a$ is the $SU(2)_L$ adjoint index and $\sigma^a$ are the Pauli matrices. For simplicity, we assume that the quark flavor structure is consistent with the minimally broken $U(2)^3$ quark flavor symmetry. In particular, we set $\lambda_i = \lambda\,(\kappa V_{td}, \kappa V_{ts}, 1)$ where $\kappa$ is an $\mathcal{O}(1)$ parameter. The lepton flavor symmetry is broken minimally by the lepton Yukawa $Y_l$ implying no lepton flavor violation and approximate lepton universality due to  $Y_l \ll 1$. Consider replacing $m^2 \to m^2 + \delta m^2 Y_l Y_l^\dagger$. For $\delta m^2 \lesssim m^2$, this model predicts no sizeable deviations in $R_{K^{(*)}}$. However, the symmetry-allowed interactions in Eq.~\eqref{eq:LagLQLFU} lead to important universal effects in $b \to s \ell^+ \ell^-$ transitions.

The tree-level matching of this model to the SMEFT, neglecting $Y_l$ breaking, gives
\begin{align}
    [C_{lq}^{(3)}]_{\alpha \beta i j} &=  \frac{\delta_{\alpha\beta} \lambda_i^\ast \lambda_j}{4 m^2}~, \\
[C_{lq}^{(1)}]_{\alpha \beta i j} &= 3 [C_{lq}^{(3)}]_{\alpha \beta i j}~.
\end{align}
Matching the model at the one-loop level will give rise to neutral meson mixing $\Delta F =2 $ and $e^+e^- \to \ell^+ \ell^-$. However, the loop suppression is enough to make these observables marginally relevant in the parameter space of interest (see the discussion in Section~\ref{sec:modelII}). This is in contrast to the case of a $Z'$ mediator where those observables are induced at the tree level and play a prominent role in shaping the preferred parameter space.

The results of the fit are presented in the $(m/|\lambda|, \kappa)$ plane, see Fig.~\ref{fig:LQ_LFU}. There exists a parameter space where all the constraints are compatible with each other. The global fit is colored red and shown at $1$ and $2\sigma$. The high-mass Drell-Yan tails, in this case, are not constraining enough since the dominant coupling is with the third quark generation. This is arguably the best-performing model among our four examples to fit $b \to s \ell^+ \ell^-$ angular distributions and branching fractions while predicting SM-like LFU ratios $R_{K^{(*)}}$ and staying compatible with the complementary bounds. The crucial aspect is the absence of four-quark and four-lepton operators at the tree level.

Again, this mechanism requires at least a doublet of leptoquarks under $U(2)_l$. A single leptoquark coupled both to electrons and muons with the same strength would lead to charged lepton flavor violation!\footnote{Consider a single $S_3$ leptoquark field with couplings in the down-quark and charged-lepton mass eigenbases satisfying $\lambda_{b \mu} \lambda^*_{s \mu} = \lambda_{b e} \lambda^*_{s e}$ such that LFU ratios are SM-like. These minimal couplings would induce too excessive $\mu \to e \gamma$ decay at the one-loop level without the possibility to cancel the $b$ quark diagram against the $s$ quark diagram, $\Gamma \propto (|\lambda_{s \mu} \lambda^*_{s e}| + |\lambda_{b \mu} \lambda^*_{b e}|)^2$, see also~\cite{Crivellin:2017dsk}. Despite the chirality flips on the muon leg, the limit from~\cite{MEG:2016leq} on $\mathcal{L} \supset \frac{e m_\mu}{16 \pi^2 \Lambda^2} \bar e_L \sigma^{\mu \nu} \mu_R F_{\mu \nu}$ is $\Lambda \gtrsim 70 $\,TeV.  In addition, $\mu \to e$ conversion on heavy nuclei induced at tree-level through a small misalignment (of the order of the Cabbibo angle) between $s_L$ and $d_L$  would be too much (see bounds in Table 8 of Ref.~\cite{Feruglio:2015gka}), while an effect due to an RGE running~\cite{Crivellin:2017rmk} from $(\bar b_L \gamma^\mu b_L)(\bar e_L \gamma^\mu \mu_L)$ is in the right ballpark. Finally, searches for $b \to s \mu e$ decays also probe the interesting parameter space, see Table 3 of Ref.~\cite{Davidson:2018rqt}. }

\subsection{MFV leptoquark}
\label{sec:modelII}

\begin{figure}[t]
         \centering
         \includegraphics[width=0.5\textwidth]{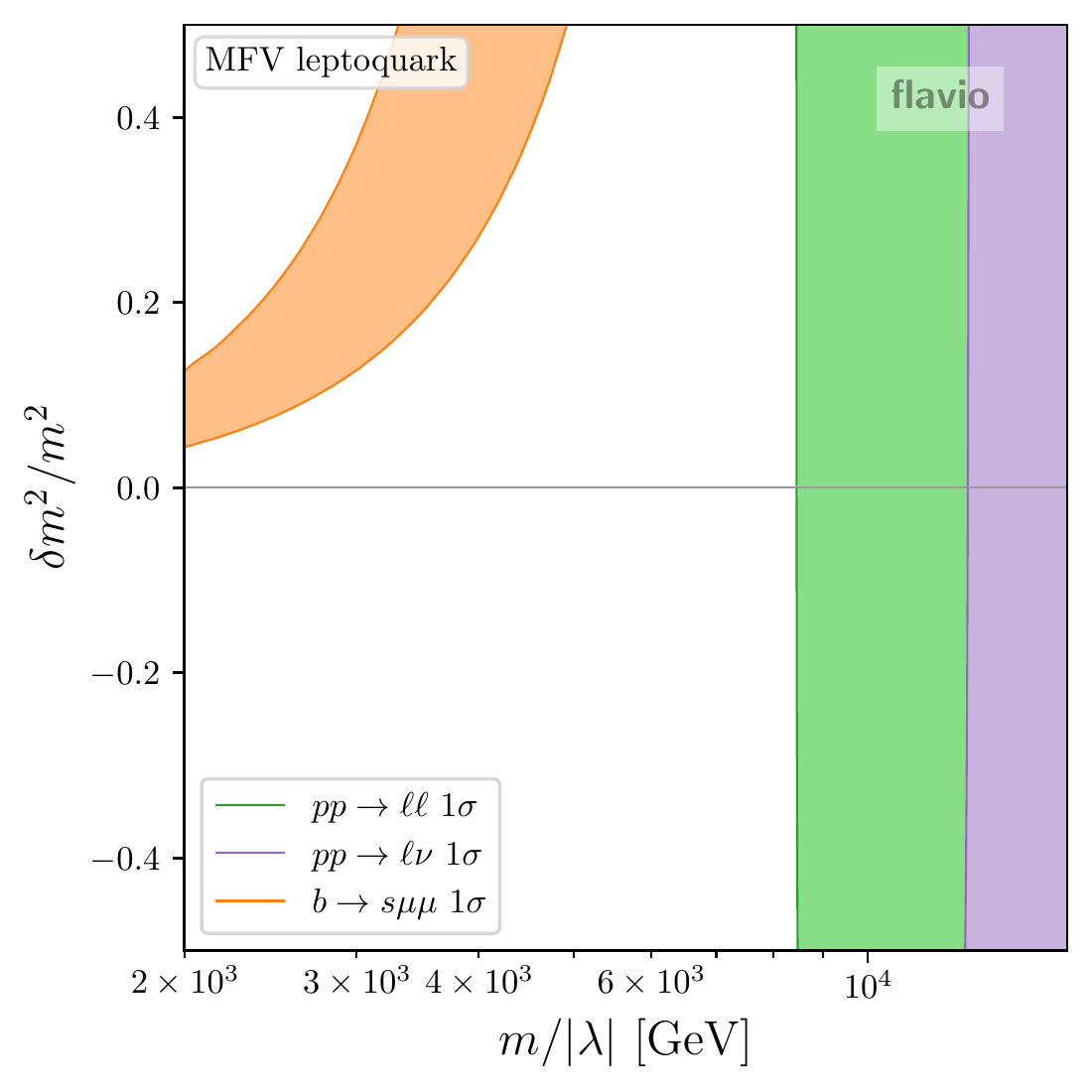}
        \caption{The best-fit regions at $1\sigma$ for all relevant constraints for the MFV leptoquark model. See Section~\ref{sec:modelII} for details.}
        \label{fig:LQ_MFV}
\end{figure}

We introduce a bitriplet of scalar leptoquark fields $S^{i \alpha}$ ($i,\alpha=1,2,3$) all of which are in $(\bar {\bf 3}, {\bf 3}, 1/3)$ representation under the SM gauge group. The extra indices $i$ and $\alpha$ are flavor indices and refer to the global symmetry of the SM kinetic term, $SU(3)_q \times SU(3)_l$, under which leptoquarks form a $({\bf \bar 3},{\bf \bar 3})$ representation.\footnote{We could have chosen more minimal LFU leptoquark options as discussed in the previous section. While here we are interested in LFU $b \to q \ell^+ \ell^-$ transitions, a different case would be to consider leptoquarks carrying a muon number -1 (muoquark) or electron number -1 (electroquark) which give corrections to $R_{K^{(*)}}$ and are still MFV in the quark sector~\cite{Davidson:2010uu,Aloni:2017ixa}.} The SM gauge symmetry and the global flavor symmetry restrict the renormalizable Lagrangian to
\begin{equation}\label{eq:LagLQ}
    \mathcal{L}_{0} \supset (D_\mu S)^\dagger (D^\mu S) - m^2 S^{\dagger} S - ( \lambda \, \bar q^c_i l_\alpha S^{i\alpha} + {\rm h.c.})~,
\end{equation}
where we omitted scalar self-interactions and the Higgs portal. Summation over the repeated quark (lepton) flavor indices $i$ ($\alpha$) is assumed.

Let us consider the case in which the global symmetry of Eq.~\eqref{eq:LagLQ} is only softly broken by the known SM sources.\footnote{One could alternatively consider $Y_u Y_u^\dagger$ directly in the interaction term. However, the soft-breaking option is easier to realize in the UV where the symmetry gets restored.} The leading correction is
\begin{equation}\label{eq:LagLQbreaking}
    \mathcal{L}_{I} \supset - \delta \tilde m^2 \, S \,Y_u Y_u^\dagger \,S^{ \dagger} ~,
\end{equation}
where $Y_u$ is the up-quark Yukawa matrix. We neglect the contributions from the down-quark Yukawa. In this model, quark flavor violation proceeds minimally through the SM sources, while there is no charged lepton flavor violation.\footnote{As a speculation, the UV origin of assumed global symmetries could be traced back to a gauged flavor symmetry spontaneously broken at some high-energy scale. The breaking term in Eq.~\eqref{eq:LagLQbreaking} might come from the cross-quartic scalar interaction between the leptoquark field and flavon fields after the latter condensate to break $SU(3)_q$ spontaneously and generate quark Yukawas. The details of the UV completion are beyond the purpose of this work.}

Working in the down-aligned quark mass basis, $q^i = ( V^*_{ji} u^j_L, d^i_L)^T$, the up-quark Yukawa matrix takes the following form $Y_u = V^\dagger \hat Y_u$, where $\hat Y_u$ is a diagonal Yukawa matrix while $V$ is the CKM mixing matrix. The leptoquark mass eigenstates are $\mathcal{S}^{i\alpha}=V^*_{i j} S^{j\alpha}$, and the quark-lepton interaction is $\mathcal{L}\supset - \lambda \mathcal{S} V \bar q^c l$. Neglecting up-sector Yukawa couplings but $(\hat Y_u)_{33} \equiv y_t$, the leptoquark masses are $m^2_{1,2} = m^2$ and $m^2_{3} = m^2 + \delta m^2 $ where $\delta m^2 = \delta \tilde m^2 y^2_t$. When the breaking parameter $\delta m^2 / m^2 \ll 1$, the $SU(3)_q$ symmetry of the leptoquark sector is still approximate, while for $\delta m^2 / m^2 \sim \mathcal{O}(1)$ the symmetry is completely broken down to the $U(2)_q$ subgroup, see~\cite{Kagan:2009bn,Barbieri:2011ci}. In our numerical analysis, we consider $|\delta m^2 / m^2| < 0.5$, which we denote as the linear MFV, while the $U(2)_q$ case was considered in Section~\ref{sec:modelIIb} in the limit in which the doublet states decouple.

We further assume $m \gtrsim 2$\,TeV, which is consistent with the LHC exclusion limits from direct searches for pair-produced leptoquarks~\cite{ATLAS:2020dsk, CMS:2022nty}. In this case, the EFT description of the Drell-Yan tails approximates well the $t$-channel exchange of a leptoquark. Under these assumptions, we can define the effective scale $m / {|\lambda|}$, which together with $\delta m^2 / m^2$ defines the parameter space of the model to be confronted against the experimental data.

The tree-level matching to the SMEFT gives
\begin{align}
    [C_{lq}^{(3)}]_{\alpha \beta i j} &=  \frac{|\lambda|^2 \delta_{\alpha\beta}}{4 m^2} \left ( \delta_{ij} + V_{tj} V_{ti}^* \left (\frac{1}{1+\delta m^2 / m^2} -1 \right) \right)~,\\
[C_{lq}^{(1)}]_{\alpha \beta i j} &= 3 [C_{lq}^{(3)}]_{\alpha \beta i j}~.
\end{align}
To first order in small $\delta m^2 / m^2$, one finds $ [C_{lq}^{(3)}]_{\alpha \beta i j} \propto \delta_{ij} - V_{tj} V_{ti}^* \, \delta m^2/ m^2$. Thus, when the symmetry is restored ($\delta m^2 = 0$)  there are no FCNCs in the quark sector.

The most relevant one-loop matching contribution is to the neutral meson mixing from the box diagrams with leptoquarks ($\Delta F = 2$ transitions). Following~\cite{Gherardi:2020det,Gherardi:2020qhc} and summing over all leptoquark flavors, we find for $i\neq j$
\begin{align}
[C_{qq}^{(3)}]_{ijij} &= - \frac{3|\lambda|^4 (V_{tj} V_{t i}^*)^2}{256 \pi^2 m^2} \left ( 1 + \frac{1}{1+\delta m^2 / m^2} - \frac{2}{\delta m^2 / m^2} \ln \left (1+\frac{\delta m^2}{m^2} \right) \right)~,\\
[C_{qq}^{(1)}]_{ijij} &= 9 [C_{qq}^{(3)}]_{ijij}~.
\end{align}
Series expansion around $\delta m^2 = 0$ gives $\approx \delta m^4 / (3 m^4)$ for the bracket in the above equation. Thus, the flavor-conserving limit is restored more quickly when the symmetry-breaking parameter is sent to zero (double GIM mechanism). We checked that the mixing bounds are not constraining enough in the relevant parameter space.

We show the bounds on the model parameter space $(m/|\lambda|, \delta m^2/m^2)$ in Fig.~\ref{fig:LQ_MFV}, from $B$ decays at low energies and both NC and CC high-mass Drell-Yan tails. Firstly, note there is a parameter space with a preferred positive value of $\delta m^2/m^2$ in which the low-energy $b \to s \mu^+ \mu^-$ data can be accommodated (corresponding to negative $C_9^{bs\ell\ell} = -C_{10}^{bs\ell\ell}$ at low energies). Note that, as in all other models in this paper, $R_{K^{(*)}}=R^{{\rm SM}}_{K^{(*)}}$. In the limit of $\delta m^2/m^2\to 0$, the model has no flavor violation and no impact on $B$ decays. However, in the limit of large $\delta m^2/m^2$, we are no longer in the linear MFV regime. Hence we cut off the plot at $\delta m^2/m^2=0.5$. Similar to the results presented in the MFV section, the high-mass DY bounds are highly relevant for flavor-conserving interactions. Hence we observe a stringent bound on $M$, with only values of $M\gtrsim 8~\mathrm{TeV}$ allowed. There is an apparent incompatibility between the low-energy and high-energy data, namely the $B$-physics data, containing the $b \to s \mu^+ \mu^-$ anomalies, which can not be explained without violating constraints from the high-mass Drell-Yan, both charged and neutral currents.

\section{Conclusions}
\label{sec:conc}

Despite undergoing multiple experimental tests in high-energy colliders without any indications of its failure, the Standard Model of particle physics must continue to be scrutinized with greater precision and nuance in order to identify potential weaknesses that could help to address unresolved questions and reveal unexplored realms. Should new physics emerge in the UV, it would be manifested in the IR through a multipole expansion of the Standard Model, known as the SMEFT. The discovery and study of higher-dimensional operators of the SMEFT, as well as their correlations, would represent a significant step toward a reductionist approach to achieving a paradigm shift.

The high dimensionality of the WC parameter space, the vast landscape of physical observables, the quantum field theory complexity of predicting observables, and the data analysis challenge of constructing, combining, and exploring likelihoods, are some of the problems that need to be coherently addressed when interpreting data in the SMEFT framework. This paper reported on an implementation of rare $b$ meson decays (Section~\ref{sec:low_energy}) and high-mass Drell-Yan production (Section~\ref{sec:implementation}) in the SMEFT at dimension 6 level in the {\tt flavio} framework. Clever approximations, optimizations, and algorithms are developed to allow efficient scans of the multidimensional SMEFT parameter space while maintaining reliable and sufficiently precise theory predictions. This work is meant to facilitate the comparison of new physics models against data and will be made publicly available in future versions of the \texttt{flavio}\footnote{\url{https://github.com/flav-io/flavio}} Python package and the global SMEFT likelihood~\cite{Aebischer:2018iyb} implemented in the \texttt{smelli} Python package\footnote{\url{https://github.com/smelli/smelli}}.

With such a tool at hand, we conduct thorough SMEFT interpretations of the existing data sets, including the latest measurements such as the LHCb update on $R_{K^{(*)}}$~\cite{LHCb:2022qnv}. The main physics results are discussed in Section~\ref{sec:pheno} and summarised in Table~\ref{tab:1dbounds} and Figs.~\ref{fig:minim-bdee-smeft}, \ref{fig:MFV_lq1C0vslq1Cu}, \ref{fig:MFV_lq1C0=+lq1Cuvslq3C0=+lq3Cu_universal} and \ref{fig:MFV_ledqCdvsledqCduu}. Those include various operator(s) and flavor structure choices that are presumably a low-energy outcome of a particular class of NP models, illustrating the interplay between different data sets.

While SMEFT analyses are helpful to draw general lessons, specific models address concrete questions, and best demonstrate the usefulness of the new {\tt flavio} functionalities. Motivated by the current trends in the data, in Section~\ref{sec:models} we construct and thoroughly investigate $Z'$ and leptoquark models, which predict LFU effects in $b \to s \ell^+ \ell^-$ transitions explaining the tension in $P_5'$ and related observables while predicting $R_{K^{(*)}} \approx R_{K^{(*)}}^{{\rm SM}}$. The main results are summarised in Figs.~\ref{fig:B-L}, \ref{fig:3B3-L}, \ref{fig:LQ_LFU} and \ref{fig:LQ_MFV}, where the global data is projected onto the parameter space of the model, finding either tension or compatibility.

There are several promising directions for future work. The SMEFT implementation of Drell-Yan with final states including $\tau$ leptons, as well as possibly (soft) $b$-jets in {\tt flavio} would be crucial for studying NP models interacting dominantly with the third family~\cite{Faroughy:2016osc,Greljo:2018tzh,Marzocca:2020ueu,Haisch:2022lkt}. We urge experimental collaborations to report unfolded differential cross sections in the high-energy tails to facilitate interpreting such final states. In addition, the LFU $Z'$ models highlighted the importance of the contact interaction searches at LEP-II in $e^+ e^- \to \ell^+ \ell^-$. The full SMEFT implementation of those, as well as $e^+ e^- \to j j$, in {\tt flavio}, requires a separate publication. In particular, the dijet production (possibly containing $b$ jets) can be directly compared with the rare $b$ decays and the high-mass Drell-Yan, again via the crossing symmetry. In addition, the FCC-ee sensitivity projection for such processes would be very informative when compared to the future forecast on rare $b$ decays. Such studies might impact the upcoming course of flavor physics in decades to come.

\section*{Acknowledgements}
We thank David Straub for providing an initial implementation of NC high-mass DY tails in {\tt flavio}. We thank Marzia Bordone, Matthew Kirk, and Felix Wilsch for useful discussions.
This work received funding from the Swiss National Science Foundation (SNF) through the Eccellenza Professorial Fellowship ``Flavor Physics at the High Energy Frontier'' project number 186866. AG is also partially supported by the European Research Council (ERC) under the European Union’s Horizon 2020 research and innovation program, grant agreement 833280 (FLAY).

\appendix

\section{Validation of the Drell-Yan implementation}
\label{app:implementation}
In this Appendix, we comment on the various validations of the high-mass Drell-Yan implementation in \texttt{flavio}, which was presented in Section~\ref{sec:implementation}. We used MadGraph5\_aMC@NLO v.3.1.0 \cite{Alwall:2014hca} for validations against Monte Carlo simulations.

\begin{table}[h]
    \centering
    \begin{tabular}{c|c|c}
        parameter & value & comment \\ \hline
        $\alpha_{EW}^{-1}$ & 127.94 & NC \\
        $G_F$ & $1.1663787\times10^{-5}$ GeV$^{-2}$ & NC \& CC \\
        $m_Z$ & 91.1876 GeV & NC \& CC\\
        $\Gamma_Z$ & 2.4952 GeV & NC\\
        $m_W$ & 80.379 GeV & CC\\
        $\Gamma_W$ & 2.08 GeV & CC\\
        $\lambda$ & 0.2248 & NC \& CC\\
        $A$ & 0.8353 & NC \& CC\\
        $\rho$ & 0.1135 & NC \& CC\\
        $\eta$ & 0.3660 & NC \& CC
    \end{tabular}
    \caption{Numerical values of SM inputs for MadGraph simulations used for the validation of our implementation. }
    \label{tab:SMinput_NC}
\end{table}

\subsection{Neutral current cross sections}
In the neutral current case, we simulated SM events in full phase space for the process $p\, p \to \mu^+\, \mu^-$.
The \textit{no\_b\_mass} model restriction of the default SM UFO implementation was used.
The NNPDF30\_nnlo\_as\_0118 PDF set \cite{NNPDF:2014otw} was used with the factorization scale set to the partonic center of mass energy\footnote{The validation was done before the release of NNPDF 4.0, hence the use of a different version of PDFs compared to the official implementation.}.
In order to increase the statistics at the high-mass tail, we simulated $10^5$ events in each of the following $m_{\ell\ell}$ bins: 10 - 40 GeV, 40 - 80 GeV, 80 - 100 GeV, 100 - 200 GeV, 200 - 500 GeV, 500 - 1000 GeV, 1000 - 2000 GeV and 2000 - 3000 GeV.
The numerical values for the SM inputs were matched between MadGraph and \texttt{flavio}.
These are reported in Table \ref{tab:SMinput_NC}.
The events from different simulated $m_{\ell\ell}$ bins were combined into one distribution of 100 bins and compared to the \texttt{flavio} prediction. An agreement at the level of few $\%$ was found, in line with the statistical uncertainties associated with Monte Carlo simulations.

\begin{figure}
    \centering
    \includegraphics[width=0.9\textwidth]{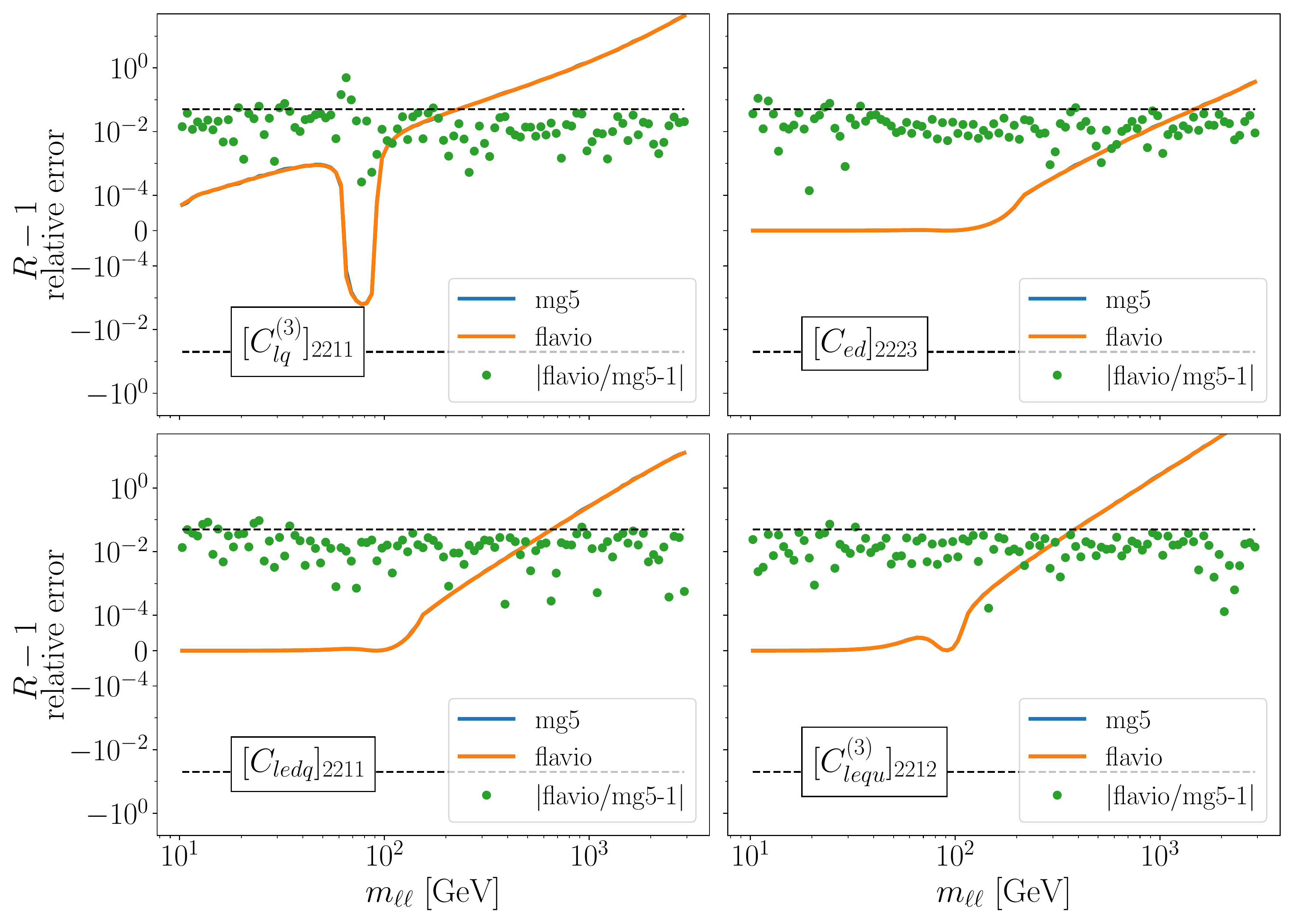}
    \caption{Validation of the neutral current NP cross-section for the process $p\, p \to \mu^+\, \mu^-$ for some selected NP operators. The prediction from MadGraph (blue) and \texttt{flavio} (orange) for the values $R_{\rm bin} - 1$ are compared. The absolute value of the relative error between the two predictions is depicted in green. The dashed black lines mark the 5\% relative error. The $y-$axis is set to a linear scale between the values $-10^{-4}$ and $10^{-4}$.}
    \label{fig:validation_NP_NC}
\end{figure}

After validating the SM cross-section, we simulated $10^5$ events in the same $m_{\ell\ell}$ bins as in the SM case for a large number of selected single Wilson coefficients set to the value 0.1 TeV$^{-2}$.
The model file used for this simulation was general SMEFTsim in the alpha scheme \cite{Brivio:2020onw}.
The SM input parameters were set to the values reported in Tab. \ref{tab:SMinput_NC}. For each of the selected coefficients $R - 1$ was compared between MadGraph and \texttt{flavio}, with $R$ the ratio of SM+NP and SM only cross sections in a particular bin, see Eq.~\eqref{eq:R-ratio}. A few representative results are shown in Fig. \ref{fig:validation_NP_NC}, demonstrating an agreement between MC simulations and \texttt{flavio} predictions in line with the statistical errors. Furthermore, we arrived at the same conclusions by considering random combinations of multiple Wilson coefficients.

\subsection{Charged current cross sections}
In the charged current case, the SM events were simulated with the general SMEFTsim model in the $M_W$ scheme with the SM limit restriction modified to include CKM.
The CKM parameters were taken from \texttt{flavio} and are summarized in Table~\ref{tab:SMinput_NC}.
In order to increase statistics in high-$p_T$ tails, $10^5$ events were simulated in the following $p_T$ bins: 50 - 100 GeV, 100 - 200 GeV, 200 - 500 GeV, 500 - 1000 GeV, and 1000 - 2000 GeV.
The {\tt NNPDF30\_nnlo\_as\_0118} PDF set was used again with the factorization scale set to the sum of the transverse mass of the final state particles.
Events from different $p_T$ bins were then combined into one distribution with 100 bins in $m_T$ and compared to prediction from \texttt{flavio}. Again an excellent agreement was found, in line with the MC statistical error of a few percent.

\begin{figure}
    \centering
    \includegraphics[width=0.9\textwidth]{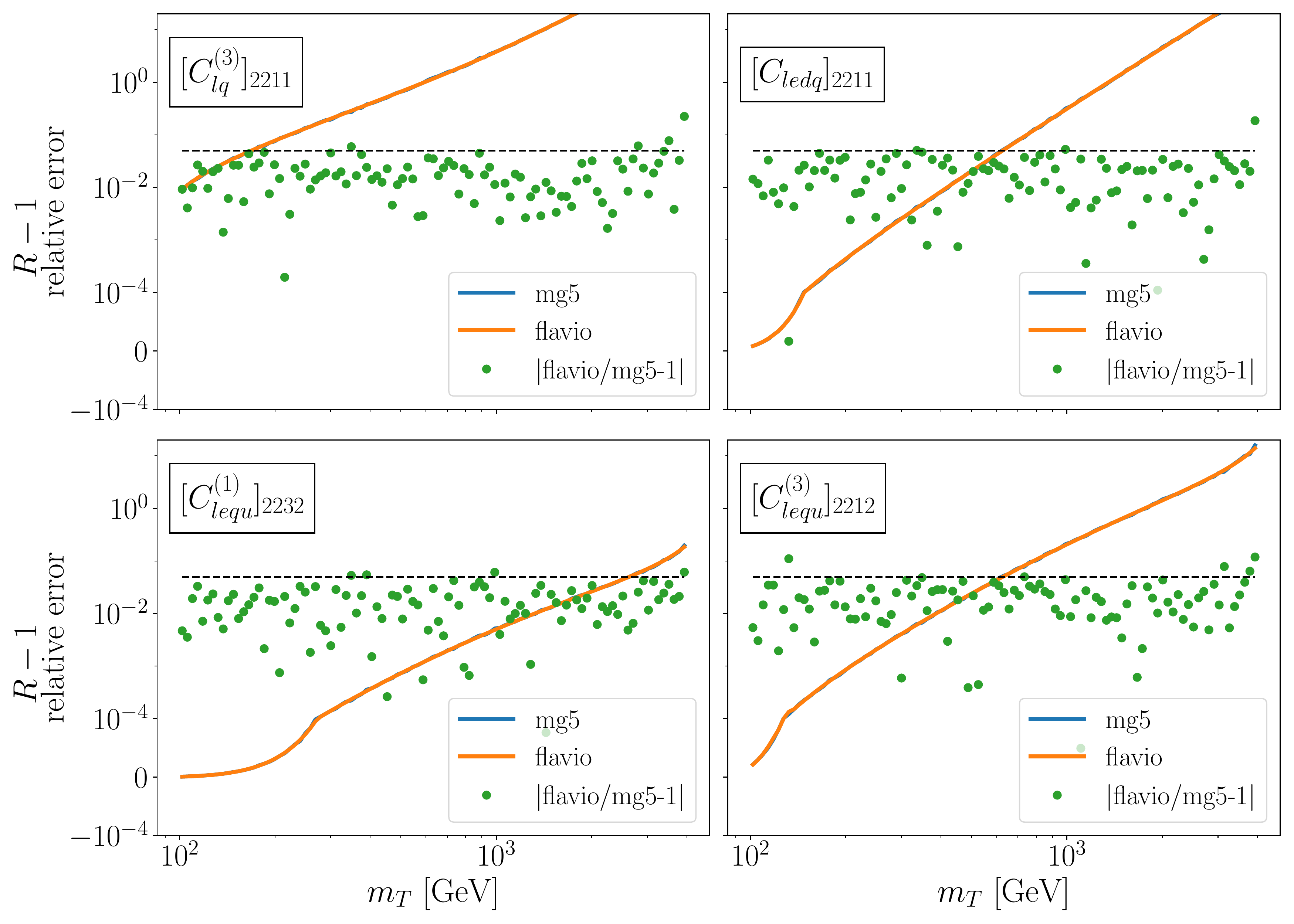}
    \caption{Validation of the charged current NP cross-section for the process $p\, p \to \mu\, \nu_\mu$ for some selected NP operators. The prediction from MadGraph (blue) and \texttt{flavio} (orange) for the values $R_{\rm bin} - 1$ are compared. The relative error between the two predictions is depicted in green. The dashed black lines mark the 5\% relative error. The $y-$axis is set to a linear scale between the values $-10^{-4}$ and $10^{-4}$.}
    \label{fig:validation_NP_CC}
\end{figure}

After validating the SM cross-section, we simulated $10^5$ events in the same $p_T$ bins for selected single Wilson coefficients set to the value 0.1 TeV$^{-2}$.
The new physics and interference contributions were simulated simultaneously again. The ratio $R- 1$ was compared between MadGraph and \texttt{flavio}, with results for selected scenarios shown in Fig. \ref{fig:validation_NP_CC}, yet again demonstrating excellent agreement. We furthermore validated the implementation by explicitly changing the values of the CKM matrix to have all entries of $\mathcal{O}(1)$, again achieving excellent agreement between $\texttt{flavio}$ predictions and MC simulations.

\subsection{Effects of phase space cuts}
\begin{figure}
    \centering
    \includegraphics[width=0.6\textwidth]{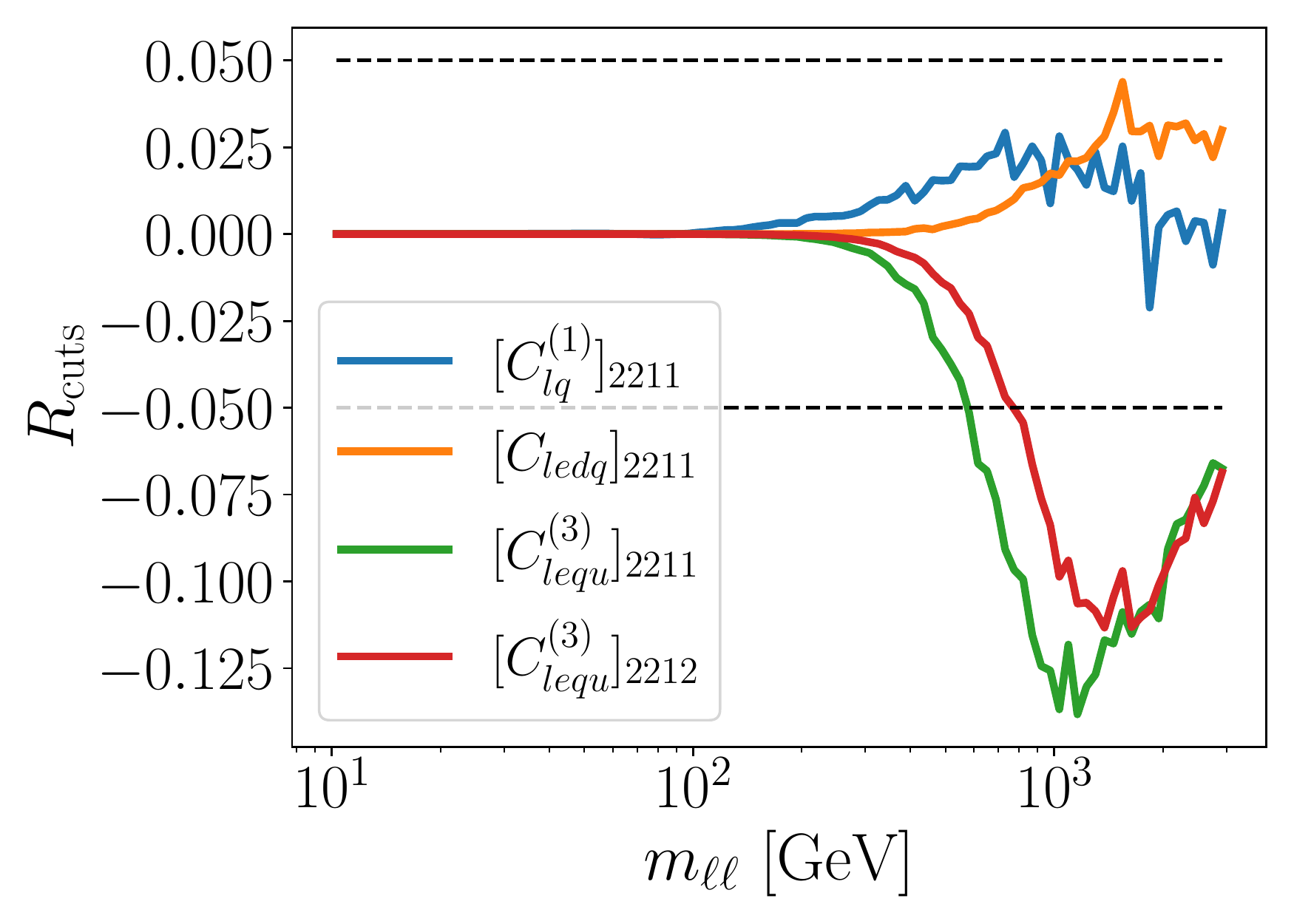}
    \caption{The relative differences between $R_{\rm bin}$ ratio before (full phase space) and after (cuts) applying the cuts in the simulated events for some selected operators. The dashed black lines mark the 5\% relative error.}
    \label{fig:validation_cuts}
\end{figure}

In order to quantify the effects of phase space cuts on the $R$-ratio, defined in Eq.~\eqref{eq:R-ratio}, we simulate events both without phase space cuts, and by including generic experimental cuts on $p_T > 20$ and $|\eta| < 2.5$.
We then calculate the ratio $R$  before (full phase space) and after applying the cuts and calculated their relative difference $R_{\rm cuts} = R^{\rm cuts}/R^{\rm fullPS} - 1$.
The results for selected operators are shown in Fig.~\ref{fig:validation_cuts}.
The effect of the cuts on the neutral current ratio $R$ is well within 5\% for all vector and scalar operators with the largest differences observed at higher values of $m_{\ell\ell}$ and for SMEFT operators involving valence quarks.  The largest difference of about 13\% was found for tensor operators involving valence quarks (as shown in the lower right plot of Fig. \ref{fig:validation_cuts}), dominated by the $\eta$ cut.
The ratio $R_{\rm cuts}$ deviates from zero at high-mass tails where the NP contribution to the cross-section becomes dominant and the angular distributions of NP are shifted inside (positive $R_{\rm cuts}$) or outside (negative $R_{\rm cuts}$) of the detector acceptance region relative to the SM.
We note that the size of this effect depends quadratically on the value of the Wilson coefficient and that the specific values of the Wilson coefficients used in our simulations are relatively large. The bounds on operators involving valence quarks are expected to be an order of magnitude smaller. Finally, the effects of cuts on the charged current new physics ratio were found to be negligible for all selected operators.

\newpage

\clearpage
\section{One-parameter $b \to s \ell \ell$ fits in WET}
\label{appendix:WEFTfit}

In this Appendix, we perform several different WET fits closely following the methodology in Ref.~\cite{Altmannshofer:2021qrr}. The results are shown in Table~\ref{tab:1d}. The WCs are defined such that the SM corresponds to $C^{(')X}_i = 0$.

\begin{table}[htbp]
\centering
\renewcommand{\arraystretch}{1.5}
\rowcolors{2}{gray!15}{white}
\addtolength{\tabcolsep}{-1pt} 
\begin{tabularx}{\textwidth}{c|cc|cc|cX}
\toprule
\rowcolor{white}
        & \multicolumn{2}{c|}{$b\to s\mu\mu$} & \multicolumn{2}{c|}{LFU, $B_s\to\mu\mu$} & \multicolumn{2}{c}{all rare $B$ decays}\\
Wilson coefficient  & best fit & pull & best fit & pull & best fit & pull\\
\midrule
$C_9^{bs\mu\mu}$ & $-0.77_{-0.21}^{+0.21}$ & $3.6\sigma$ & $-0.21_{-0.19}^{+0.17}$ & $1.2\sigma$ & $-0.42_{-0.14}^{+0.13}$ & $3.2\sigma$ \\
$C_9^{\prime bs\mu\mu}$ & $+0.29_{-0.25}^{+0.25}$ & $1.2\sigma$ & $-0.22_{-0.18}^{+0.17}$ & $1.3\sigma$ & $-0.04_{-0.13}^{+0.13}$ & $0.3\sigma$ \\
$C_{10}^{bs\mu\mu}$ & $+0.33_{-0.24}^{+0.24}$ & $1.3\sigma$ & $+0.16_{-0.11}^{+0.12}$ & $1.4\sigma$ & $+0.17_{-0.10}^{+0.10}$ & $1.8\sigma$ \\
$C_{10}^{\prime bs\mu\mu}$ & $-0.05_{-0.15}^{+0.16}$ & $0.3\sigma$ & $+0.04_{-0.12}^{+0.11}$ & $0.3\sigma$ & $+0.02_{-0.09}^{+0.09}$ & $0.2\sigma$ \\
$C_9^{bs\mu\mu}=C_{10}^{bs\mu\mu}$ & $-0.27_{-0.15}^{+0.15}$ & $1.7\sigma$ & $+0.17_{-0.18}^{+0.18}$ & $1.0\sigma$ & $-0.08_{-0.11}^{+0.11}$ & $0.7\sigma$ \\
$C_9^{bs\mu\mu}=-C_{10}^{bs\mu\mu}$ & $-0.53_{-0.13}^{+0.13}$ & $3.6\sigma$ & $-0.10_{-0.07}^{+0.07}$ & $1.4\sigma$ & $-0.17_{-0.06}^{+0.06}$ & $2.7\sigma$ \\
\midrule
$C_9^{bs\ell\ell}$ & $-0.77_{-0.21}^{+0.21}$ & $3.6\sigma$ &  &  & $-0.78_{-0.21}^{+0.21}$ & $3.7\sigma$ \\
$C_9^{\prime bs\ell\ell}$ & $+0.29_{-0.25}^{+0.25}$ & $1.2\sigma$ &  &  & $+0.30_{-0.25}^{+0.25}$ & $1.2\sigma$ \\
$C_{10}^{bs\ell\ell}$ & $+0.33_{-0.24}^{+0.24}$ & $1.3\sigma$ & $+0.21_{-0.19}^{+0.19}$ & $1.1\sigma$ & $+0.23_{-0.15}^{+0.15}$ & $1.6\sigma$ \\
$C_{10}^{\prime bs\ell\ell}$ & $-0.05_{-0.15}^{+0.16}$ & $0.3\sigma$ & $-0.21_{-0.19}^{+0.19}$ & $1.1\sigma$ & $-0.08_{-0.12}^{+0.11}$ & $0.7\sigma$ \\
$C_9^{bs\ell\ell}=C_{10}^{bs\ell\ell}$ & $-0.27_{-0.15}^{+0.15}$ & $1.7\sigma$ & $+0.21_{-0.19}^{+0.19}$ & $1.1\sigma$ & $-0.09_{-0.11}^{+0.11}$ & $0.8\sigma$ \\
$C_9^{bs\ell\ell}=-C_{10}^{bs\ell\ell}$ & $-0.53_{-0.13}^{+0.13}$ & $3.6\sigma$ & $-0.21_{-0.19}^{+0.19}$ & $1.1\sigma$ & $-0.40_{-0.11}^{+0.11}$ & $3.5\sigma$ \\
\midrule
$\left(C_S^{bs\mu\mu}=-C_P^{bs\mu\mu}\right)\times\text{GeV}$ &  &  & $-0.002_{-0.002}^{+0.001}$ & $1.1\sigma$ & $-0.001_{-0.001}^{+0.001}$ & $0.7\sigma$ \\
$\left(C_S^{\prime bs\mu\mu}=C_P^{\prime bs\mu\mu}\right)\times\text{GeV}$ &  &  & $-0.002_{-0.002}^{+0.001}$ & $1.1\sigma$ & $-0.001_{-0.001}^{+0.001}$ & $0.7\sigma$ \\
\bottomrule
\end{tabularx}
\addtolength{\tabcolsep}{-4pt} 
\caption{WET fits for scenarios involving a single real Wilson coefficient. Label $\ell$ stands for LFU NP contributions while $\mu$ for NP only in decays to muons. The best-fit values with their corresponding $1\sigma$ ranges and the pulls (in sigma) between the best-fit point and the SM point are reported. The fits are divided into three columns: ``$b\to s \mu\mu$", which includes only branching ratios and angular observables for this process, ``LFU, $B_s\to\mu\mu$", which includes only the LFU observables and the branching ratio of $B_s \to \mu^+\mu^-$, and ``all rare $B$ decays", which presents the results of the combined fit. For the scalar Wilson coefficients, both the SM-like solution and a sign-flipped solution are allowed~\cite{Altmannshofer:2017wqy}.
}
\label{tab:1d}
\end{table}
\clearpage

\section{Combining the measurements of $B_q\to\mu^+\mu^-$ branching ratios}
\label{appendix:Bsmumu}

The $B^0\to\mu^+\mu^-$ and $B_s\to\mu^+\mu^-$ branching ratios have been measured by ATLAS~\cite{ATLAS:2018cur}, CMS~\cite{CMS:2022mgd}, and LHCb~\cite{LHCb:2021awg, LHCb:2021vsc}.
However, their combined measurement presented in~\cite{LHCb:2020zud} does not include the latest results from LHCb and CMS. A combination including the latest results from LHCb was performed in~\cite{Altmannshofer:2021qrr} before CMS released an update of their measurement. Closely following~\cite{Altmannshofer:2021qrr} and~\cite{Aebischer:2019mlg}, we provide an updated combination which also includes the latest CMS results~\cite{CMS:2022mgd}.

\begin{figure}
\centering
\includegraphics[width=0.8\textwidth]{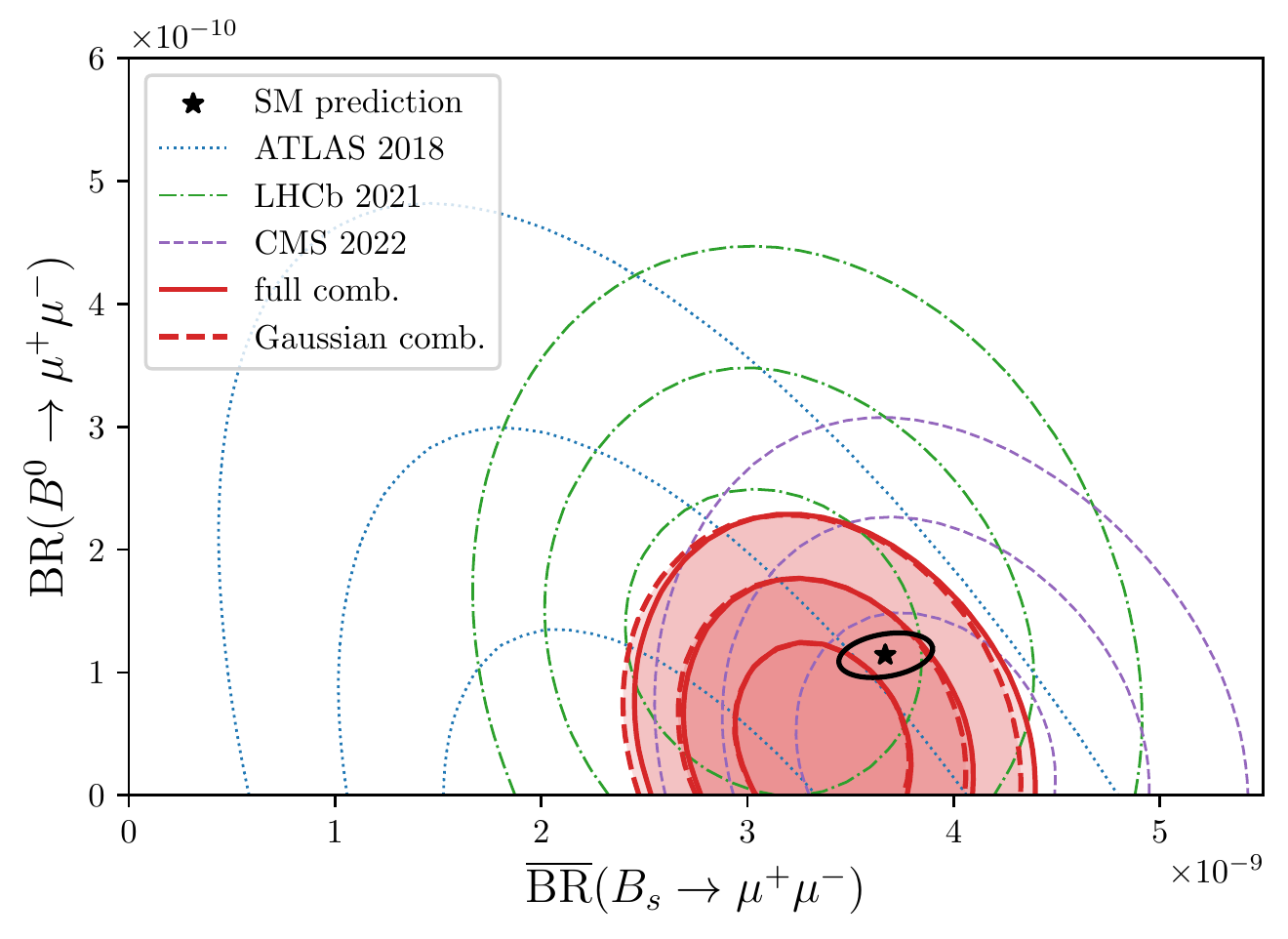}
\caption{Likelihood contours in the BR$(B^0\to\mu^+\mu^-)$ and BR$(B_s\to\mu^+\mu^-)$ plane for individual measurements from ATLAS, CMS and LHCb (thin contours), as well as our full combination (thick solid contours) and the Gaussian approximation (thick dashed contours). In addition, the SM predictions and their 1$\sigma$ correlated uncertainties are shown.}
\label{fig:Bsmumu}
\end{figure}

Due to the similar masses of the $B^0$ and $B_s$ mesons, the measurements of their branching ratios to $\mu^+\mu^-$ are correlated two-dimensional likelihoods.
We reconstruct the full non-Gaussian likelihoods from the digitised contour plots provided by the experiments and combine them, assuming that the different experiments are uncorrelated.
We approximate the full non-Gaussian combination by fitting a two-dimensional Gaussian to it.
Fig.~\ref{fig:Bsmumu} shows the individual likelihoods as thin lines, our combined likelihood as a thick solid red line, and the Gaussian approximation as a thick dashed red line. The SM predictions are also shown for comparison.

\pagebreak

Our result for the two-dimensional Gaussian approximation is
\begin{align}
  \overline{\text{BR}}(B_s\to\mu^+\mu^-)_\text{exp}
  &= (3.36\pm0.28) \times 10^{-9},
 \\
 {\text{BR}}(B^0\to\mu^+\mu^-)_\text{exp}
 &= (0.42\pm0.54) \times 10^{-10},
\end{align}
with a correlation coefficient of $\rho = -0.18$.

For the SM predictions, we use (see~\cite{Altmannshofer:2021qrr} for more details)
\begin{align}
  \overline{\text{BR}}(B_s\to\mu^+\mu^-)_\text{SM}
  &= (3.67\pm0.15) \times 10^{-9},
 \\
 {\text{BR}}(B^0\to\mu^+\mu^-)_\text{SM}
 &= (1.14\pm0.12) \times 10^{-10},
\end{align}
with a correlation coefficient of $\rho = +0.28$.

We obtain the following one-dimensional pulls by comparing the experimental likelihood with the SM prediction\footnote{%
The "one-dimensional pull" is defined as $\sqrt{-2\ln(\mathcal{L}(\vec{O}_{\rm SM})/\mathcal{L}(\vec{O}_{\rm exp}))}$, where $\mathcal{L}(\vec{O}_{\rm exp})$ and $\mathcal{L}(\vec{O}_{\rm SM})$ are the likelihoods at the experimental and SM points, respectively. The likelihood $\mathcal{L}$ is obtained by convolving the experimental likelihood with the SM uncertainties.
}:

\begin{itemize}
  \item if both branching ratios are SM-like, $1.7\sigma$,\footnote{%
  Computing the pull with two degrees of freedom from the likelihood ratio gives $1.2\sigma$.
  }
  \item if $B_s\to\mu^+\mu^-$ is SM-like and
  $B^0\to\mu^+\mu^-$ profiled over, $1.0\sigma$,
  \item if $B^0\to\mu^+\mu^-$ is SM-like and
  $B_s\to\mu^+\mu^-$ profiled over, $1.3\sigma$.
\end{itemize}

The confidence regions for the $B_s\to\mu^+\mu^-$ branching ratio when we either profile over $\text{BR}(B^0\to\mu^+\mu^-)$ or fix it to its SM central value are
\begin{align}
\overline{\text{BR}}(B_s\to\mu^+\mu^-)
&= (3.32_{-0.25}^{+0.32}) \times 10^{-9}~,
&& \text{BR}(B^0\to\mu^+\mu^-) \text{ profiled,}
 \\
 \overline{\text{BR}}(B_s\to\mu^+\mu^-)
&= (3.27_{-0.23}^{+0.32}) \times 10^{-9}~,
 && \text{BR}(B^0\to\mu^+\mu^-) \text{ SM-like.}
\end{align}
Analogously, for $B^0\to\mu^+\mu^-$ we get
\begin{align}
{\text{BR}}(B^0\to\mu^+\mu^-)
&= (0.48_{-0.31}^{+0.52}) \times 10^{-10}~,
&& \overline{\text{BR}}(B_s\to\mu^+\mu^-) \text{ profiled,}
 \\
 {\text{BR}}(B^0\to\mu^+\mu^-)
&= (0.37_{-0.24}^{+0.54}) \times 10^{-10}~,
 && \overline{\text{BR}}(B_s\to\mu^+\mu^-) \text{ SM-like.}
\end{align}

\newpage

\section{$B_d$ and $B_s$ bag parameters}
\label{appendix:BdBsBag}
\subsection{Definitions}

The hadronic matrix elements relevant for $B_q$ mixing (with $q=d,s$) can be expressed in terms of so-called bag parameters $B^{(i)}_{B_q}$ defined by~\cite{Gabbiani:1996hi}.
In the following, the convention from~\cite{FermilabLattice:2016ipl} is adapted, which reads
\begin{align}
   \langle\mathcal{O}^q_1\rangle(\mu) &= c_1 f^2_{B_q} M^2_{B_q} B^{(1)}_{B_q}(\mu) ,
    \label{eq:Bq_1} \\
    \langle\mathcal{O}^q_i\rangle(\mu) &= c_i \left( \frac{M_{B_q}}{m_b(\mu)+m_q(\mu)} \right)^2
        f^2_{B_q} M^2_{B_q} B^{(i)}_{B_q}(\mu) ,  \quad i = 2,3,
    \label{eq:Bq_23} \\
    \langle\mathcal{O}^q_i\rangle(\mu) &= c_i \left[ \left( \frac{M_{B_q}}{m_b(\mu)+m_q(\mu)}\right)^2 + d_i \right]
        f^2_{B_q} M^2_{B_q} B^{(i)}_{B_q}(\mu) , \quad i = 4,5,
    \label{eq:Bq_45}\\
    \langle\tilde{\mathcal{O}}^q_i\rangle(\mu) &= \langle\mathcal{O}^q_i\rangle(\mu)  ,  \quad i = 1,2,3,
\end{align}
where $c_i = \{2/3, -5/12, 1/12, 1/2, 1/6\}$, $d_4=1/6$, and $d_5=3/2$,
and the operators $\mathcal{O}_i^q$ and $\tilde{\mathcal{O}}_i^q$ are given by
\begin{subequations} \label{eq:Oi}
\begin{align}
    \mathcal{O}_1^q &=(\bar{b}^\alpha\gamma_\mu P_L q^\alpha) \, (\bar{b}^\beta\gamma_\mu P_L q^\beta) , \\
    \mathcal{O}_2^q &=(\bar{b}^\alpha P_L q^\alpha) \, (\bar{b}^\beta P_L q^\beta) , \\
    \mathcal{O}_3^q &=(\bar{b}^\alpha P_L q^\beta) \, (\bar{b}^\beta P_L q^\alpha) , \\
    \mathcal{O}_4^q &=(\bar{b}^\alpha P_L q^\alpha) \, (\bar{b}^\beta P_R q^\beta) ,  \\
    \mathcal{O}_5^q &=(\bar{b}^\alpha P_L q^\beta) \, (\bar{b}^\beta P_R q^\alpha) , \\
    \tilde{\mathcal{O}}_1^q&=(\bar{b}^\alpha\gamma_\mu P_R q^\alpha) \ (\bar{b}^\beta\gamma_\mu P_R q^\beta) ,  \\
    \tilde{\mathcal{O}}_2^q&=(\bar{b}^\alpha P_R q^\alpha) \ (\bar{b}^\beta P_R q^\beta) , \\
    \tilde{\mathcal{O}}_3^q&=(\bar{b}^\alpha P_R q^\beta) \ (\bar{b}^\beta P_R q^\alpha) .
\end{align}
\end{subequations}

\subsection{Results from lattice QCD and sum rules}

The most recent values of the bag parameters have been determined from Lattice QCD by the FNAL/MILC~\cite{FermilabLattice:2016ipl} and HPQCD~\cite{Dowdall:2019bea} collaborations and from sum rules by King, Lenz, and Rauh~(KLR)~\cite{King:2019lal}.

FNAL/MILC provides results for both the $B_d$ and $B_s$ bag parameters including their correlations in~\cite{FermilabLattice:2016ipl}.
For convenience, we show these results in table~\ref{tab:FNAL/MILC}.
\begin{table}[h]
\centering
\small
\renewcommand{\arraystretch}{1.7}
\addtolength{\tabcolsep}{-3pt} 
\begin{tabular}{l|r}
\multicolumn{2}{c}{FNAL/MILC 2016}\\
\hline
$B^{(1)}_{B_d}$ & $0.913 \pm 0.086$ \\
$B^{(2)}_{B_d}$ & $0.761 \pm 0.076$ \\
$B^{(3)}_{B_d}$ & $1.070 \pm 0.216$ \\
$B^{(4)}_{B_d}$ & $1.040 \pm 0.087$ \\
$B^{(5)}_{B_d}$ & $0.964 \pm 0.102$ \\
$B^{(1)}_{B_s}$ & $0.952 \pm 0.066$ \\
$B^{(2)}_{B_s}$ & $0.806 \pm 0.059$ \\
$B^{(3)}_{B_s}$ & $1.100 \pm 0.155$ \\
$B^{(4)}_{B_s}$ & $1.022 \pm 0.066$ \\
$B^{(5)}_{B_s}$ & $0.943 \pm 0.075$ \\
\end{tabular}\quad
\begin{tabular}{l|rrrrrrrrrr}
& $B^{(1)}_{B_d}$ & $B^{(2)}_{B_d}$ & $B^{(3)}_{B_d}$ & $B^{(4)}_{B_d}$ & $B^{(5)}_{B_d}$ & $B^{(1)}_{B_s}$ & $B^{(2)}_{B_s}$ & $B^{(3)}_{B_s}$ & $B^{(4)}_{B_s}$ & $B^{(5)}_{B_s}$ \\
\hline
$B^{(1)}_{B_d}$ & $1\phantom{.000}$ & $0.504$ & $0.162$ & $0.494$ & $0.422$ & $0.754$ & $0.311$ & $0.077$ & $0.264$ & $0.246$ \\
$B^{(2)}_{B_d}$ &  & $1\phantom{.000}$ & $0.282$ & $0.494$ & $0.389$ & $0.311$ & $0.766$ & $0.201$ & $0.283$ & $0.226$ \\
$B^{(3)}_{B_d}$ &  &  & $1\phantom{.000}$ & $0.225$ & $0.148$ & $0.068$ & $0.179$ & $0.929$ & $0.115$ & $0.063$ \\
$B^{(4)}_{B_d}$ &  &  &  & $1\phantom{.000}$ & $0.528$ & $0.279$ & $0.297$ & $0.134$ & $0.705$ & $0.332$ \\
$B^{(5)}_{B_d}$ &  &  &  &  & $1\phantom{.000}$ & $0.247$ & $0.228$ & $0.074$ & $0.318$ & $0.797$ \\
$B^{(1)}_{B_s}$ &  &  &  &  &  & $1\phantom{.000}$ & $0.575$ & $0.215$ & $0.560$ & $0.486$ \\
$B^{(2)}_{B_s}$ &  &  &  &  &  &  & $1\phantom{.000}$ & $0.329$ & $0.581$ & $0.466$ \\
$B^{(3)}_{B_s}$ &  &  &  &  &  &  &  & $1\phantom{.000}$ & $0.278$ & $0.195$ \\
$B^{(4)}_{B_s}$ &  &  &  &  &  &  &  &  & $1\phantom{.000}$ & $0.581$ \\
$B^{(5)}_{B_s}$ &  &  &  &  &  &  &  &  &  & $1\phantom{.000}$ \\
\end{tabular}
\addtolength{\tabcolsep}{+2pt} 
\caption{$B_d$ and $B_s$ bag parameters including their uncertainties and correlations from FNAL/MILC~\cite{FermilabLattice:2016ipl}.}
\label{tab:FNAL/MILC}
\end{table}

HPQCD does not provide combined results including correlations for both the $B_d$ and $B_s$ bag parameters.
However, they provide results for the $B_s$ bag parameters and for the ratios $B^{(i)}_{B_s}/B^{(i)}_{B_d}$, for which the correlations are given separately.
From this information, we obtain combined results including correlations for both the $B_d$ and $B_s$ bag parameters in the following way:
\begin{itemize}
 \item From the central values, uncertainties, and correlations given in~\cite{Dowdall:2019bea}, we construct two five-dimensional multivariate normal distributions, one for the $B_s$ bag parameters and one for the ratios $B^{(i)}_{B_s}/B^{(i)}_{B_d}$.
 \item We draw $N=5\times 10^6$ samples from the distributions.
 \item For each sample we compute $B^{(i)}_{B_d}$ to obtain combined samples of $B^{(i)}_{B_d}$ and $B^{(i)}_{B_s}$.
 \item From the combined samples, we compute the ten-dimensional mean and covariance matrix.
 \item From the covariance matrix, we obtain the uncertainties and correlations.
\end{itemize}
The result of this procedure is shown in table~\ref{tab:HPQCD}. As expected, the central values and uncertainties of $B^{(i)}_{B_s}$ as well as the correlations among the $B^{(i)}_{B_s}$ are (up to few rounding errors on the last digit) exactly those given in~\cite{Dowdall:2019bea}.
\begin{table}[h]
\centering
\small
\renewcommand{\arraystretch}{1.7}
\addtolength{\tabcolsep}{-3pt} 
\begin{tabular}{l|r}
\multicolumn{2}{c}{HPQCD 2019}\\
\hline
$B^{(1)}_{B_d}$ & $0.807 \pm 0.040$ \\
$B^{(2)}_{B_d}$ & $0.769 \pm 0.044$ \\
$B^{(3)}_{B_d}$ & $0.748 \pm 0.057$ \\
$B^{(4)}_{B_d}$ & $1.078 \pm 0.055$ \\
$B^{(5)}_{B_d}$ & $0.973 \pm 0.046$ \\
$B^{(1)}_{B_s}$ & $0.813 \pm 0.035$ \\
$B^{(2)}_{B_s}$ & $0.817 \pm 0.043$ \\
$B^{(3)}_{B_s}$ & $0.816 \pm 0.057$ \\
$B^{(4)}_{B_s}$ & $1.033 \pm 0.047$ \\
$B^{(5)}_{B_s}$ & $0.941 \pm 0.038$ \\
\end{tabular}\quad
\begin{tabular}{l|rrrrrrrrrr}
& $B^{(1)}_{B_d}$ & $B^{(2)}_{B_d}$ & $B^{(3)}_{B_d}$ & $B^{(4)}_{B_d}$ & $B^{(5)}_{B_d}$ & $B^{(1)}_{B_s}$ & $B^{(2)}_{B_s}$ & $B^{(3)}_{B_s}$ & $B^{(4)}_{B_s}$ & $B^{(5)}_{B_s}$ \\
\hline
$B^{(1)}_{B_d}$ & $1\phantom{.000}$ & $0.108$ & $0.002$ & $0.043$ & $0.041$ & $0.865$ & $0.053$ & $0.010$ & $0.032$ & $0.033$ \\
$B^{(2)}_{B_d}$ &  & $1\phantom{.000}$ & $0.172$ & $0.204$ & $0.127$ & $0.056$ & $0.917$ & $0.163$ & $0.213$ & $0.137$ \\
$B^{(3)}_{B_d}$ &  &  & $1\phantom{.000}$ & $0.146$ & $0.088$ & $0.010$ & $0.162$ & $0.913$ & $0.156$ & $0.098$ \\
$B^{(4)}_{B_d}$ &  &  &  & $1\phantom{.000}$ & $0.276$ & $0.033$ & $0.209$ & $0.153$ & $0.898$ & $0.230$ \\
$B^{(5)}_{B_d}$ &  &  &  &  & $1\phantom{.000}$ & $0.033$ & $0.129$ & $0.093$ & $0.220$ & $0.861$ \\
$B^{(1)}_{B_s}$ &  &  &  &  &  & $1\phantom{.000}$ & $0.061$ & $0.012$ & $0.037$ & $0.038$ \\
$B^{(2)}_{B_s}$ &  &  &  &  &  &  & $1\phantom{.000}$ & $0.178$ & $0.233$ & $0.150$ \\
$B^{(3)}_{B_s}$ &  &  &  &  &  &  &  & $1\phantom{.000}$ & $0.170$ & $0.108$ \\
$B^{(4)}_{B_s}$ &  &  &  &  &  &  &  &  & $1\phantom{.000}$ & $0.256$ \\
$B^{(5)}_{B_s}$ &  &  &  &  &  &  &  &  &  & $1\phantom{.000}$ \\
\end{tabular}
\addtolength{\tabcolsep}{+2pt} 
\caption{$B_d$ and $B_s$ bag parameters including their uncertainties and correlations obtained from the $B_s$ bag parameters and the ratios $B^{(i)}_{B_s}/B^{(i)}_{B_d}$ from HPQCD~\cite{Dowdall:2019bea}, using the procedure described in the text.}
\label{tab:HPQCD}
\end{table}

KLR does not provide combined results including correlations for both the $B_d$ and $B_s$ bag parameters.
They provide central values and uncertainties for the $B_s$ bag parameters and for the ratios $B^{(i)}_{B_s}/B^{(i)}_{B_d}$, but they do not give any correlations.
However, when combining the results for $B_s$ bag parameters and $B^{(i)}_{B_s}/B^{(i)}_{B_d}$ ratios in order to obtain results for the $B_d$ bag parameters, the fact that the ratios $B^{(i)}_{B_s}/B^{(i)}_{B_d}$ have significantly smaller uncertainties than the $B_s$ bag parameters leads to strong correlations between the resulting uncertainties of $B_d$ and $B_s$ bag parameters.
We obtain combined results including correlations for both the $B_d$ and $B_s$ bag parameters in the following way:
\begin{itemize}
 \item The results in~\cite{King:2019lal} are given with asymmetric uncertainties arising from the non-linear dependence of the output quantities on the input parameters.
 As explained in~\cite{DAgostini:2004kis}, in the presence of such asymmetries, the central values of the output quantities obtained from the central values of the input parameters are shifted relative to the actual expectation values.
 To obtain expectation values and standard deviations, we use the procedure described in~\cite{DAgostini:2004kis} for treating such asymmetric uncertainties.
 \item From the expectation values and standard deviations we construct ten normal distributions, five for the $B_s$ bag parameters and five for the ratios $B^{(i)}_{B_s}/B^{(i)}_{B_d}$.
 \item We draw $N=5\times 10^6$ samples from the distributions.
 \item For each sample we compute $B^{(i)}_{B_d}$ to obtain combined samples of $B^{(i)}_{B_d}$ and $B^{(i)}_{B_s}$.
 \item From the combined samples, we compute the ten-dimensional mean and covariance matrix.
 \item From the covariance matrix, we obtain the uncertainties and correlations.
\end{itemize}
\begin{table}[h]
\centering
\small
\renewcommand{\arraystretch}{1.7}
\addtolength{\tabcolsep}{-2pt} 
\begin{tabular}{l|r}
\multicolumn{2}{c}{KLR 2019}\\
\hline
$B^{(1)}_{B_d}$ & $0.867 \pm 0.053$ \\
$B^{(2)}_{B_d}$ & $0.845 \pm 0.075$ \\
$B^{(3)}_{B_d}$ & $0.821 \pm 0.150$ \\
$B^{(4)}_{B_d}$ & $1.054 \pm 0.089$ \\
$B^{(5)}_{B_d}$ & $1.078 \pm 0.080$ \\
$B^{(1)}_{B_s}$ & $0.856 \pm 0.052$ \\
$B^{(2)}_{B_s}$ & $0.856 \pm 0.076$ \\
$B^{(3)}_{B_s}$ & $0.914 \pm 0.159$ \\
$B^{(4)}_{B_s}$ & $1.043 \pm 0.088$ \\
$B^{(5)}_{B_s}$ & $1.054 \pm 0.077$ \\
\end{tabular}\hskip6pt
\addtolength{\tabcolsep}{-2pt} 
\begin{tabular}{l|rrrrrrrrrr}
& $B^{(1)}_{B_d}$ & $B^{(2)}_{B_d}$ & $B^{(3)}_{B_d}$ & $B^{(4)}_{B_d}$ & $B^{(5)}_{B_d}$ & $B^{(1)}_{B_s}$ & $B^{(2)}_{B_s}$ & $B^{(3)}_{B_s}$ & $B^{(4)}_{B_s}$ & $B^{(5)}_{B_s}$ \\
\hline
$B^{(1)}_{B_d}$ & $1\phantom{.000}$ & $0.000$ & $0.001$ & $0.000$ & $0.001$ & $0.992$ & $0.000$ & $0.000$ & $0.000$ & $0.001$ \\
$B^{(2)}_{B_d}$ &  & $1\phantom{.000}$ & $0.001$ & $0.000$ & $0.000$ & $0.000$ & $0.995$ & $0.000$ & $0.000$ & $0.000$ \\
$B^{(3)}_{B_d}$ &  &  & $1\phantom{.000}$ & $0.001$ & $-0.001$ & $0.001$ & $0.000$ & $0.952$ & $0.001$ & $-0.001$ \\
$B^{(4)}_{B_d}$ &  &  &  & $1\phantom{.000}$ & $0.000$ & $0.000$ & $0.000$ & $0.001$ & $0.996$ & $0.000$ \\
$B^{(5)}_{B_d}$ &  &  &  &  & $1\phantom{.000}$ & $0.001$ & $0.000$ & $-0.001$ & $0.000$ & $0.987$ \\
$B^{(1)}_{B_s}$ &  &  &  &  &  & $1\phantom{.000}$ & $0.000$ & $0.000$ & $0.000$ & $0.001$ \\
$B^{(2)}_{B_s}$ &  &  &  &  &  &  & $1\phantom{.000}$ & $0.000$ & $0.000$ & $0.000$ \\
$B^{(3)}_{B_s}$ &  &  &  &  &  &  &  & $1\phantom{.000}$ & $0.001$ & $-0.001$ \\
$B^{(4)}_{B_s}$ &  &  &  &  &  &  &  &  & $1\phantom{.000}$ & $0.001$ \\
$B^{(5)}_{B_s}$ &  &  &  &  &  &  &  &  &  & $1\phantom{.000}$ \\
\end{tabular}
\addtolength{\tabcolsep}{+4pt} 
\caption{$B_d$ and $B_s$ bag parameters including their uncertainties and correlations obtained from the $B_s$ bag parameters and the ratios $B^{(i)}_{B_s}/B^{(i)}_{B_d}$ from KLR~\cite{King:2019lal}, using the procedure described in the text.}
\label{tab:KLR}
\end{table}
The result of this procedure is shown in table~\ref{tab:KLR}. The central values of  the $B^{(i)}_{B_s}$ are slightly shifted compared to those given in~\cite{King:2019lal} due to our treatment of asymmetric uncertainties following~\cite{DAgostini:2004kis}.
As expected from the small uncertainties of the ratios $B^{(i)}_{B_s}/B^{(i)}_{B_d}$, the uncertainties of $B^{(i)}_{B_d}$ and $B^{(i)}_{B_s}$ bag parameters are strongly correlated.

\subsection{Combination}
Having brought the results of FNAL/MILC~\cite{FermilabLattice:2016ipl}, HPQCD~\cite{Dowdall:2019bea} and KLR~\cite{King:2019lal} all into the same form of values with uncertainties and correlations, it is now straightforward to combine them.
The combination taking into account all correlations is given in table~\ref{tab:combination}.

The results of FNAL/MILC~\cite{FermilabLattice:2016ipl} are based on a lattice study with only $N_f = 2 + 1$ dynamic quark flavours, missing effects of a dynamical charm quark. These potentially important effects are taken into account by HPQCD~\cite{Dowdall:2019bea}, which uses $N_f = 2 + 1 + 1$. Therefore, we also perform a combination of only HPQCD~\cite{Dowdall:2019bea} and KLR~\cite{King:2019lal}, which excludes $N_f = 2 + 1$ results. This combination is shown in table~\ref{tab:combination_HPQCD_KLR}.

In~\cite{DiLuzio:2019jyq}, a combination similar to the one shown in table~\ref{tab:combination} was performed, but without taking correlations into account. Its central values and uncertainties differ significantly from those shown in table~\ref{tab:combination}.
For comparison, we perform a weighted average of the bag parameters given in tables~\ref{tab:FNAL/MILC}, \ref{tab:HPQCD}, and \ref{tab:KLR} neglecting all correlations.
This yields
\begin{equation}
 \begin{aligned}
B^{(1)}_{B_d}\big|_\text{no\ corr.} &= 0.839 \pm 0.030\,, &\qquad&&
B^{(1)}_{B_s}\big|_\text{no\ corr.} &= 0.846 \pm 0.026\,, \\
B^{(2)}_{B_d}\big|_\text{no\ corr.} &= 0.783 \pm 0.034\,, &\qquad&&
B^{(2)}_{B_s}\big|_\text{no\ corr.} &= 0.821 \pm 0.032\,, \\
B^{(3)}_{B_d}\big|_\text{no\ corr.} &= 0.775 \pm 0.052\,, &\qquad&&
B^{(3)}_{B_s}\big|_\text{no\ corr.} &= 0.857 \pm 0.051\,, \\
B^{(4)}_{B_d}\big|_\text{no\ corr.} &= 1.064 \pm 0.041\,, &\qquad&&
B^{(4)}_{B_s}\big|_\text{no\ corr.} &= 1.032 \pm 0.035\,, \\
B^{(5)}_{B_d}\big|_\text{no\ corr.} &= 0.995 \pm 0.037\,, &\qquad&&
B^{(5)}_{B_s}\big|_\text{no\ corr.} &= 0.960 \pm 0.031\,, \\
 \end{aligned}
\end{equation}
which, up to small differences that might be related to our treatment of the asymmetric uncertainties of~\cite{King:2019lal}, is very similar to the result of~\cite{DiLuzio:2019jyq}. In particular, it is also significantly different from our full combination that takes into account all correlations.
This suggests that the correlations are crucial and cannot be neglected.
%
\begin{figure}[t]
\centering
\includegraphics[width=\textwidth]{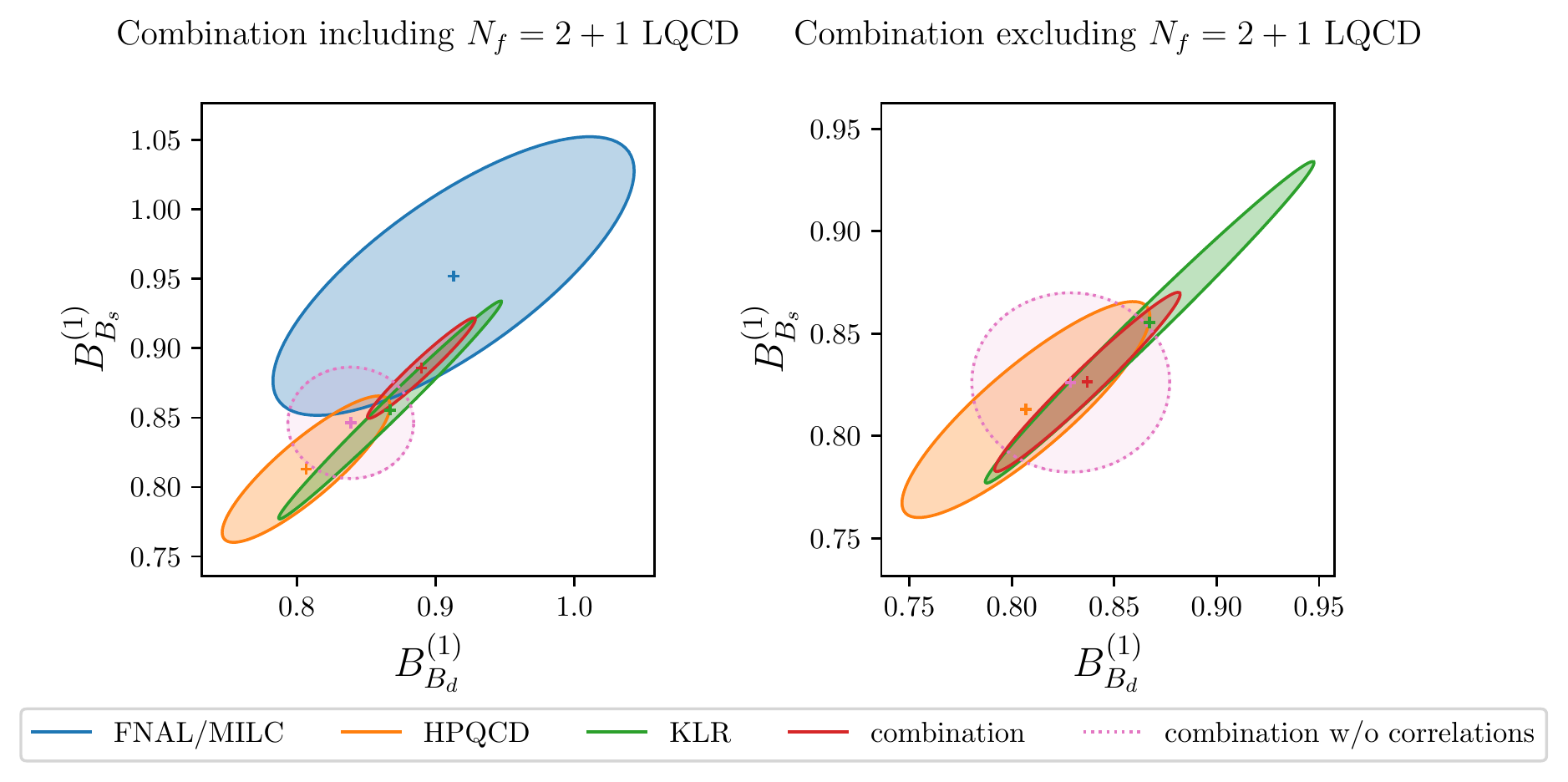}
\caption{Individual results for bag parameters $B_{B_{d,s}}^{(1)}$ from FNAL/MILC, HPQCD, and KLR as well their combinations either taking into account all correlations (solid red) or neglecting correlations (dotted pink). The combinations in the left column include the ${N_f=2+1}$~LQCD results, while the combinations in the right column exclude them. All contours represent $1\sigma$ regions with two degrees of freedom.}
\label{fig:bag_parameters_1}
\end{figure}
%
To illustrate this point, we show the strong correlations between the $B^{(1)}_{B_d}$ and $B^{(1)}_{B_s}$ bag parameters in Fig.~\ref{fig:bag_parameters_1} and those between the $B^{(i)}_{B_d}$ and $B^{(i)}_{B_s}$ bag parameters with $i\in\{2,3,4,5\}$ in Fig.~\ref{fig:bag_parameters_2-4}. In both figures, the left (right) column represents the combinations including (excluding) $N_f = 2 + 1$ LQCD results.
It can clearly be seen that the combinations that neglect correlations (dashed pink contours) not only fail to capture the shape of the confidence intervals, but also have central values that are significantly different from those of the full combination (solid red contours).

\begin{table}[h]
\centering
\small
\renewcommand{\arraystretch}{1.7}
\addtolength{\tabcolsep}{-3pt} 
\begin{tabular}{l|r}
\multicolumn{2}{c}{Full combination}\\
\hline
$B^{(1)}_{B_d}$ & $0.890 \pm 0.026$ \\
$B^{(2)}_{B_d}$ & $0.773 \pm 0.027$ \\
$B^{(3)}_{B_d}$ & $0.779 \pm 0.049$ \\
$B^{(4)}_{B_d}$ & $0.998 \pm 0.032$ \\
$B^{(5)}_{B_d}$ & $0.955 \pm 0.032$ \\
$B^{(1)}_{B_s}$ & $0.886 \pm 0.024$ \\
$B^{(2)}_{B_s}$ & $0.788 \pm 0.025$ \\
$B^{(3)}_{B_s}$ & $0.852 \pm 0.044$ \\
$B^{(4)}_{B_s}$ & $0.979 \pm 0.029$ \\
$B^{(5)}_{B_s}$ & $0.926 \pm 0.028$ \\
\end{tabular}\quad
\begin{tabular}{l|rrrrrrrrrr}
& $B^{(1)}_{B_d}$ & $B^{(2)}_{B_d}$ & $B^{(3)}_{B_d}$ & $B^{(4)}_{B_d}$ & $B^{(5)}_{B_d}$ & $B^{(1)}_{B_s}$ & $B^{(2)}_{B_s}$ & $B^{(3)}_{B_s}$ & $B^{(4)}_{B_s}$ & $B^{(5)}_{B_s}$ \\
\hline
$B^{(1)}_{B_d}$ & $1\phantom{.000}$ & $0.243$ & $0.062$ & $0.198$ & $0.135$ & $0.971$ & $0.270$ & $0.092$ & $0.218$ & $0.166$ \\
$B^{(2)}_{B_d}$ &  & $1\phantom{.000}$ & $0.181$ & $0.295$ & $0.170$ & $0.272$ & $0.973$ & $0.227$ & $0.319$ & $0.203$ \\
$B^{(3)}_{B_d}$ &  &  & $1\phantom{.000}$ & $0.142$ & $0.083$ & $0.076$ & $0.190$ & $0.929$ & $0.151$ & $0.097$ \\
$B^{(4)}_{B_d}$ &  &  &  & $1\phantom{.000}$ & $0.263$ & $0.227$ & $0.327$ & $0.181$ & $0.975$ & $0.304$ \\
$B^{(5)}_{B_d}$ &  &  &  &  & $1\phantom{.000}$ & $0.153$ & $0.187$ & $0.105$ & $0.274$ & $0.950$ \\
$B^{(1)}_{B_s}$ &  &  &  &  &  & $1\phantom{.000}$ & $0.308$ & $0.110$ & $0.253$ & $0.190$ \\
$B^{(2)}_{B_s}$ &  &  &  &  &  &  & $1\phantom{.000}$ & $0.251$ & $0.356$ & $0.226$ \\
$B^{(3)}_{B_s}$ &  &  &  &  &  &  &  & $1\phantom{.000}$ & $0.196$ & $0.126$ \\
$B^{(4)}_{B_s}$ &  &  &  &  &  &  &  &  & $1\phantom{.000}$ & $0.327$ \\
$B^{(5)}_{B_s}$ &  &  &  &  &  &  &  &  &  & $1\phantom{.000}$ \\
\end{tabular}
\addtolength{\tabcolsep}{+2pt} 
\caption{$B_d$ and $B_s$ bag parameters including their uncertainties and correlations obtained from combining the results of FNAL/MILC~\cite{FermilabLattice:2016ipl}, HPQCD~\cite{Dowdall:2019bea}, and KLR~\cite{King:2019lal} given in tables~\ref{tab:FNAL/MILC}, \ref{tab:HPQCD}, and \ref{tab:KLR}.}
\label{tab:combination}
\end{table}

\begin{table}[h]
\centering
\small
\renewcommand{\arraystretch}{1.7}
\addtolength{\tabcolsep}{-3pt} 
\begin{tabular}{l|r}
\multicolumn{2}{c}{HPQCD + KLR}\\
\hline
$B^{(1)}_{B_d}$ & $0.837 \pm 0.030$ \\
$B^{(2)}_{B_d}$ & $0.808 \pm 0.037$ \\
$B^{(3)}_{B_d}$ & $0.761 \pm 0.053$ \\
$B^{(4)}_{B_d}$ & $1.054 \pm 0.042$ \\
$B^{(5)}_{B_d}$ & $0.986 \pm 0.037$ \\
$B^{(1)}_{B_s}$ & $0.826 \pm 0.029$ \\
$B^{(2)}_{B_s}$ & $0.824 \pm 0.037$ \\
$B^{(3)}_{B_s}$ & $0.830 \pm 0.053$ \\
$B^{(4)}_{B_s}$ & $1.040 \pm 0.041$ \\
$B^{(5)}_{B_s}$ & $0.964 \pm 0.034$ \\
\end{tabular}\quad
\begin{tabular}{l|rrrrrrrrrr}
& $B^{(1)}_{B_d}$ & $B^{(2)}_{B_d}$ & $B^{(3)}_{B_d}$ & $B^{(4)}_{B_d}$ & $B^{(5)}_{B_d}$ & $B^{(1)}_{B_s}$ & $B^{(2)}_{B_s}$ & $B^{(3)}_{B_s}$ & $B^{(4)}_{B_s}$ & $B^{(5)}_{B_s}$ \\
\hline
$B^{(1)}_{B_d}$ & $1\phantom{.000}$ & $0.045$ & $0.003$ & $0.023$ & $0.025$ & $0.976$ & $0.044$ & $0.006$ & $0.023$ & $0.025$ \\
$B^{(2)}_{B_d}$ &  & $1\phantom{.000}$ & $0.133$ & $0.167$ & $0.099$ & $0.043$ & $0.983$ & $0.135$ & $0.170$ & $0.105$ \\
$B^{(3)}_{B_d}$ &  &  & $1\phantom{.000}$ & $0.120$ & $0.070$ & $0.005$ & $0.131$ & $0.921$ & $0.121$ & $0.073$ \\
$B^{(4)}_{B_d}$ &  &  &  & $1\phantom{.000}$ & $0.186$ & $0.023$ & $0.170$ & $0.129$ & $0.984$ & $0.195$ \\
$B^{(5)}_{B_d}$ &  &  &  &  & $1\phantom{.000}$ & $0.025$ & $0.101$ & $0.075$ & $0.186$ & $0.952$ \\
$B^{(1)}_{B_s}$ &  &  &  &  &  & $1\phantom{.000}$ & $0.043$ & $0.006$ & $0.024$ & $0.026$ \\
$B^{(2)}_{B_s}$ &  &  &  &  &  &  & $1\phantom{.000}$ & $0.138$ & $0.173$ & $0.106$ \\
$B^{(3)}_{B_s}$ &  &  &  &  &  &  &  & $1\phantom{.000}$ & $0.132$ & $0.079$ \\
$B^{(4)}_{B_s}$ &  &  &  &  &  &  &  &  & $1\phantom{.000}$ & $0.198$ \\
$B^{(5)}_{B_s}$ &  &  &  &  &  &  &  &  &  & $1\phantom{.000}$ \\
\end{tabular}
\addtolength{\tabcolsep}{+2pt} 
\caption{$B_d$ and $B_s$ bag parameters including their uncertainties and correlations obtained from combining the results of HPQCD~\cite{Dowdall:2019bea} and KLR~\cite{King:2019lal} given in tables~\ref{tab:HPQCD} and~\ref{tab:KLR}.}
\label{tab:combination_HPQCD_KLR}
\end{table}

\afterpage{%
\begin{figure}[h]
\centering
\includegraphics[width=\textwidth]{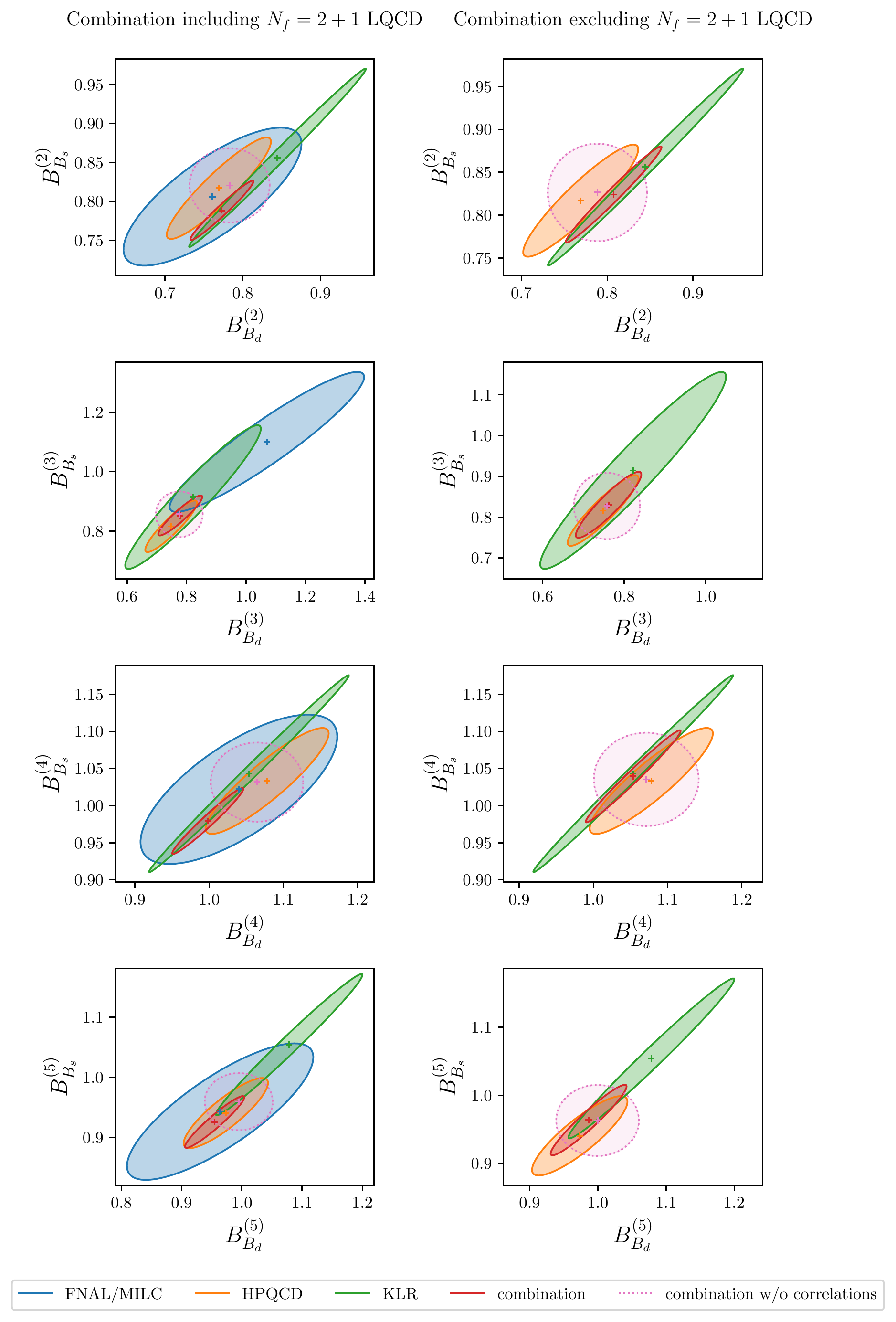}
\caption{Bag parameters $B_{B_{d,s}}^{(i)}$ with $i\in\{2,3,4,5\}$ as described in caption of Fig.~\ref{fig:bag_parameters_1}.}
\label{fig:bag_parameters_2-4}
\end{figure}
\clearpage
}

\newpage

\bibliography{references}
\bibliographystyle{JHEP}

\end{document}